\documentclass[iop]{emulateapj}

\UseRawInputEncoding

\usepackage{natbib,aas_macros,amsmath}
\usepackage{multirow,color}

\usepackage[backref,breaklinks,colorlinks,citecolor=blue,urlcolor=blue,linkcolor=red]{hyperref} 
\usepackage[all]{hypcap}

\begin{document}
\shorttitle{Machine Learning-Aided Selection of $\sim 9$,$300$ LAEs at $\lowercase{z} =2.2$--$7.0$ from HSC SSP and CHORUS}
\shortauthors{Ono et al.}
%\slugcomment{Ver. \today}
\slugcomment{Accepted for publication in ApJ}

\title{%
SILVERRUSH X: Machine Learning-Aided Selection of $9$,$318$ LAEs \\ 
at $\lowercase{z} =2.2$, $3.3$, $4.9$, $5.7$, $6.6$, and $7.0$ 
from the HSC SSP and CHORUS Survey Data 
}

\author{%
Yoshiaki Ono\altaffilmark{1}, 
Ryohei Itoh\altaffilmark{1,2}, 
Takatoshi Shibuya\altaffilmark{3}, 
Masami Ouchi\altaffilmark{4,1,5}, 
Yuichi Harikane\altaffilmark{1,6},  
Satoshi Yamanaka\altaffilmark{7,8}, 
Akio K. Inoue\altaffilmark{9,7}, 
Toshiyuki Amagasa\altaffilmark{10,11}, 
Daichi Miura\altaffilmark{10}, 
Maiki Okura\altaffilmark{10}, 
Kazuhiro Shimasaku\altaffilmark{12,13}, \\
Ikuru Iwata\altaffilmark{4},  
Yoshiaki Taniguchi\altaffilmark{14},  
Seiji Fujimoto\altaffilmark{15}, 
Masanori Iye\altaffilmark{4},  
Anton T. Jaelani\altaffilmark{16,17}, \\
Nobunari Kashikawa\altaffilmark{12,13}, 
Shotaro Kikuchihara\altaffilmark{1,12}, 
Satoshi Kikuta\altaffilmark{11}, 
Masakazu A.R. Kobayashi\altaffilmark{18}, \\
Haruka Kusakabe\altaffilmark{19}, 
Chien-Hsiu Lee\altaffilmark{20}, 
Yongming Liang\altaffilmark{4},  
Yoshiki Matsuoka\altaffilmark{8},  
Rieko Momose\altaffilmark{12}, \\
Tohru Nagao\altaffilmark{8},  
Kimihiko Nakajima\altaffilmark{4},  
and 
Ken-ichi Tadaki\altaffilmark{4}
}

%\email{ono@icrr.u-tokyo.ac.jp}

\altaffiltext{1}{Institute for Cosmic Ray Research, The University of Tokyo, 5-1-5 Kashiwanoha, Kashiwa, Chiba 277-8582, Japan}
\altaffiltext{2}{Department of Physics, Graduate School of Science, The University of Tokyo, 7-3-1 Hongo, Bunkyo-ku, Tokyo 113-0033, Japan}
\altaffiltext{3}{Kitami Institute of Technology, 165 Koen-cho, Kitami, Hokkaido 090-8507, Japan}
\altaffiltext{4}{National Astronomical Observatory of Japan, 2-21-1 Osawa, Mitaka, Tokyo 181-8588, Japan}
\altaffiltext{5}{Kavli Institute for the Physics and Mathematics of the Universe (WPI), University of Tokyo, Kashiwa, Chiba 277-8583, Japan}
\altaffiltext{6}{Department of Physics and Astronomy, University College London, Gower Street, London WC1E 6BT, UK}
\altaffiltext{7}{Waseda Research Institute for Science and Engineering, Faculty of Science and Engineering, Waseda University, 3-4-1, Okubo, Shinjuku, Tokyo 169-8555, Japan}
\altaffiltext{8}{Research Center for Space and Cosmic Evolution, Ehime University, Bunkyo-cho, Matsuyama, Ehime 790-8577, Japan}
\altaffiltext{9}{Department of Physics, School of Advanced Science and Engineering, Faculty of Science and Engineering, Waseda University, 3-4-1, Okubo, Shinjuku, Tokyo 169-8555, Japan}
\altaffiltext{10}{Graduate School of Systems and Information Engineering, University of Tsukuba, Ibaraki 305-8573, Japan}
\altaffiltext{11}{Center for Computational Sciences, University of Tsukuba, Ibaraki 305-8577, Japan}
\altaffiltext{12}{Department of Astronomy, Graduate School of Science, The University of Tokyo, 7-3-1 Hongo, Bunkyo-ku, Tokyo 113-0033, Japan}
\altaffiltext{13}{Research Center for the Early Universe, Graduate School of Science, The University of Tokyo, 7-3-1 Hongo, Bunkyo-ku, Tokyo 113-0033, Japan}
\altaffiltext{14}{The Open University of Japan, 2-11 Wakaba, Mihama-ku, Chiba, Chiba 261-8586, Japan}
\altaffiltext{15}{Cosmic DAWN Center, Copenhagen, Denmark; Niels Bohr Institute, University of Copenhagen, Lyngbyvej 2, DK-2100, Copenhagen, Denmark}
\altaffiltext{16}{Astronomy Research Division and Bosscha Observatory, FMIPA, Institut Teknologi Bandung, Jl. Ganesha 10, Bandung, 40132, Indonesia}
\altaffiltext{17}{Department of Physics, Kindai University, 3-4-1 Kowakae, Higashi-Osaka, Osaka 577-8502, Japan}
\altaffiltext{18}{Faculty of Natural Sciences, National Institute of Technology, Kure College, 2-2-11, Agaminami, Kure, Hiroshima 737-8506, Japan}
\altaffiltext{19}{Observatoire de Gen\`{e}ve, Universit\'e de Gen\`{e}ve, 51 chemin de P\'egase, 1290 Versoix, Switzerland}
\altaffiltext{20}{NSF's National Optical-Infrared Astronomy Research Laboratory, Tucson, AZ 85719, USA}

%---------------------------------------------------------------------
\begin{abstract}
We present a new catalog of $9$,$318$ Ly$\alpha$ emitter (LAE) candidates  
at $z = 2.2$, $3.3$, $4.9$, $5.7$, $6.6$, and $7.0$ 
that are photometrically selected by the SILVERRUSH program with a machine learning technique  
from large area (up to $25.0$ deg$^2$) imaging data with six narrowband filters 
taken by 
the Subaru Strategic Program with Hyper Suprime-Cam (HSC SSP) 
and a Subaru intensive program, 
Cosmic HydrOgen Reionization Unveiled with Subaru (CHORUS). 
We construct a convolutional neural network 
that distinguishes between real LAEs and contaminants 
with a completeness of $94${\%} and a contamination rate of $1${\%}, 
enabling us to efficiently remove contaminants from the photometrically selected LAE candidates. 
We confirm that our LAE catalogs include 
$177$ LAEs that have been spectroscopically identified 
in our SILVERRUSH programs and previous studies, 
ensuring the validity of our machine learning selection.  
In addition, 
we find that the object-matching rates between our LAE catalogs and our previous results 
are $\simeq 80$--$100${\%} at bright NB magnitudes of $\lesssim 24$ mag. 
We also confirm that 
the surface number densities of our LAE candidates 
are consistent with previous results. 
Our LAE catalogs will be made public on our project webpage. 
\end{abstract}

%---------------------------------------------------------------------
\keywords{%
galaxies: formation ---
galaxies: evolution ---
galaxies: high-redshift 
}
%---------------------------------------------------------------------
%---------------------------------------------------------------------

%%%%%%%%%%%%%%%%%%%%%%%%%%%%%%%%%%%%%%%%%%%%%%%%%%%%%%%%%%%%%%%%%
%%%%%%%%%%%%%%%%%%%%%%%%%%%%%%%%%%%%%%%%%%%%%%%%%%%%%%%%%%%%%%%%%
\section{Introduction} \label{sec:introduction}
%%%%%%%%%%%%%%%%%%%%%%%%%%%%%%%%%%%%%%%%%%%%%%%%%%%%%%%%%%%%%%%%%
%%%%%%%%%%%%%%%%%%%%%%%%%%%%%%%%%%%%%%%%%%%%%%%%%%%%%%%%%%%%%%%%%

Large panoramic surveys over cosmological volumes 
such as 
the Subaru Hyper Suprime-Cam (HSC) survey \citep{2018PASJ...70S...4A} 
and the Dark Energy Survey (DES; \citealt{2018ApJS..239...18A}) 
are providing massive multiwavelength data sets 
for unprecedentedly large numbers of astronomical sources. 
Such big data are key to understanding
galaxy formation and evolution,  
allowing us to investigate statistical properties of high-$z$ galaxies 
(\citealt{2018PASJ...70S..10O}; \citealt{2018PASJ...70S..11H}; \citealt{2018PASJ...70S..13O}; 
\citealt{2018PASJ...70S..16K}; \citealt{2018ApJ...863...63S}; \citealt{2019MNRAS.483.3060G}: 
\citealt{2020MNRAS.494.1771A}; \citealt{2020MNRAS.494.1894M})
and pinpoint very rare interesting structures and objects 
such as 
galaxy overdensities (\citealt{2018PASJ...70S..12T}; \citealt{2019ApJ...878...68I}; 
\citealt{2019ApJ...879...28H}; \citealt{2019ApJ...883..142H}), 
very luminous Ly$\alpha$ emitters (LAEs)
and Ly$\alpha$ blobs 
(\citealt{2018PASJ...70S..14S}; \citealt{2018ApJ...859...91S}; \citealt{2019PASJ...71L...2K}; \citealt{2020ApJ...891..177Z}; \citealt{2020ApJ...895..132T}; 
see also the recent review of \citealt{2020ARA&A..58..617O}). 
In the near future, further significant progress is expected to be made by next generation telescopes  
such as 
Vera C. Rubin Observatory (VRO), 
\textit{Euclid}, 
and 
the \textit{Nancy Grace Roman Space Telescope}.

One of the fundamental problems in analyzing such big data is that 
the amount of obtained data can be too large to handle. 
For example, 
in our last LAE search based on the HSC survey data \citep{2018PASJ...70S..14S}, 
we have conducted visual inspection of the photometrically selected candidates 
to remove contaminants in a similar way to previous studies 
(e.g., \citealt{2008ApJS..176..301O}; \citealt{2010ApJ...723..869O}). 
However, 
due to the very large area of the HSC survey, 
it took about two person-months to complete the visual inspection task, 
suggesting that it could be almost impossible 
to obtain scientific results in a timely manner 
by taking advantage of big data obtained by future large surveys
with reasonable human and computational resources.

Recently, advances in machine learning techniques 
have revolutionized the field of image recognition 
and shown a wide variety of successful applications 
(\citealt{2015Natur.521..436L}). 
Because observational astronomical studies also utilize imaging data sets, 
machine learning methods are becoming prevalent in astronomy (\citealt{2019arXiv190407248B}),  
such as for 
source detections 
(\citealt{2019MNRAS.484.2793V}; \citealt{2019MNRAS.490.3952B}), 
galaxy morphological classifications and measurements 
(\citealt{2015MNRAS.450.1441D}; \citealt{2015ApJS..221....8H}; \citealt{2018MNRAS.476.3661D}; 
\citealt{2020MNRAS.491.1408M};
\citealt{2020ApJS..248...20H}; \citealt{2020MNRAS.493.4209C}; \citealt{2020ApJ...895..112G}; 
\citealt{2020arXiv201209081T}; 
see also, \citealt{2020MNRAS.496.4276T}), 
photometric redshift estimates 
(\citealt{2016A&C....16...34H}; \citealt{2018A&A...609A.111D}; \citealt{2019A&A...621A..26P}), 
gravitationally lensed image searches 
(\citealt{2017Natur.548..555H}; \citealt{2018MNRAS.473.3895L}; \citealt{2018A&A...611A...2S}; 
\citealt{2019ApJS..243...17J}; \citealt{2020MNRAS.497..556H}; \citealt{2020ApJ...899...30L}; 
\citealt{2020arXiv200413048C}; \citealt{2020arXiv200504730H}), 
extremely metal-poor galaxy searches (\citealt{2020ApJ...898..142K}), 
transient source detections (\citealt{2016PASJ...68..104M}; \citealt{2020arXiv201015425P}), 
light curve classifications 
(\citealt{2017ApJ...837L..28C}; \citealt{2018A&A...611A..97P}; \citealt{2018AJ....155...94S}; \citealt{2020PASJ..tmp..240T}), 
optical depth reconstructions of Ly$\alpha$ forest absorption 
(\citealt{2020arXiv200910673H}), 
stellar spectrum classifications 
(\citealt{2017ApJS..228...24T}; \citealt{2018MNRAS.478.4513B}), 
and anomaly detections 
(\citealt{2019MNRAS.484..834G}; \citealt{2019MNRAS.487.2874I}; 
\citealt{2021MNRAS.500.4849C}; \citealt{2020arXiv200608235D}; \citealt{2020arXiv201011202L}).  
Convolutional neural networks (CNNs) are frequently adopted for image classification problems 
as well as various other computer vision related tasks 
such as image segmentation and object recognition. 
Similar to other machine learning techniques, 
CNNs are trained to learn how to directly extract representative features from imaging data   
and to make intelligent judgements or estimates.

\setcounter{footnote}{0}

In this work, we apply machine learning in the form of CNNs 
to image classifications between high-$z$ LAEs and contaminants 
to efficiently construct samples of LAE candidates at $z=2.2$, $3.3$, $4.9$, $5.7$, $6.6$, and $7.0$
based on the HSC imaging data 
obtained by the Subaru Strategic Program (HSC SSP; \citealt{2018PASJ...70S...4A}; 
see also \citealt{2012SPIE.8446E..0ZM}; \citealt{2018PASJ...70S...1M}; \citealt{2018PASJ...70S...2K}; \citealt{2018PASJ...70...66K}; \citealt{2018PASJ...70S...3F})  
as well as a Subaru intensive program, 
Cosmic HydrOgen Reionization Unveiled with Subaru 
(CHORUS; \citealt{2020PASJ...72..101I}; 
see also \citealt{2018ApJ...867...46I}; \citealt{2020ApJ...891..177Z}). 
This paper is included in a series of papers on LAEs based on the HSC SSP survey, 
Systematic Identification of LAEs for Visible Exploration and Reionization Research Using Subaru HSC 
(SILVERRUSH; \citealt{2018PASJ...70S..13O}; \citealt{2018PASJ...70S..14S}; \citealt{2018PASJ...70S..15S}; 
\citealt{2018PASJ...70S..16K}; \citealt{2018ApJ...859...84H}; \citealt{2018PASJ...70...55I}; 
\citealt{2019ApJ...879...28H}; \citealt{2019ApJ...883..142H}; \citealt{2019arXiv190600173K}; 
\textcolor{blue}{H. Goto et al. in prep.}; \textcolor{blue}{S. Kikuchihara et al. in prep.}). 
The major improvements from our previous work are 
the advent of a new narrowband (NB) filter, $NB387$, 
and deeper depths of broadband (BB) data, 
as well as a compilation of results from four NB data of 
$NB387$, $NB527$, $NB718$, and $NB973$ 
taken by the CHORUS program.  
Our new SILVERRUSH LAE catalogs based on new HSC SSP and CHORUS data 
with the aid of the machine learning technique 
will be made public on our project webpage.\footnote{\url{http://cos.icrr.u-tokyo.ac.jp/rush.html}}

This paper is organized as follows. 
In Section \ref{sec:data}, 
we describe our HSC SSP and CHORUS data, 
and photometric selections for high-$z$ LAE candidates 
based on multiwavelength catalogs with HSC NB and BB data. 
Our machine learning selection is described in Section \ref{sec:deep_learning_selection}, 
and the constructed LAE catalogs and their statistical properties are presented in Section \ref{sec:results}. 
A summary is described in Section \ref{sec:summary}. 
In this paper, 
we use magnitudes in the AB system \citep{1983ApJ...266..713O} 
and 
assume a flat universe with 
$\Omega_{\rm m} = 0.3$, 
$\Omega_\Lambda = 0.7$, 
and $H_0 = 70$ km s$^{-1}$ Mpc$^{-1}$.

%ttttttttttttttttttttttttttttttttttttttttttttttttttttttttttttttttttttttttt%
\capstartfalse
\begin{deluxetable*}{ccccc} 
\tablecolumns{5} 
\tablewidth{0pt} 
\tablecaption{Summary of the NB filter properties used in this study 
\label{tab:NB_properties}}
\tablehead{
    \colhead{Filter}     
    &  \colhead{$\lambda_{\rm c}$}
    &  \colhead{FWHM}
    &  \colhead{$z_{{\rm Ly}\alpha}$}
    &  \colhead{References}
   \\
    \colhead{ }
    &  \colhead{({\AA})}
    &  \colhead{({\AA})}
    &  \colhead{ }
    &  \colhead{ }
    \\
    \colhead{(1)}
    &  \colhead{(2)}
    &  \colhead{(3)}
    &  \colhead{(4)}
    &  \colhead{(5)}
}
\startdata 
$NB387$ & 3863 &  55 & $2.178 \pm 0.023$ & \cite{2018PASJ...70S...4A}; \cite{2020PASJ...72..101I} \\
$NB527$ & 5260 &  79 & $3.327 \pm 0.032$ & \cite{2020PASJ...72..101I} \\
$NB718$ & 7171 & 111 & $4.899 \pm 0.046$ & \cite{2020PASJ...72..101I} \\
$NB816$ & 8177 & 113 & $5.726 \pm 0.046$ & \cite{2018PASJ...70S...4A}; \cite{2018PASJ...70S..13O} \\
$NB921$ & 9215 & 135 & $6.580 \pm 0.056$ & \cite{2018PASJ...70S...4A}; \cite{2018PASJ...70S..13O} \\
$NB973$ & 9712 & 112 & $6.989 \pm 0.046$ & \cite{2020PASJ...72..101I} 
\enddata 
\tablecomments{
(1) Filter name. 
(2) Central wavelength in {\AA}. 
(3) FWHM of the transmission function in {\AA}. 
(4) Redshift range of Ly$\alpha$ emission that can be probed within the NB FWHM.  
(5) Papers that present the detailed properties of the NBs. 
}
\end{deluxetable*} 
%ttttttttttttttttttttttttttttttttttttttttttttttttttttttttttttttttttttttttt%

%FFFFFFFFFFFFFFFFFFFFFFFFFFFFFFFFFFFFFFFFFFFFFFFFFFFFFFFFFFFFFFFF%
\begin{figure}[h]
\begin{center}
   \includegraphics[scale=0.3]{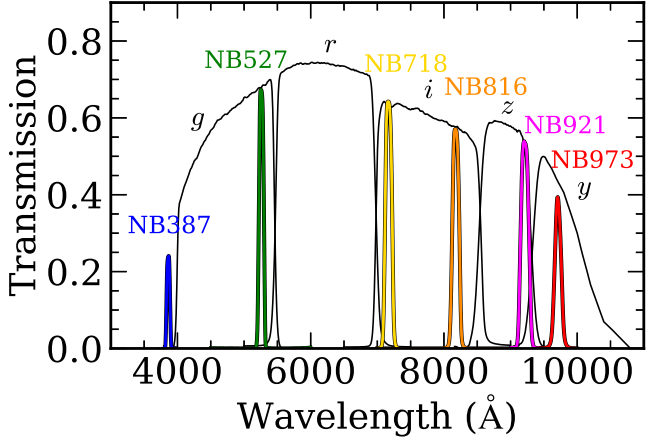}
\caption[]
{
Transmission curves of the NB and BB filters. 
The blue, green, yellow, orange, magenta, and red curves 
correspond to 
the $NB387$, $NB527$, $NB718$, $NB816$, $NB921$, and $NB973$ filters, respectively. 
The black curves represent the $g$-, $r$-, $i$-, $z$-, and $y$-band filters from left to right. 
These transmission values are obtained 
by multiplying the area-weighted mean values of the filters 
by 
the quantum efficiency of CCDs, 
the transmittance values of the dewar window and the primary focus unit of HSC, 
and 
the reflectivity of the primary mirror. 
}
\label{fig:filters}
\end{center}
\end{figure}
%FFFFFFFFFFFFFFFFFFFFFFFFFFFFFFFFFFFFFFFFFFFFFFFFFFFFFFFFFFFFFFFF%

%ttttttttttttttttttttttttttttttttttttttttttttttttttttttttttttttttttttttttt%
\capstartfalse
\begin{deluxetable*}{cccccccccccc} 
\tablecolumns{12} 
\tablewidth{0pt} 
\tablecaption{HSC SSP Survey data used in this study
\label{tab:hsc_data}}
\tablehead{
    \colhead{Field}     
    &  \colhead{Area ($NB387$)}
    &  \colhead{Area ($NB816$)}
    &  \colhead{Area ($NB921$)}
    &  \colhead{$NB387$}
    &  \colhead{$NB816$}
    &  \colhead{$NB921$}
    &  \colhead{$g$}
    &  \colhead{$r$}
    &  \colhead{$i$}
    &  \colhead{$z$}
    &  \colhead{$y$}
   \\
    \colhead{ }
    &  \colhead{(deg$^2$)}
    &  \colhead{(deg$^2$)}
    &  \colhead{(deg$^2$)}
    &  \colhead{(mag)}
    &  \colhead{(mag)}
    &  \colhead{(mag)}
    &  \colhead{(mag)}
    &  \colhead{(mag)}
    &  \colhead{(mag)}
    &  \colhead{(mag)}
    &  \colhead{(mag)}
    \\
    \colhead{(1)}
    &  \colhead{(2)}
    &  \colhead{(3)}
    &  \colhead{(4)}
    &  \colhead{(5)}
    &  \colhead{(6)}
    &  \colhead{(7)}
    &  \colhead{(8)}
    &  \colhead{(9)}
    &  \colhead{(10)}
    &  \colhead{(11)}
    &  \colhead{(12)}
}
\startdata 
UD-SXDS    & ---   & 2.278 & 2.278 & --- & 25.61 & 25.31 & 26.83 & 26.42 & 26.24 & 25.74 & 24.96 \\
UD-COSMOS  & ---   & 2.261 & 2.278 & --- & 25.75 & 25.48 & 26.93 & 26.52 & 26.50 & 26.30 & 25.60 \\
D-SXDS     & 4.928 & 4.789 & ---   & 24.29 & 25.16 & ---   & 26.65 & 26.03 & 25.62 & 25.51 & 24.27 \\
D-COSMOS   & 7.296 & ---   & 6.198 & 24.71 & ---   & 25.07 & 26.58 & 26.32 & 26.23 & 25.90 & 25.02 \\
D-ELAIS-N1 & ---   & 7.145 & 7.215 & ---   & 25.23 & 25.02 & 26.63 & 26.16 & 26.00 & 25.57 & 24.68 \\
D-DEEP2-3  & 6.802 & 7.197 & 6.980 & 24.25 & 25.26 & 24.86 & 26.51 & 26.10 & 25.81 & 25.52 & 24.80 \\ 
\hline
Total & 19.03 & 23.67 & 24.95 & & & & & & & & 
\enddata 
\tablecomments{
(1) Field name. 
(2)--(4) Effective area in deg$^2$. 
(5)--(12) Typical $5\sigma$ limiting magnitudes
measured with $2\farcs0$ diameter circular apertures
in $NB387$, $NB816$, $NB921$, $g$, $r$, $i$, $z$, and $y$.
For $NB387$, we consider the 0.45 mag offset described in Section \ref{sec:data}. 
The standard deviation is typically $\simeq 0.5$ mag,
if the edge of the field of view, where few objects are selected, is also taken into account.
It is much smaller if we focus on more central regions.  
Note that the limiting magnitudes are measured for different aperture sizes 
when compared to those in Table 1 of \cite{2018PASJ...70S..14S}. 
}
\end{deluxetable*} 
%ttttttttttttttttttttttttttttttttttttttttttttttttttttttttttttttttttttttttt%

%ttttttttttttttttttttttttttttttttttttttttttttttttttttttttttttttttttttttttt%
\capstartfalse
\begin{deluxetable*}{ccccccccc} 
\tablecolumns{9} 
\tablewidth{0pt} 
\tablecaption{CHORUS NB data used in this study
\label{tab:hsc_data_chorus}}
\tablehead{
    \colhead{Field}     
    &  \colhead{Area ($NB387$)}
    &  \colhead{Area ($NB527$)}
    &  \colhead{Area ($NB718$)}
    &  \colhead{Area ($NB973$)}
    &  \colhead{$NB387$}
    &  \colhead{$NB527$}
    &  \colhead{$NB718$}
    &  \colhead{$NB973$}
   \\
    \colhead{ }
    &  \colhead{(deg$^2$)}
    &  \colhead{(deg$^2$)}
    &  \colhead{(deg$^2$)}
    &  \colhead{(deg$^2$)}
    &  \colhead{(mag)}
    &  \colhead{(mag)}
    &  \colhead{(mag)}
    &  \colhead{(mag)}
    \\
    \colhead{(1)}
    &  \colhead{(2)}
    &  \colhead{(3)}
    &  \colhead{(4)}
    &  \colhead{(5)}
    &  \colhead{(6)}
    &  \colhead{(7)}
    &  \colhead{(8)}
    &  \colhead{(9)}
}
\startdata 
UD-COSMOS  & 1.561 & 1.613 & 1.575 & 1.603 & 25.67 & 26.39 & 25.59 & 24.90 
\enddata 
\tablecomments{
(1) Field name. 
(2)--(5) Effective area in deg$^2$ 
after removing the masked regions 
(\citealt{2020PASJ...72..101I}). 
(6)--(9) Typical $5\sigma$ limiting magnitudes measured with $2\farcs0$ diameter circular apertures 
in $NB387$, $NB527$, $NB718$, and $NB973$. 
For $NB387$, we consider the 0.45 mag offset described in Section \ref{sec:data}. 
See Figure 4 of \cite{2020PASJ...72..101I} for the spatial variations of the limiting magnitudes. 
}
\end{deluxetable*} 
%ttttttttttttttttttttttttttttttttttttttttttttttttttttttttttttttttttttttttt%

%%%%%%%%%%%%%%%%%%%%%%%%%%%%%%%%%%%%%%%%%%%%%%%%%%%%%%%%%%%%%%%%%
%%%%%%%%%%%%%%%%%%%%%%%%%%%%%%%%%%%%%%%%%%%%%%%%%%%%%%%%%%%%%%%%%
\section{Data and Photometric Selection} \label{sec:data}
%%%%%%%%%%%%%%%%%%%%%%%%%%%%%%%%%%%%%%%%%%%%%%%%%%%%%%%%%%%%%%%%%
%%%%%%%%%%%%%%%%%%%%%%%%%%%%%%%%%%%%%%%%%%%%%%%%%%%%%%%%%%%%%%%%%

%FFFFFFFFFFFFFFFFFFFFFFFFFFFFFFFFFFFFFFFFFFFFFFFFFFFFFFFFFFFFFFFF%
\begin{figure*}
\begin{center}
   \includegraphics[scale=0.4]{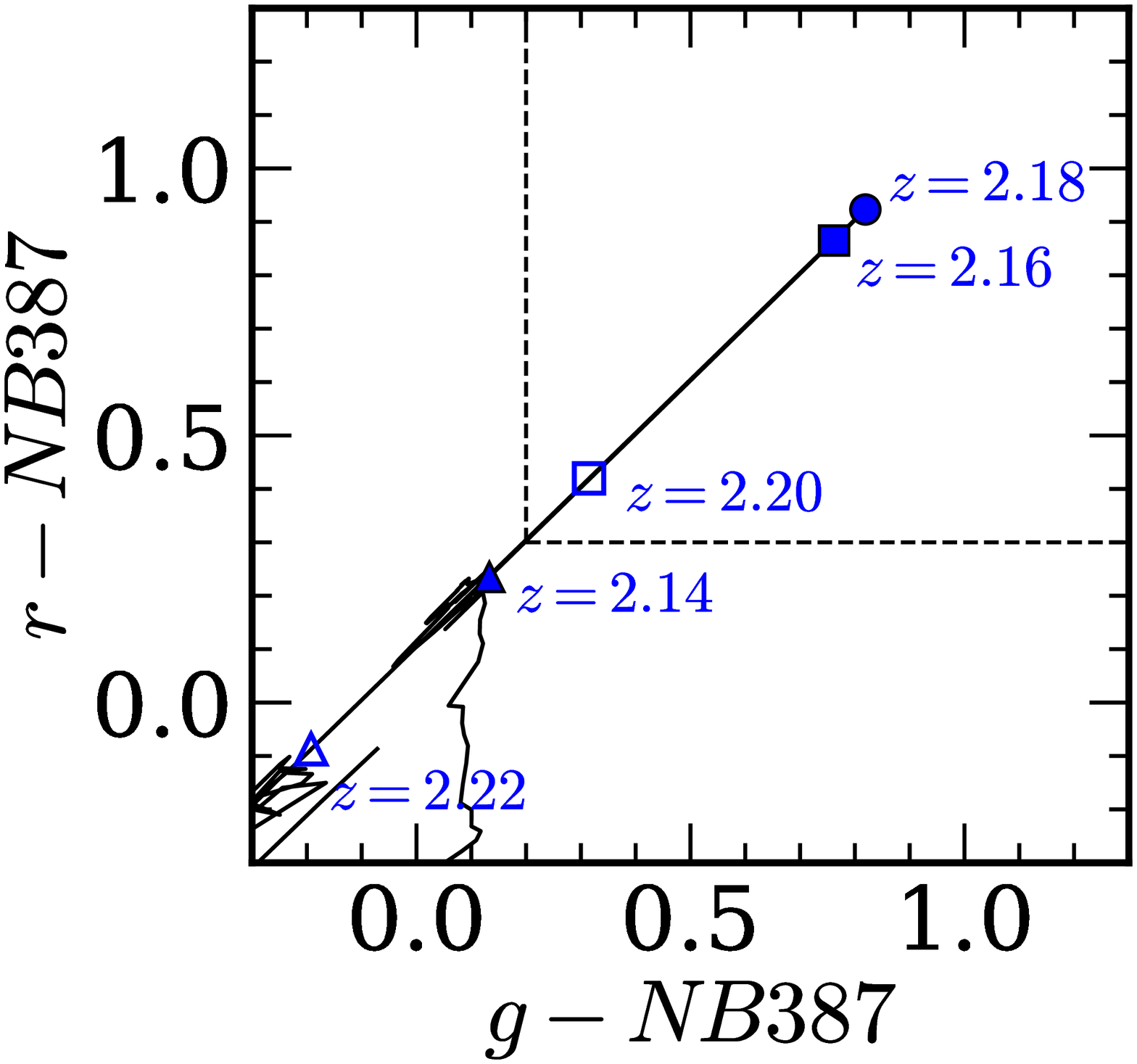}
   \includegraphics[scale=0.4]{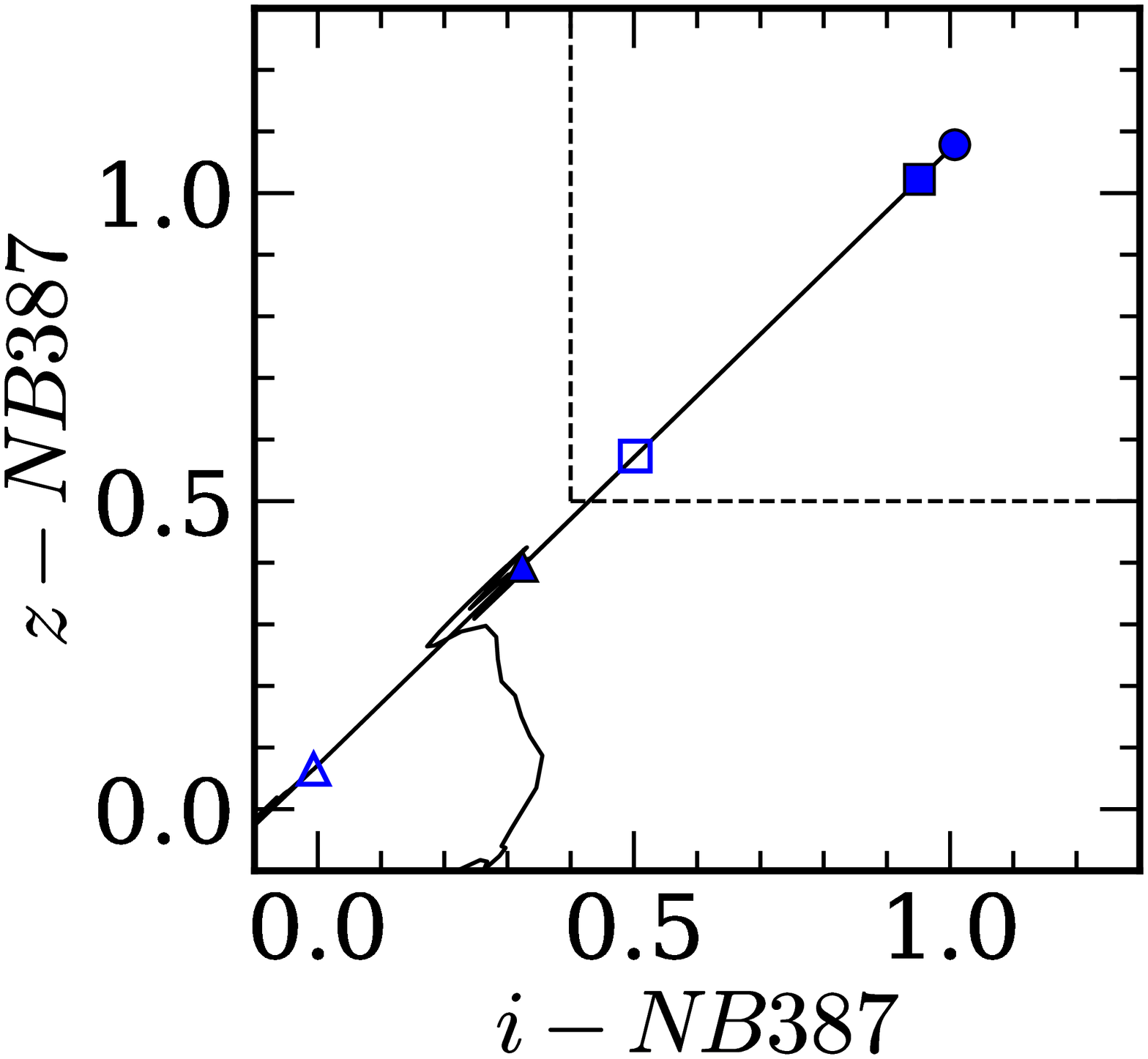}
   \includegraphics[scale=0.4]{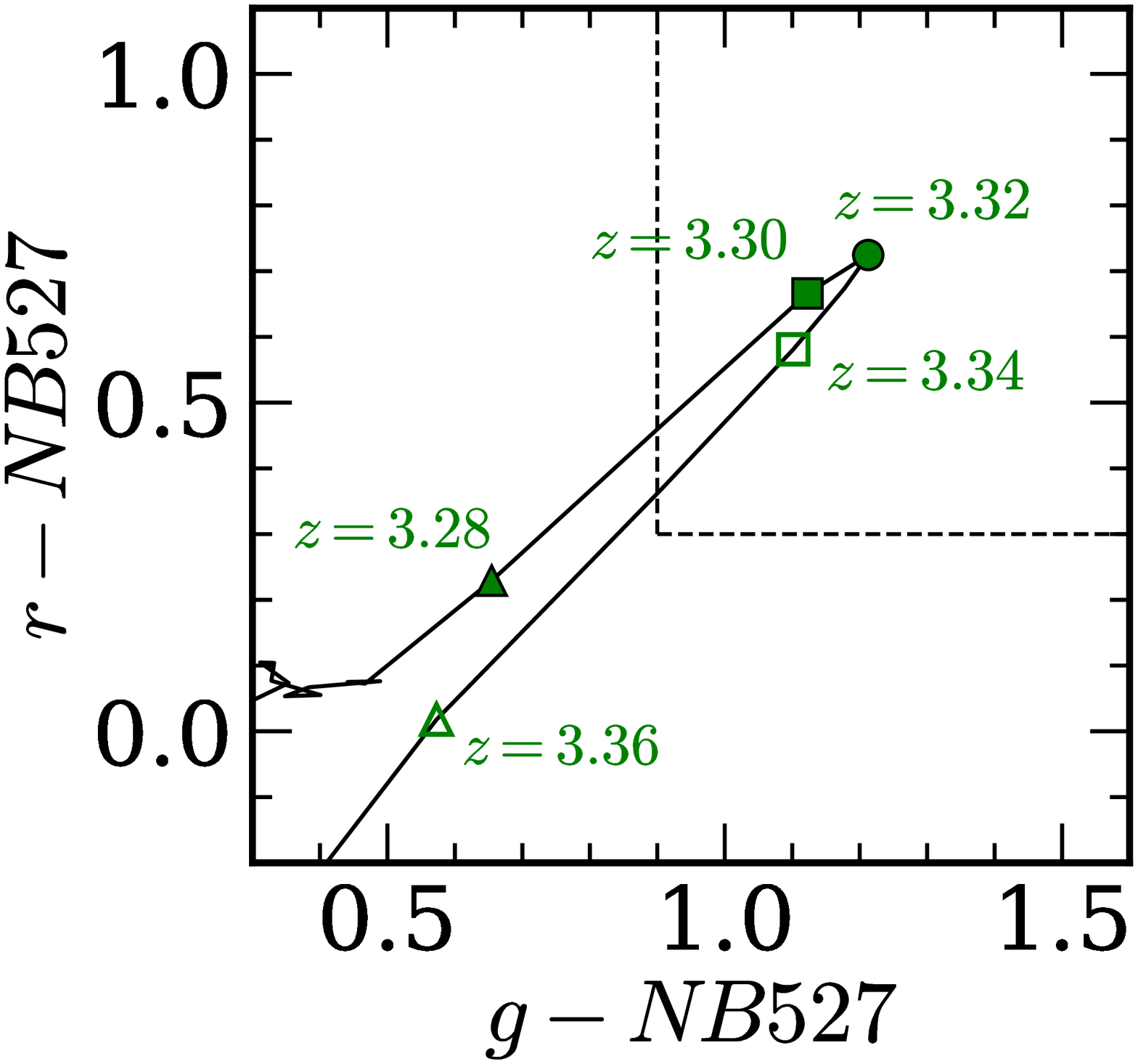}
   \includegraphics[scale=0.4]{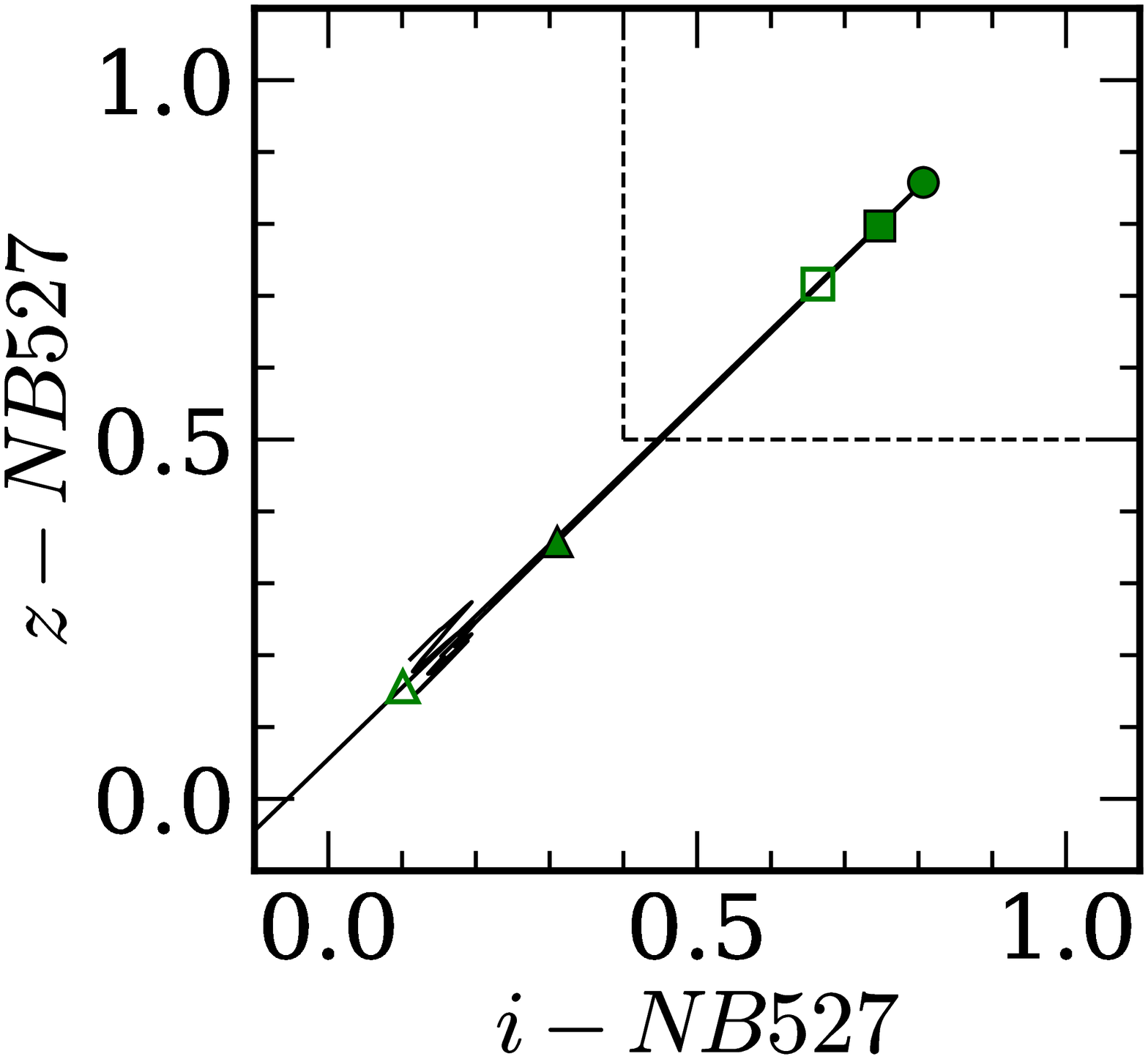}
\caption[]
{
BB$-$NB color diagrams for photometric selections of LAEs at $z=2.2$ (top) and $z=3.3$ (bottom): 
$r - NB387$ vs. $g - NB387$ (top left); 
$z - NB387$ vs. $i - NB387$ (top right); 
$r - NB527$ vs. $g - NB527$ (bottom left); 
$z - NB527$ vs. $i - NB527$ (bottom right). 
The solid curve corresponds to the color track of 
a redshifted galaxy SED with Ly$\alpha$ EW of $20${\AA} in the rest frame. 
We use the \cite{2003MNRAS.344.1000B} stellar population synthesis model 
with a \cite{1955ApJ...121..161S} initial mass function (IMF)  
and adopt a $30$ Myr simple stellar population 
whose UV continuum slope is sufficiently blue, 
consistent with previous observational results 
(e.g., \citealt{2010MNRAS.408.1628S}; \citealt{2020MNRAS.493..141S}).  
We add a Ly$\alpha$ emission 
and 
take into account the \cite{1995ApJ...441...18M} prescription 
of the intergalactic medium (IGM) absorption. 
The SED is redshifted from $z=0.0$ to $z=3.50$ with a step of $\Delta z = 0.01$. 
The symbols on these model tracks 
correspond to 
$z = 2.14$ (filled triangles), $z = 2.16$ (filled squares), 
$z = 2.18$ (filled circles), $z = 2.20$ (open squares), and $z = 2.22$ (open triangles) 
in the top panels, 
and 
$z = 3.28$ (filled triangles), $z = 3.30$ (filled squares), 
$z = 3.32$ (filled circles), $z = 3.34$ (open squares), and $z = 3.36$ (open triangles) 
in the bottom panels. 
The dashed lines denote the color criteria to photometrically select our LAE candidates. 
Note that 
the model track slightly changes with Ly$\alpha$ EW 
only in the two color diagram of $g-NB527$ vs. $r-NB527$. 
In the other diagrams, 
the model tracks with larger EW values overlap and extend to the upper right, 
since there is no wavelength overlap between NB and BB.
}
\label{fig:NB_BB_colors}
\end{center}
\end{figure*}
%FFFFFFFFFFFFFFFFFFFFFFFFFFFFFFFFFFFFFFFFFFFFFFFFFFFFFFFFFFFFFFFF%

%FFFFFFFFFFFFFFFFFFFFFFFFFFFFFFFFFFFFFFFFFFFFFFFFFFFFFFFFFFFFFFFF%
\begin{figure*}
\begin{center}
   \includegraphics[scale=0.21]{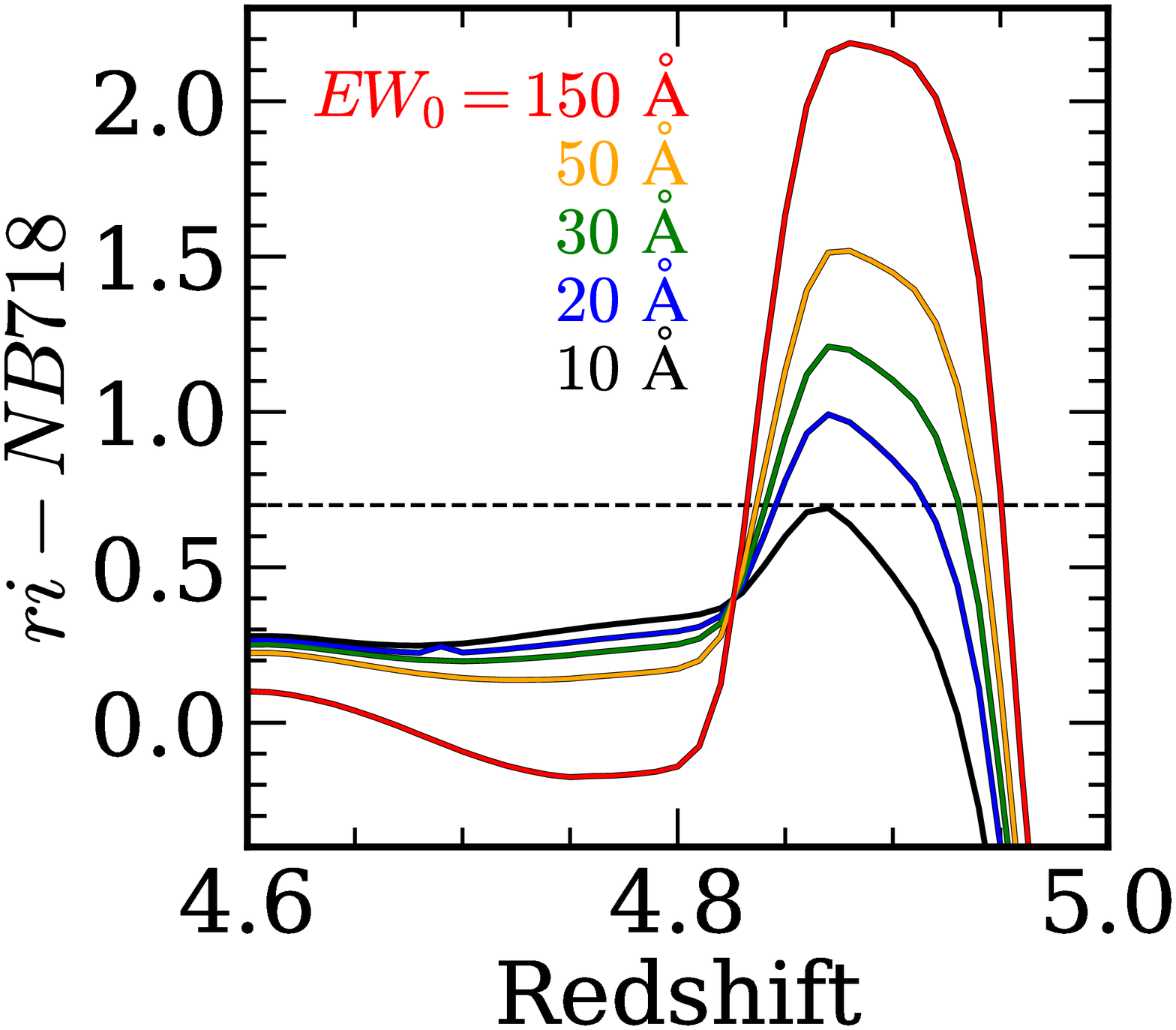}
   \includegraphics[scale=0.21]{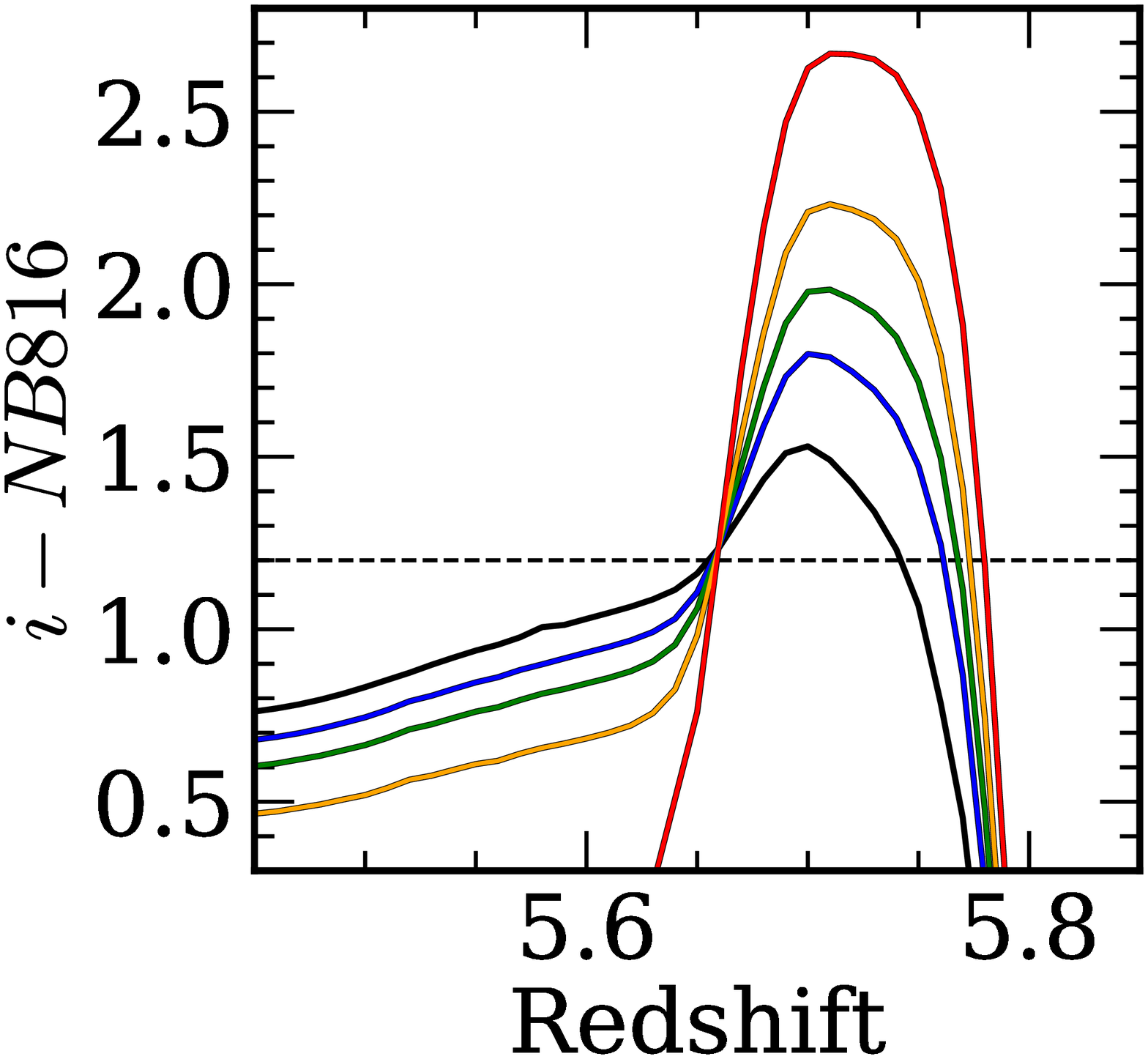}
   \includegraphics[scale=0.21]{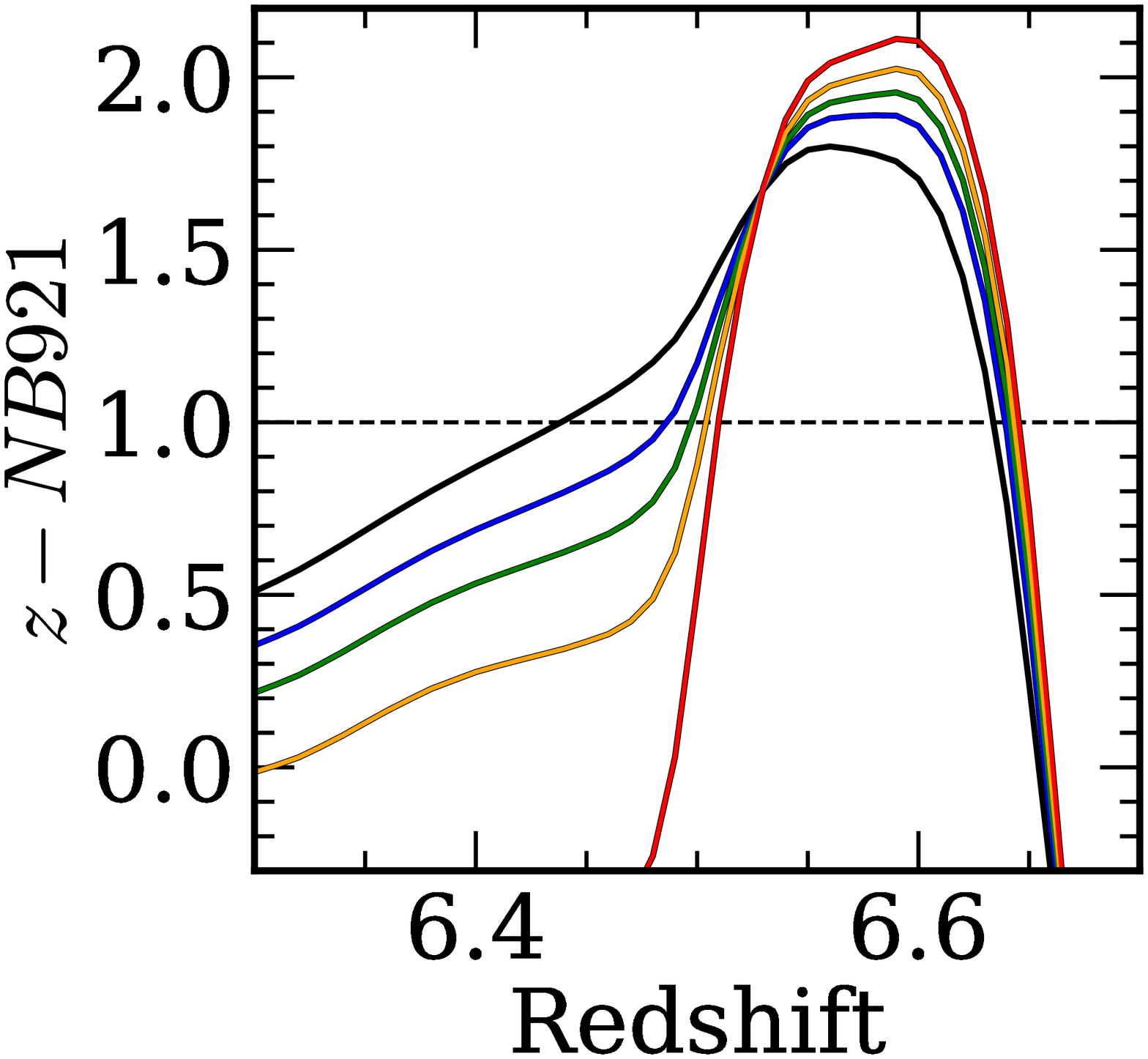}
   \includegraphics[scale=0.21]{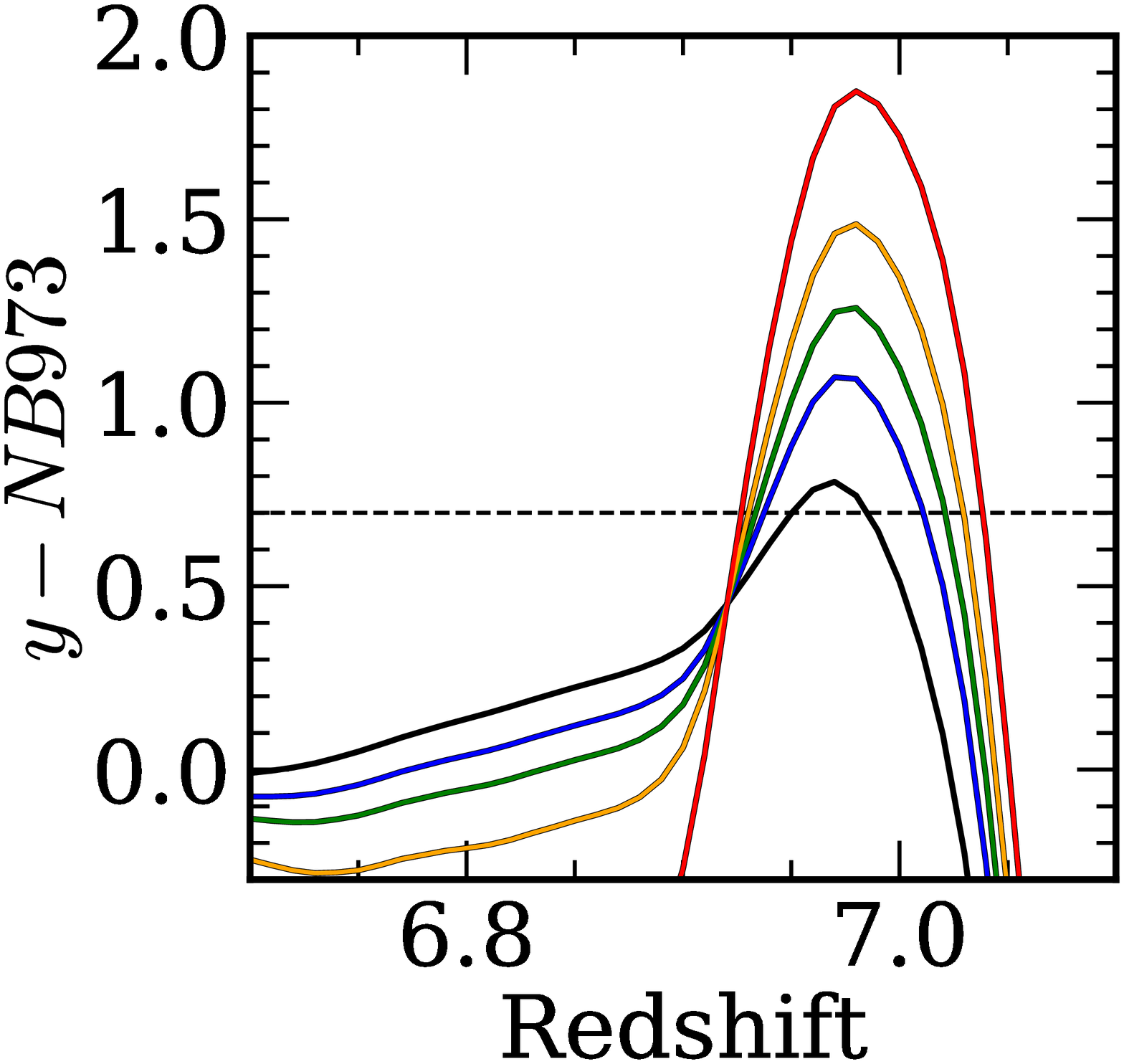}
\caption[]
{
BB$-$NB colors as a function of redshift for $z=4.9$, $z=5.7$, $z=6.6$, and $z=7.0$ LAEs: 
$ri - NB718$, $i - NB816$,  $z - NB921$,  and $y - NB973$ from left to right. 
The solid blue curves denote the track of the redshifted galaxy SED 
with Ly$\alpha$ EW of $20${\AA} in the rest frame, 
which is the same as that of Figure \ref{fig:NB_BB_colors}. 
The solid black, green, orange, and red curves are the same as the solid blue curve 
but for other Ly$\alpha$ EWs of $10$, $30$, $50$, and $150${\AA} in the rest frame, respectively, 
as presented in the legend in the left panel.
The horizontal dashed lines denote the color criteria to photometrically select our LAE candidates. 
}
\label{fig:NB_BB_colors_highz}
\end{center}
\end{figure*}
%FFFFFFFFFFFFFFFFFFFFFFFFFFFFFFFFFFFFFFFFFFFFFFFFFFFFFFFFFFFFFFFF%

We use data products of the on-going HSC SSP survey 
that were obtained between 2014 March and 2018 January 
(\citealt{2019PASJ...71..114A}).  
Specifically, we use the internal data release of S18A, 
which is basically identical to the version of Public Data Release 2.\footnote{\url{https://hsc.mtk.nao.ac.jp/ssp/data-release/}} 
The HSC survey consists of three layers: Ultra-deep (UD), Deep (D), and Wide. 
Among these three layers, 
the UD (D) layer is covered with two (three) NB filters of 
($NB387$,) $NB816$, and $NB921$ 
that are designed to efficiently identify emission line sources,  
as well as five BB filters of $g$, $r$, $i$, $z$, and $y$. 
Thus, in this study, 
we use the HSC SSP data for the UD and D layers. 
The basic properties of the $NB387$, $NB816$, and $NB921$ filters 
are summarized in Table \ref{tab:NB_properties} 
and their transmission curves are shown 
in Figure \ref{fig:filters}.\footnote{The filter transmission curve data 
are downloaded from \url{https://www.subarutelescope.org/Observing/Instruments/HSC/sensitivity.html}} 
The UD layer has two fields: UD-SXDS and UD-COSMOS. 
The D layer consists of four fields: 
D-SXDS, D-COSMOS, D-ELAIS-N1, and D-DEEP2-3. 
The observed areas with the NBs are shown in Table \ref{tab:hsc_data}.

In addition, we use HSC NB data taken by the CHORUS program 
between 2017 January and 2018 March 
(\citealt{2020PASJ...72..101I}). 
In the CHORUS program, the COSMOS field in the UD layer 
is covered with four NB filters of $NB387$, $NB527$, $NB718$, and $NB973$, 
as well as an intermediate filter, $IB945$. 
In this study, we use these NB data to efficiently select LAE candidates. 
The basic properties of these NB filters are also presented in Table \ref{tab:NB_properties} 
and their transmission curves are also shown in Figure \ref{fig:filters}.
The effective survey area of the NB data 
are summarized in Table \ref{tab:hsc_data_chorus}.

These HSC data were reduced with hscPipe 6.7,
which exploits codes from the VRO software pipeline 
(\citealt{2018PASJ...70S...5B}; 
see also 
\citealt{2010SPIE.7740E..15A}; 
\citealt{2017ASPC..512..279J}; 
\citealt{2019ApJ...873..111I}).
Full details of the HSC SSP survey and the CHORUS program 
such as observations and data reduction
are described in \cite{2019PASJ...71..114A} 
(see also \citealt{2018PASJ...70S...4A}) 
and 
\cite{2020PASJ...72..101I}, respectively. 
Note that we correct a systematic offset 
in the zero-point magnitude of the $NB387$ data 
by subtracting $0.45$ mag, 
following the previous studies 
(\citealt{2020arXiv200801733L}; 
\citealt{2020PASJ...72..101I}).

To photometrically select LAE candidates based on NB color excess features, 
we create multiwavelength catalogs of NB-detected sources in the HSC data. 
We use the $NB387$, $NB527$, $NB718$, $NB816$, $NB921$, and $NB973$ data as detection images 
for $z=2.2$, $3.3$, $4.9$, $5.7$, $6.6$, and $7.0$ LAE candidate selections, respectively.  
First we smooth the HSC BB images and each NB image for each redshift sample 
with Gaussian kernels 
so that their point spread function (PSF) full width at half maximums (FWHMs) are matched to the largest ones. 
For this purpose, for the HSC SSP data, 
we download a PSF image for each filter for each $\simeq 12' \times 12'$ patch 
from the PSF Picker website\footnote{\url{https://hsc-release.mtk.nao.ac.jp/psf/pdr2/}} 
and smooth the multiwavelength images patch by patch. 
For the CHORUS data, 
we create a PSF image for each filter by stacking $\sim 200$ point-source-like bright objects 
detected in the UD-COSMOS field.  
Typical PSF FWHMs of the smoothed HSC images are about 0\farcs8--1\farcs1. 
We then run SExtractor \citep{1996A&AS..117..393B} in the dual image mode 
for the PSF-matched images, 
to make multiwavelength catalogs. 
The main reason why we use SExtractor instead of hscPipe 
to construct multiwavelength catalogs for our photometric selections of LAEs 
is because it is easier for us 
to change the priority of the detection filters.
We set DETECT\_MINAREA to 5 and DETECT\_THRESH to 2.0. 
We adopt MAG\_AUTO as total magnitudes, 
while we use $2\farcs0$ diameter circular aperture magnitudes to measure colors of detected sources. 
The limiting magnitudes within $2\farcs0$ diameter apertures 
for the HSC SSP and CHORUS data 
are summarized in Tables \ref{tab:hsc_data} and \ref{tab:hsc_data_chorus}, 
respectively.

From the multiwavelength catalogs, 
we photometrically select LAE candidates at $z=2.2$, $3.3$, $4.9$, $5.7$, $6.6$, and $7.0$  
that show clear NB excesses compared to BB magnitudes. 
First, we select sources with signal-to-noise ratio (S/N) $>5$ 
within $2\farcs0$ diameter apertures 
in the corresponding NB data for each redshift sample. 
We then select LAE candidates 
by using their NB and BB magnitudes and colors. 
For $z=2.2$ LAEs, we adopt 
\begin{align}
g - NB387 &\geq 0.2, \\ 
r - NB387 &\geq 0.3, \\ 
i - NB387 &\geq 0.4, \\ 
z - NB387 &\geq 0.5, 
\end{align}
and 
for $z=3.3$ LAEs, we apply the following criteria 
\begin{align}
g - NB527 &\geq 0.9, \\ 
r - NB527 &\geq 0.3, \\ 
i - NB527 &\geq 0.4, \\ 
z - NB527 &\geq 0.5,  
\end{align}
by considering the NB$-$BB colors of 
spectral energy distributions (SEDs) of young galaxies 
with strong Ly$\alpha$ emission 
(see also \citealt{2012ApJ...745...12N}; \citealt{2016ApJ...823...20K}) 
as presented in Figure \ref{fig:NB_BB_colors}. 
For $z=4.9$ LAEs, 
we apply these criteria, 
\begin{align}
&ri - NB718 > 0.7 
	\,\, \mathrm{and} \,\, 
	r - i > 0.8 \,\, \mathrm{and} \\
&ri - NB718 > \left( ri - NB718 \right)_{3\sigma}
	 \,\, \mathrm{and} \,\, 
	g > g_{2 \sigma}, 
\end{align}
where $ri$ is calculated by the linear combination of 
the fluxes in the $r$ band, $f_r$, 
and the fluxes in the $i$ band, $f_i$, 
following $f_{ri} = 0.3 f_r + 0.7 f_i$ 
\citep{2020ApJ...891..177Z}. 
The $2\sigma$ and $3\sigma$ subscripts denote 
$2 \sigma$ and $3\sigma$ 
limiting 
magnitudes, respectively. 
We use the $ri$ magnitude, 
because the wavelength of the $NB718$ band is located between the $r$ and $i$ bands (Figure \ref{fig:filters}). 
The expected $ri - NB718$ color as a function of redshift for LAEs 
with Ly$\alpha$ equivalent width (EW) in the rest frame of $EW_0 = 20${\AA} 
is presented in Figure \ref{fig:NB_BB_colors_highz}. 
Similar NB excess selections have been adopted in previous studies 
(e.g., \citealt{2003ApJ...582...60O}; \citealt{2003ApJ...586L.111S}; \citealt{2004ApJ...605L..93S}). 
For $z=5.7$ and $z=6.6$ LAEs, 
we adopt 
\begin{equation}
i - NB816 \geq 1.2, 
\end{equation}
and 
\begin{equation}
z - NB921 \geq 1.0 
\end{equation}
respectively, 
following \cite{2018PASJ...70S..14S}. 
For $z=7.0$ LAEs, 
we adopt 
\begin{align}
&\left[ \left( y < y_{3 \sigma} \,\, \mathrm{and} \,\, y - NB973 > 0.7 \right) 
	\,\, \mathrm{or} \,\, y > y_{3\sigma} \right] 
	\,\, \mathrm{and} \\
&\left[ \left( z < z_{3 \sigma} \,\, \mathrm{and} \,\, z-y > 2 \right) 
	\,\, \mathrm{or} \,\, z > z_{3\sigma} \right] 
	\,\, \mathrm{and} \\
&\hspace{0.3em}
	g > g_{2 \sigma} \,\, \mathrm{and} \,\, 
	r > r_{2 \sigma} \,\, \mathrm{and} \,\, 
	i > i_{2 \sigma},  
\end{align}
which are almost the same as those used in 
\cite{2018ApJ...867...46I} and \cite{2020ApJ...891..177Z}. 
Their expected BB$-$NB colors as a function of redshift 
are also shown in Figure \ref{fig:NB_BB_colors_highz}. 
These BB$-$NB selection criteria correspond to 
Ly$\alpha$ EW limits of $EW_0 \gtrsim 10$--$20${\AA} in the rest frame 
(see also \citealt{2020ApJ...891..177Z}; 
\citealt{2018PASJ...70S..14S}; 
\citealt{2018ApJ...867...46I}).  
The specific Ly$\alpha$ EW limit for each redshift sample 
is summarized in Section \ref{sec:results} 
together with previous studies for comparisons of the LAE number counts.

%ttttttttttttttttttttttttttttttttttttttttttttttttttttttttttttttttttttttttt%
\capstartfalse
\begin{deluxetable}{ccccc} 
\tablecolumns{5} 
\tablewidth{0pt} 
\tablecaption{Layout of our CNN
\label{tab:architecture}}
\tablehead{
    \colhead{ID}     
    &  \colhead{Type} 
    &  \colhead{Size}
    &  \colhead{$n$}
    &  \colhead{Activation}
   \\
    \colhead{(1)}
    &  \colhead{(2)}
    &  \colhead{(3)}
    &  \colhead{(4)}
    &  \colhead{(5)}
}
\startdata 
1 & input & $50 \times 50$ & --- & --- \\
2 & convolutional & $5 \times 5$ & $32$ & ReLU \\
3 & convolutional & $5 \times 5$ & $64$ & ReLU \\
4 & dropout (0.5) & --- & --- & --- \\
5 & average pooling & $2 \times 2$ & --- & --- \\
6 & convolutional & $5 \times 5$ & $128$ & ReLU \\
7 & convolutional & $5 \times 5$ & $128$ & ReLU \\
8 & dropout (0.5) & --- & --- & --- \\
9 & average pooling & $2 \times 2$ & --- & --- \\
10 & dense & $256$ & --- & ReLU \\
11 & dropout (0.5) & --- & --- & --- \\
12 & dense & $8$ & --- & softmax 
\enddata 
\tablecomments{
(1) ID of the layers. 
(2) Type of the layers. 
(3) Size of the data or the filters. 
(4) Number of the filters. 
(5) Activation function adopted in the layers. 
}
\end{deluxetable} 
%ttttttttttttttttttttttttttttttttttttttttttttttttttttttttttttttttttttttttt%

%ttttttttttttttttttttttttttttttttttttttttttttttttttttttttttttttttttttttttt%
\capstartfalse
\begin{deluxetable}{cl} 
\tablecolumns{2} 
\tablewidth{0pt} 
\tablecaption{Classes in the output layer of our CNN
\label{tab:output_classes}}
\tablehead{
    \colhead{Class}     
    &  \colhead{Comment} 
}
\startdata 
1 & LAEs with S/N = 3--4 \\
2 & LAEs with S/N = 4--10 \\
3 & LAEs with S/N = 10--30 \\
4 & LAEs with S/N = 30--200 \\
5 & bright satellite trails \\
6 & faint satellite trails \\
7 & randomly selected noise images \\
8 & randomly selected noise images with positive sky residuals  
\enddata 
\end{deluxetable} 
%ttttttttttttttttttttttttttttttttttttttttttttttttttttttttttttttttttttttttt%

%FFFFFFFFFFFFFFFFFFFFFFFFFFFFFFFFFFFFFFFFFFFFFFFFFFFFFFFFFFFFFFFF%
\begin{figure*}
\begin{center}
   \includegraphics[scale=0.42]{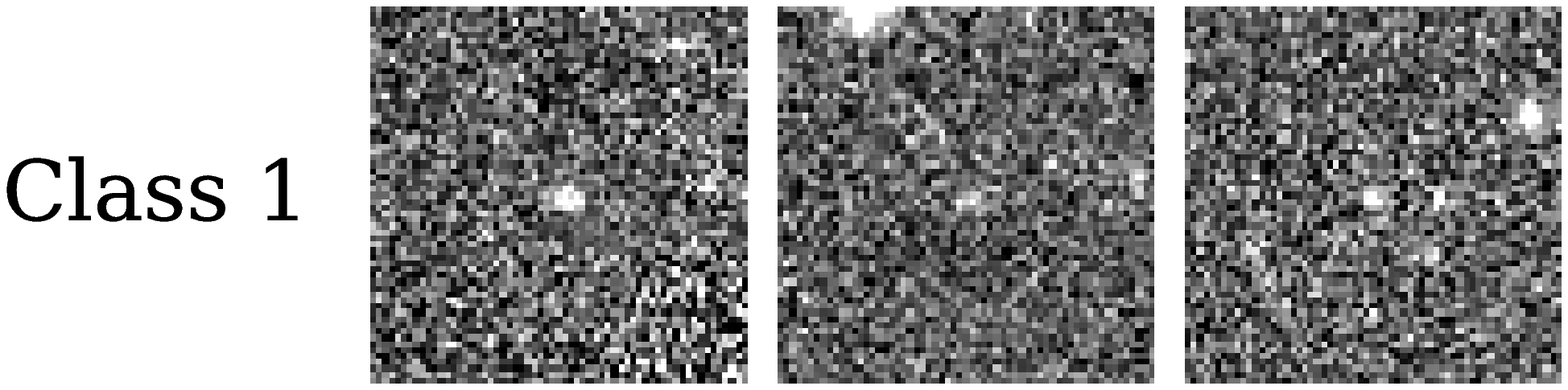}
   \includegraphics[scale=0.42]{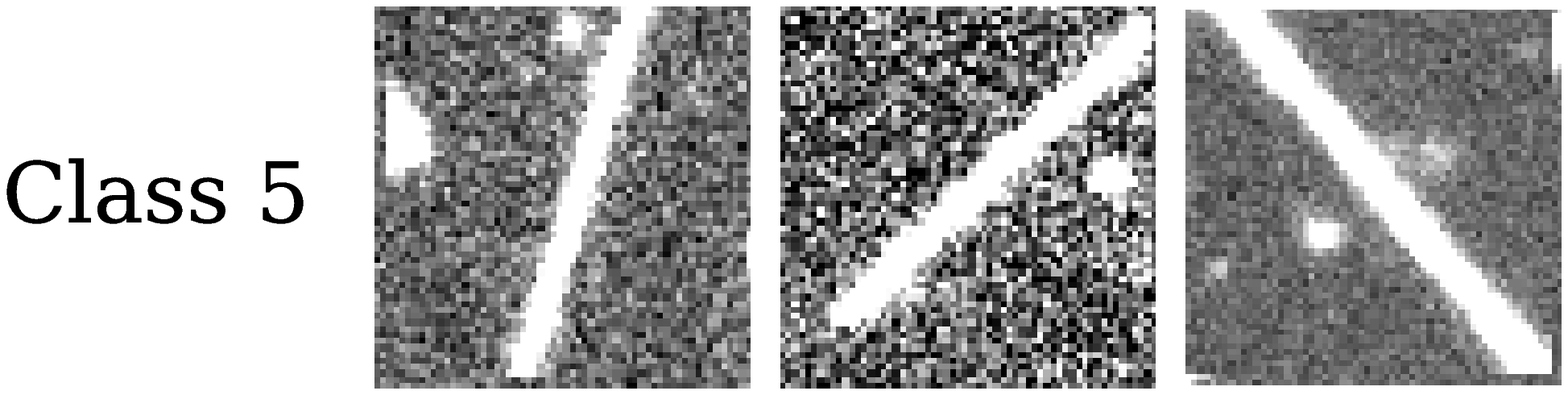}   
   \includegraphics[scale=0.42]{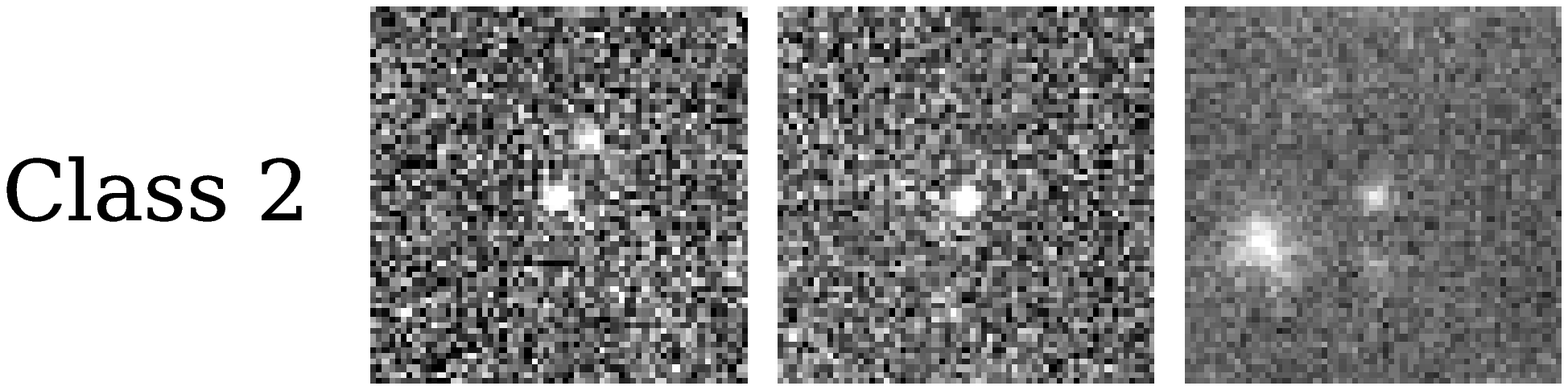}
   \includegraphics[scale=0.42]{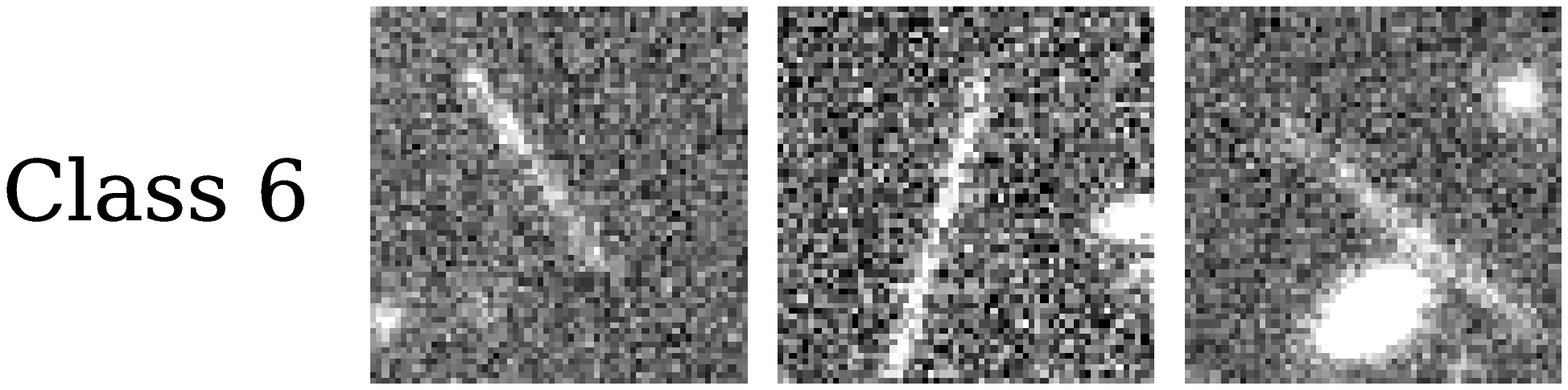}   
   \includegraphics[scale=0.42]{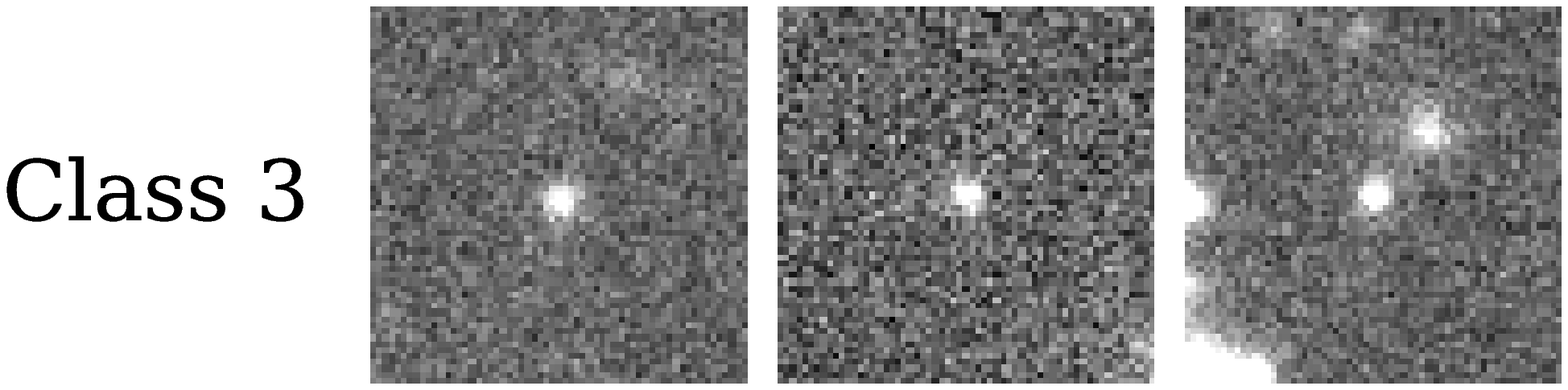}
   \includegraphics[scale=0.42]{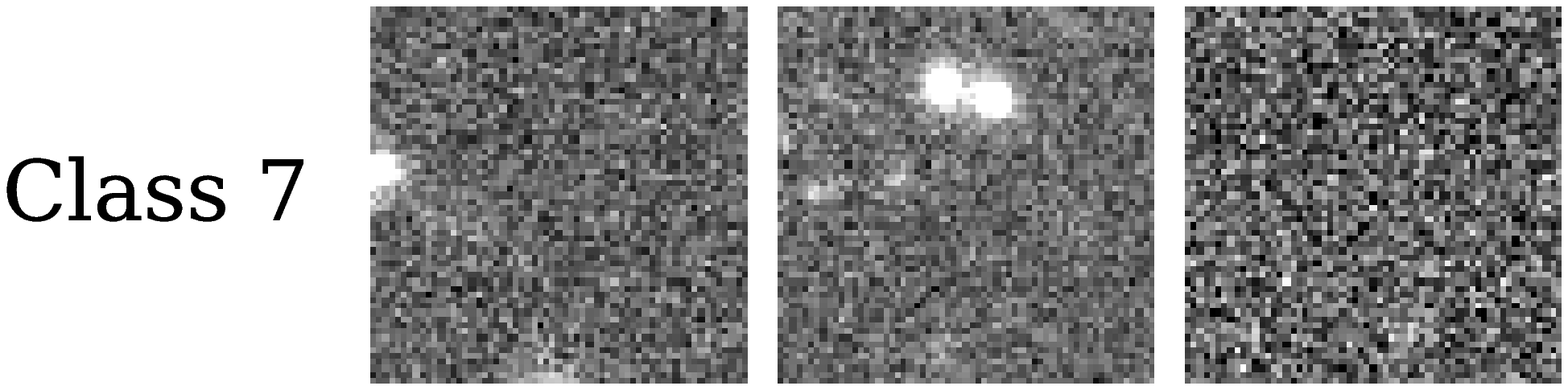}   
   \includegraphics[scale=0.42]{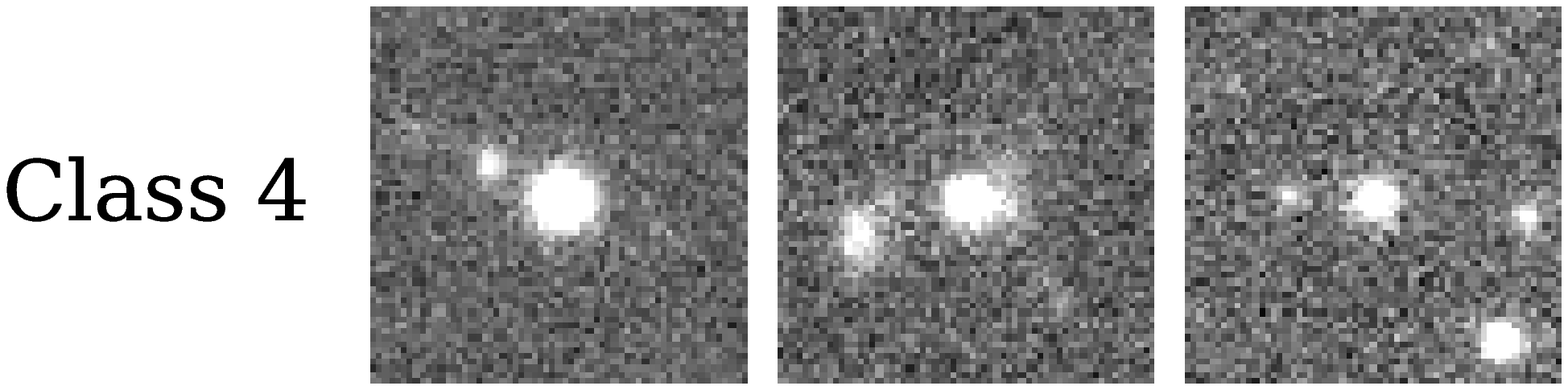}
   \includegraphics[scale=0.42]{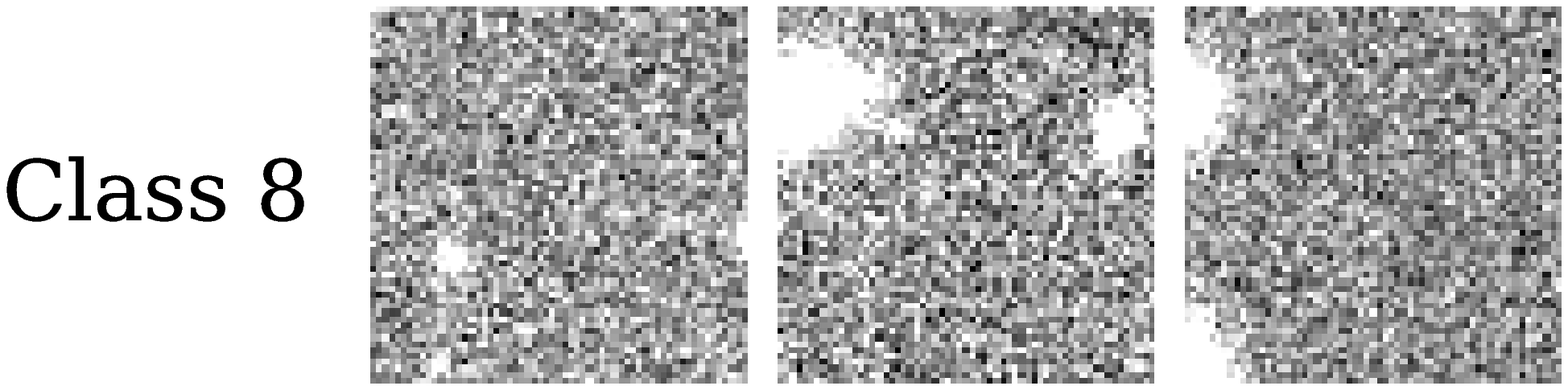}   
\caption[]
{
Example images of 
Classes 1--8. 
Classes 1, 2, 3, and 4 are simulated LAEs 
with S/Ns of $3$--$4$, $4$--$10$, $10$--$30$, and $30$--$200$, respectively. 
Classes 5 and 6 are bright and faint satellite trails, respectively. 
Classes 7 and 8 are randomly selected noise images 
without and with positive sky residuals, respectively. 
The size of the images is $\simeq 10'' \times 10''$. 
}
\label{fig:fig_example_images}
\end{center}
\end{figure*}
%FFFFFFFFFFFFFFFFFFFFFFFFFFFFFFFFFFFFFFFFFFFFFFFFFFFFFFFFFFFFFFFF%

%FFFFFFFFFFFFFFFFFFFFFFFFFFFFFFFFFFFFFFFFFFFFFFFFFFFFFFFFFFFFFFFF%
\begin{figure}[h]
\begin{center}
   \includegraphics[scale=0.3]{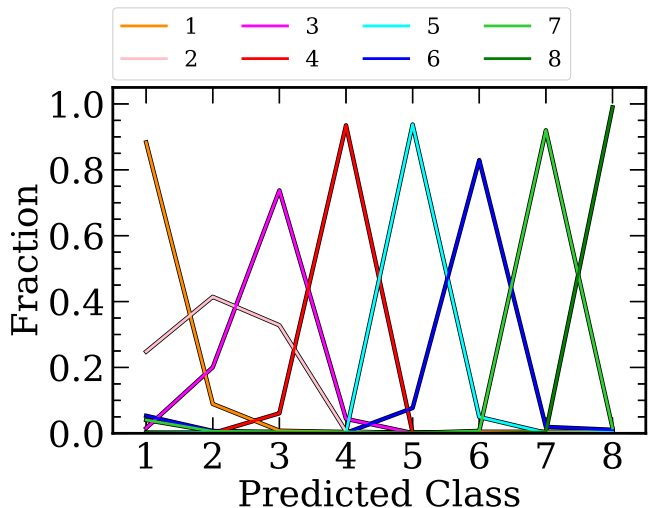}
\caption[]
{Results of the 8-label classification for the test data. 
The $x$-axis is the predicted class. 
The colored solid lines represent the fractions of classified images, 
and their colors denote the true classes  
(red: Class 1; 
light pink: Class 2; 
magenta: Class 3; 
red: Class 4; 
cyan: Class 5; 
blue: Class 6; 
green: Class 7; 
dark green: Class 8). 
}
\label{fig:accuracy}
\end{center}
\end{figure}
%FFFFFFFFFFFFFFFFFFFFFFFFFFFFFFFFFFFFFFFFFFFFFFFFFFFFFFFFFFFFFFFF%

%FFFFFFFFFFFFFFFFFFFFFFFFFFFFFFFFFFFFFFFFFFFFFFFFFFFFFFFFFFFFFFFF%
\begin{figure}[h]
\begin{center}
   \includegraphics[scale=0.36]{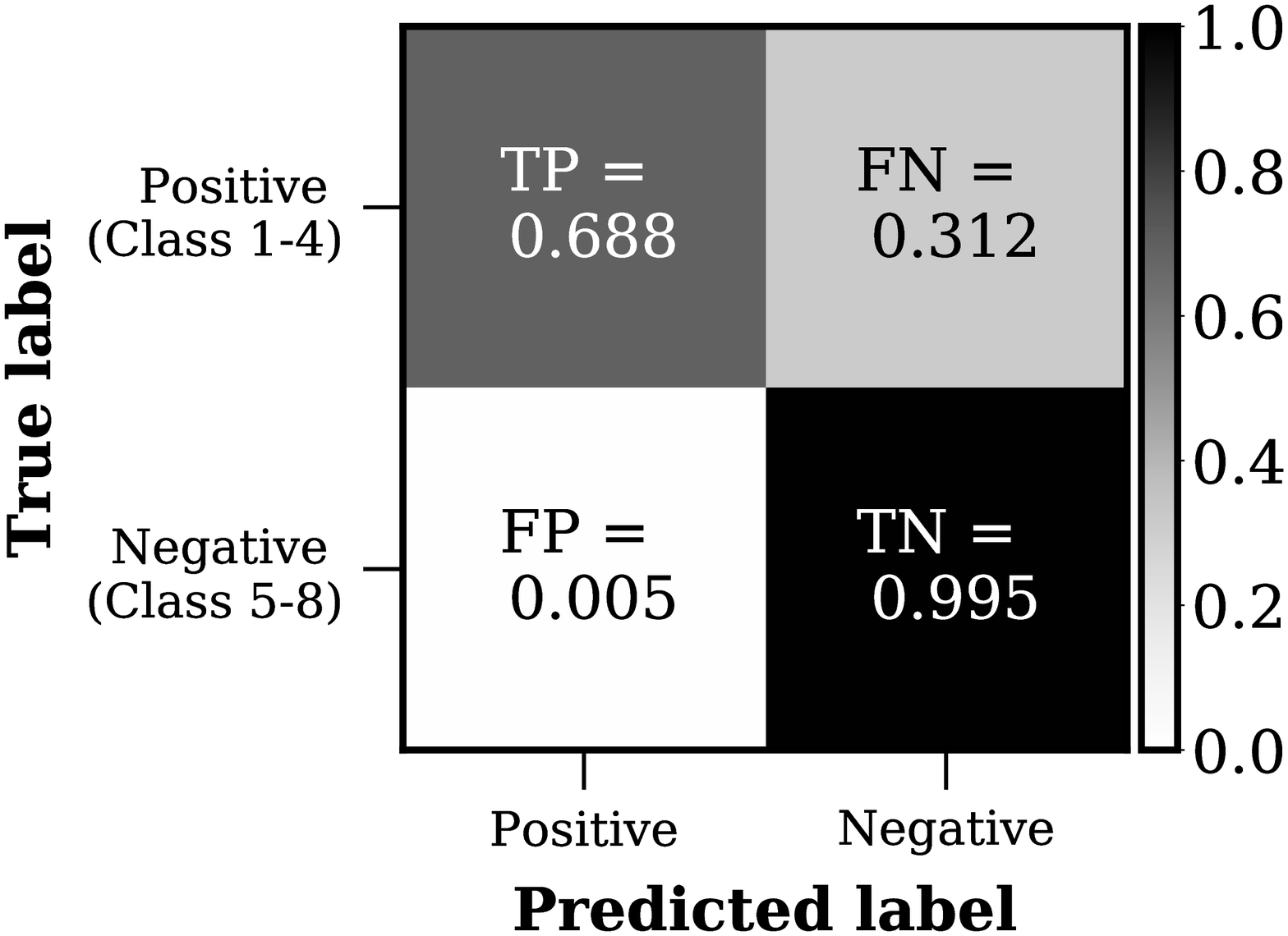}
   \includegraphics[scale=0.36]{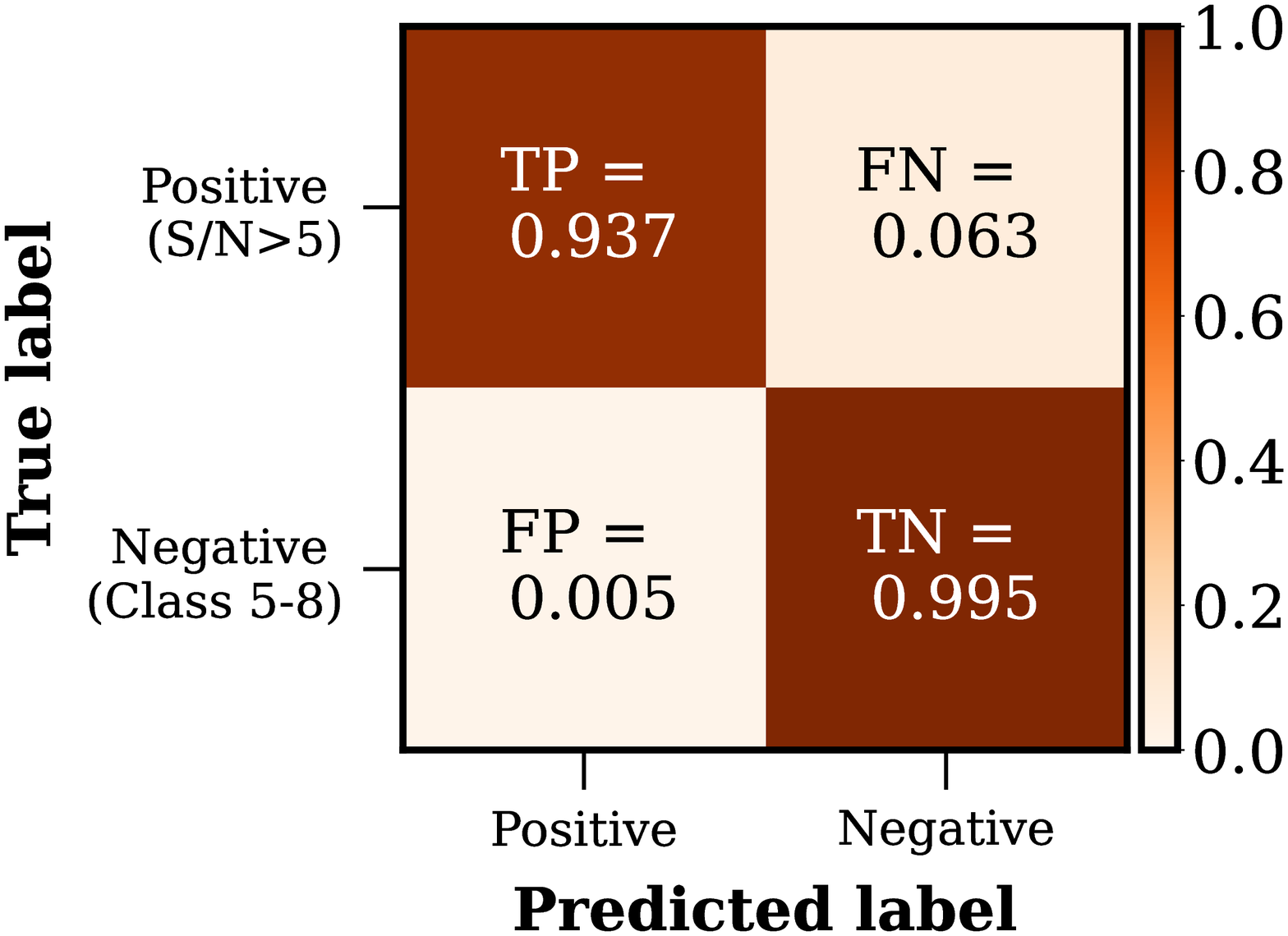}
\caption[]
{
Confusion matrix. 
In the top panel, 
Classes 1--4 (5--8) of the true labels are classified as positive (negative). 
For the predicted labels, 
we classify sources as positive 
if they show a high score of being an LAE 
(for details, see Section \ref{subsec:training}). 
The numbers in the cells denote  
true positives (TP), false negatives (FN), false positives (FP), and true negatives (TN). 
In the best case scenario, 
the confusion matrix has non-zero elements 
only in its diagonal cells and zero elements in the others.  
The bottom panel is the same as the top panel, 
except that positive for the true labels 
corresponds to simulated LAEs with S/N$>5$. 
}
\label{fig:confusion_matrix}
\end{center}
\end{figure}
%FFFFFFFFFFFFFFFFFFFFFFFFFFFFFFFFFFFFFFFFFFFFFFFFFFFFFFFFFFFFFFFF%

%FFFFFFFFFFFFFFFFFFFFFFFFFFFFFFFFFFFFFFFFFFFFFFFFFFFFFFFFFFFFFFFF%
\begin{figure}[h]
\begin{center}
   \includegraphics[scale=0.30]{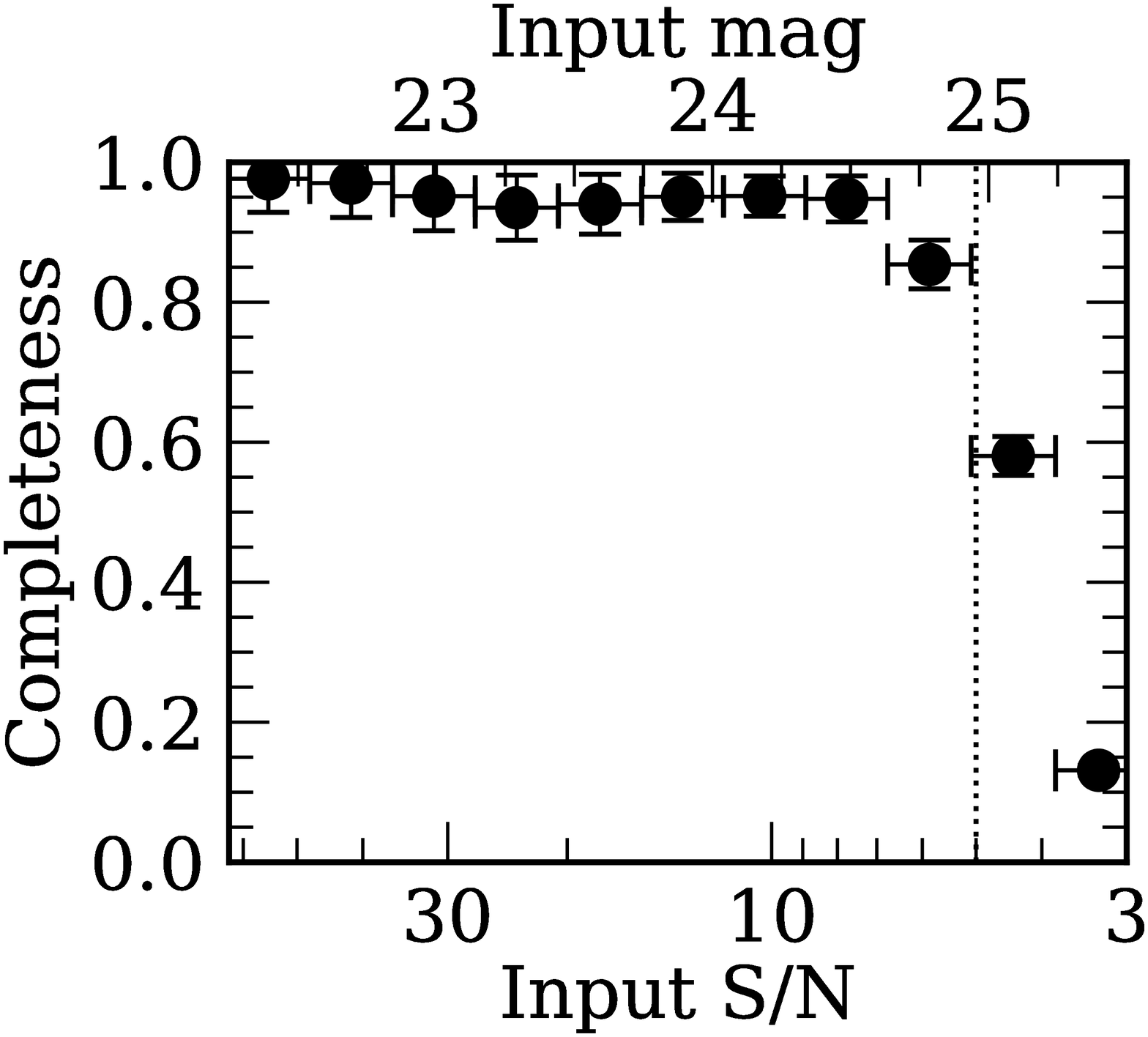}
\caption[]
{
Completeness as a function of input S/N   
with corresponding input magnitude 
where the $5\sigma$ limiting magnitude is about $25.0$ mag 
on the upper $x$-axis. 
The vertical dotted line denotes S/N $=5$. 
}
\label{fig:completeness_SNR}
\end{center}
\end{figure}
%FFFFFFFFFFFFFFFFFFFFFFFFFFFFFFFFFFFFFFFFFFFFFFFFFFFFFFFFFFFFFFFF%

The numbers of LAE candidates that are photometrically selected with the color criteria described above are 
$258670$ for $z=2.2$ LAEs, 
$13937$ for $z=3.3$ LAEs, 
$2544$ for $z=4.9$ LAEs, 
$186386$ for $z=5.7$ LAEs, 
$176129$ for $z=6.6$ LAEs, 
and 
$21789$ for $z=7.0$ LAEs.
These large numbers include 
many contaminants such as 
satellite trails and low-$z$ emission line galaxies.\footnote{Our data are significantly contaminated by satellite trails 
probably because the coadded images generated by the HSC pipeline 
are computed as a direct weighted average of CCD images 
where the weights are calculated from the inverse of the mean variance of the input images \citep{2018PASJ...70S...5B}. 
In contrast, previous studies using Subaru Suprime-Cam 
have masked out bad data areas including satellite trails before stacking of single exposure images, 
and removed outlier pixels in the process of stacking with the rejected-mean algorithm 
(e.g., \citealt{2008ApJS..176..301O}; \citealt{2010ApJ...723..869O}).} 
As mentioned in Section \ref{sec:introduction}, 
in our previous work, 
\cite{2018PASJ...70S..14S} performed visual inspection 
for the BB and NB images of their photometrically selected 121000 LAE candidates, 
which took about two person-months. 
In this study, the number of photometrically selected LAE candidates is 
$258670 + 13937 + 2544 + 186386 + 176129 + 21789 = 659455$. 
If we scale the results of \cite{2018PASJ...70S..14S}, 
the amount of human effort that is needed for visual inspection of 
all the photometrically selected LAE candidates 
is estimated to be about 11 person-months. 
In this study, 
we apply a machine learning technique 
for removal of obvious contaminants to overcome this problem. 
This technique allows us to significantly reduce the required human resources.

%%%%%%%%%%%%%%%%%%%%%%%%%%%%%%%%%%%%%%%%%%%%%%%%%%%%%%%%%%%%%%%%%
%%%%%%%%%%%%%%%%%%%%%%%%%%%%%%%%%%%%%%%%%%%%%%%%%%%%%%%%%%%%%%%%%
\section{Machine Learning Selection} \label{sec:deep_learning_selection}
%%%%%%%%%%%%%%%%%%%%%%%%%%%%%%%%%%%%%%%%%%%%%%%%%%%%%%%%%%%%%%%%%
%%%%%%%%%%%%%%%%%%%%%%%%%%%%%%%%%%%%%%%%%%%%%%%%%%%%%%%%%%%%%%%%%

%%%%%%%%%%%%%%%%%%%%%%%%%%%%%%%%%%%%%%%%%%%%%%%%%%%%%%%%%%%%%%%%%
\subsection{Implementation} \label{subsec:implementation}
%%%%%%%%%%%%%%%%%%%%%%%%%%%%%%%%%%%%%%%%%%%%%%%%%%%%%%%%%%%%%%%%%

We adopt a machine learning based selection 
to remove contaminants such as 
satellite trails and low-$z$ galaxies 
from the photometrically selected LAE candidate catalogs. 
For this purpose, 
we train a CNN 
so that it can distinguish 
between real LAEs, 
satellite trails, and 
noise images  
based on single band images.  
Our strategy is, 
with the trained CNN, 
to select sources as LAE candidates that are classified 
as real LAEs 
in NB probing Ly$\alpha$ emission 
and also classified as noise images in BB probing a wavelength range bluer than the Lyman limit 
(Section \ref{sec:results}).

%ttttttttttttttttttttttttttttttttttttttttttttttttttttttttttttttttttttttttt%
\capstartfalse
\begin{deluxetable*}{ccccccc} 
\tablecolumns{7} 
\tablewidth{0pt} 
\tablecaption{Number of selected LAEs
\label{tab:LAE_number}}
\tablehead{
    \colhead{Field}     
    &  \colhead{$z=2.2$}
    &  \colhead{$z=3.3$}
    &  \colhead{$z=4.9$}
    &  \colhead{$z=5.7$}
    &  \colhead{$z=6.6$}
    &  \colhead{$z=7.0$}
    \\
    \colhead{(1)}
    &  \colhead{(2)}
    &  \colhead{(3)}
    &  \colhead{(4)}
    &  \colhead{(5)}
    &  \colhead{(6)}
    &  \colhead{(7)}
}
\startdata 
%            z=2.2   z=3.3   z=4.9   z=5.7   z=6.6   z=7.0
UD-SXDS    &  ---  &  ---  &  ---  &  560  &   75  & ---   \\
UD-COSMOS  &  542$^{\dagger1}$  &  959$^{\dagger1}$  &  349$^{\dagger1}$  &  395  &  150  &  40$^{\dagger1}$   \\
D-SXDS     &  850  &  ---  &  ---  &  532$^{\dagger3}$  & ---   & ---   \\
D-COSMOS   & 2173$^{\dagger2}$  &  ---  &  ---  & ---   &  111$^{\dagger4}$  & ---   \\
D-ELAIS-N1 &  ---  &  ---  &  ---  &  409  &  170  & ---   \\
D-DEEP2-3  &  844  &  ---  &  ---  &  985  &  174  & ---   \\
\hline 
Total$^{\dagger5}$      & 4409  &  959  &  349  & 2881  &  680  &  40 
\enddata 
\tablecomments{
(1) Field name. 
(2)--(7) Number of selected LAEs at $z =2.2$--$7.0$. 
}
\tablenotetext{$\dagger1$}{%
Based on the CHORUS data. 
}
\tablenotetext{$\dagger2$}{%
This number includes $213$ LAEs selected in UD-COSMOS. 
}
\tablenotetext{$\dagger3$}{%
This number includes $3$ LAEs selected in UD-SXDS. 
}
\tablenotetext{$\dagger4$}{%
This number includes $4$ LAEs selected in UD-COSMOS. 
}
\tablenotetext{$\dagger5$}{%
The total number of our LAE candidates is $4409 + 959 + 349 + 2881 + 680 + 40 = 9318$.
}
\end{deluxetable*} 
%ttttttttttttttttttttttttttttttttttttttttttttttttttttttttttttttttttttttttt%

The CNN architecture used in this study is presented in Table \ref{tab:architecture}. 
We input a $50 \times 50$ pixel (corresponding to $ 8\farcs4 \times 8\farcs4$) cutout image centered on a galaxy coordinate. 
Our CNN consists of four convolutional layers with $5 \times 5$ filters,\footnote{Although 
we also try a filter size of $3 \times 3$ for the convolutional layers, which is the same as that of \cite{2018ApJ...858..114H}, 
we find that the accuracy is better with $5 \times 5$ filters, rather than $3 \times 3$. 
This is probably because of the different PSF sizes and pixel scales of the images 
used for our study compared to those for \cite{2018ApJ...858..114H}. 
We use images taken with Subaru/HSC, while \cite{2018ApJ...858..114H} use mock \textit{Hubble}/WFC3 images.}   
two average pooling layers, and two fully connected layers. 
Although many previous studies use max pooling for image classification problems, 
we find that average pooling performs better for our purpose, 
probably because background noises in our data are relatively high 
compared to those in typical image classification work 
(see also \citealt{2019A&A...621A..26P}).  
Dropout regularization is performed after two convolutional layers 
to reduce the chance of overfitting 
by randomly dropping a half of the output neurons during training. 
The hidden layers use the ReLU (linear rectifier) activation function \citep{Nair2010}, 
and the output layer uses the softmax function. 
Because the slope of the ReLU activation function is zero for negative input values, 
we add $+0.2$ to the pixel counts in the input images 
so that most of the pixels have positive values.  
The output of the fully connected layer for each category 
is a predicted score; 
a higher score for a category means that the input is more likely to be classified as the category. 
In the output layer, 
we have eight classes in total, 
depending on their types 
and S/Ns (Table \ref{tab:output_classes}; 
for details, see Section \ref{subsec:training}). 
We use the Adam optimization algorithm (\citealt{2014arXiv1412.6980K}) 
to minimize the cross-entropy error function over training data. 
The basic idea of this network architecture is motivated by previous similar studies 
(\citealt{2018ApJ...858..114H}; see also, 
\textcolor{blue}{M. Okura et al. in prep.}).
We implement this network by using the Python neural network library Keras 
(\citealt{2018ascl.soft06022C}) 
with the TensorFlow backend 
(\citealt{2016arXiv160304467A}).

%%%%%%%%%%%%%%%%%%%%%%%%%%%%%%%%%%%%%%%%%%%%%%%%%%%%%%%%%%%%%%%%%
\subsection{Training} \label{subsec:training}
%%%%%%%%%%%%%%%%%%%%%%%%%%%%%%%%%%%%%%%%%%%%%%%%%%%%%%%%%%%%%%%%%

For the training of the CNN, 
we create data sets for LAEs. 
Because the number of spectroscopically identified LAEs is limited, 
we make simulated images of LAEs 
by using the SynPipe software \citep{2018PASJ...70S...6H}, 
which utilizes GalSim v1.4 \citep{2015A&C....10..121R} 
and the HSC pipeline. 
The simulated LAE images are generated 
so that they have various S/Ns of $3$--$200$ 
in $2\farcs0$ diameter circular apertures. 
The size distribution of these simulated LAEs follows 
the observational results of \cite{2017A&A...608A...8L}. 
Specifically, we create a histogram of 
the $r_{S_\mathrm{cont}}$ values presented in Table B.1 of \cite{2017A&A...608A...8L} 
and generate random size values that follow the distribution. 
We adopt a uniform distribution of S\'ersic indices in the range of $1$--$3$ 
to consider the scatter of the measured S\'ersic indices of LAEs 
\citep{2019ApJ...871..164S}. 
The SynPipe software inserts the simulated LAE images into HSC images at random positions. 
Although these simulated images allow the CNN to learn 
what most LAEs are like, 
it should be noted that there are also rare, extended LAEs (Ly$\alpha$ blobs) 
and Ly$\alpha$-emitting merging galaxies
some of which may be excluded 
in our machine learning selection. 
We will check this issue in Section \ref{sec:results} 
by comparing our final LAE sample and previously spectroscopically identified LAEs. 
Because of the wide S/N range of the simulated LAEs, 
we divide them into four classes according to their S/Ns 
to improve the accuracy of our CNN classification. 
Specifically, Classes 1, 2, 3, and 4 consist of simulated LAEs with S/Ns of 
$3$--$4$, $4$--$10$, $10$--$30$, and $30$--$200$, respectively. 
The number of simulated images in each class is 12500.
Their example images are shown in Figure \ref{fig:fig_example_images}.

In addition, we collect data sets for contaminants. 
We create two classes for satellite trails depending on their S/Ns; 
Classes 5 and 6 are bright and faint satellite trails  
that are labelled in our previous visual inspection \citep{2018PASJ...70S..14S}. 
Since the numbers of labelled satellite trails are limited, 
we augment their images by applying $90$, $180$, and $270$ degree rotations 
and horizontal and vertical flips. 
We also create two classes for null detections. 
Class 7 corresponds to randomly selected noise images 
and Class 8 is the same as Class 7 but shows positive sky residuals of $\simeq 0.1$--$0.3$. 
Because such noise images can be selected as LAE candidates due to noise fluctuations,  
these classes are also categorized as contaminants. 
The number of these images in each class is 12500, 
and their example images are presented in Figure \ref{fig:fig_example_images}. 
Note that, in the photometrically selected sample,  
there should be some contaminants that are not considered in the training data for the CNN 
such as cosmic rays and diffuse artifacts. 
Although such contaminants would not be completely excluded 
in this machine learning selection, 
we perform visual inspection to remove the remaining ones afterwards. 
The main advantage of this machine learning selection is that 
it enables us to remove many obvious contaminants automatically.

In total, we have eight classes as summarized in Table \ref{tab:output_classes}.  
We divide the data for each class into two subsets: 
training (80{\%}) and test (20{\%}). 
We use the training data to tune model parameters of the CNN during development, 
applying a five-fold cross-validation procedure.
The test data are used for evaluation of the developed model performance.   
Note that the model performance does not depend on redshift,  
because the size distribution of \cite{2017A&A...608A...8L} 
and the S\'ersic indices considering the results of \cite{2019ApJ...871..164S} 
are adopted to create the simulated LAE images for the training, 
which are sufficient to account for morphologies of most LAEs 
in the redshift range of $z=2.2$--$7.0$.
We thus apply the trained CNN model to all the redshift samples. 
The differences of absolute noise values in different input images are taken into account 
by scaling the input images as described in Section \ref{sec:results}.

%ttttttttttttttttttttttttttttttttttttttttttttttttttttttttttttttttttttttttt%
\capstartfalse
\begin{deluxetable*}{cccccc} 
\tablecolumns{6} 
\tablewidth{0pt} 
\tablecaption{Summary of LAE searches used in our number count comparisons 
\label{tab:previou_number_count_work}}
\tablehead{
    \colhead{Reference}     
    &  \colhead{Redshift}     
    &  \colhead{Ly$\alpha$ $EW_0$ limit}
    &  \colhead{$m_{{\rm NB}, 5\sigma}$}
    &  \colhead{Area}
    &  \colhead{\# of LAE candidates}
   \\
    \colhead{ }
    &  \colhead{ }
    &  \colhead{({\AA})}
    &  \colhead{(mag)}
    &  \colhead{(deg$^2$)}
    &  \colhead{ }
    \\
    \colhead{(1)}
    &  \colhead{(2)}
    &  \colhead{(3)}
    &  \colhead{(4)}
    &  \colhead{(5)}
    &  \colhead{(6)}
}
\startdata 
\cite{2012ApJ...745...12N,2013ApJ...769....3N} & $2.2$ & $30$ &  25.6--26.5 & $1.2$$^{\dagger1}$ & 3169$^{\dagger1}$  \\
\cite{2018ApJ...864..145H} & $2.2$ & $20$--$30$ &  26.0$^{\dagger2}$ & 0.34 & 475  \\
This study & $2.2$ & $20$ &  24.3--25.7 & 20.6 & 4409  \\
\hline 
\cite{2008ApJS..176..301O} & $3.1$ & $64$ &  25.1--25.5 & 1.0 & 356$^{\dagger3}$  \\
This study & $3.3$ & $20$ &  26.4 & 1.6 & 959  \\
\cite{2008ApJS..176..301O} & $3.7$ & $44$ &  24.6--25.2 & 1.0 & 101$^{\dagger3}$  \\
\hline 
\cite{2020ApJ...891..177Z} & $4.9$ & $20$ &  26.2$^{\dagger4}$ & 1.6 & 141  \\
This study & $4.9$ & $20$ &  25.6 & 1.6 & 349  \\
\hline 
\cite{2008ApJS..176..301O} & $5.7$ & $27$ &  25.9--26.1 & 1.0 & 401$^{\dagger3}$  \\
\cite{2018PASJ...70S..14S} & $5.7$ & $10$ &  25.2--25.7$^{\dagger4}$ & 13.8 & 1077  \\
This study & $5.7$ & $10$ &  25.2--25.7 & 23.7 & 2881  \\
\hline 
\cite{2010ApJ...723..869O} & $6.6$ & $14$ &  25.6--25.8 & 0.90 & 207  \\
\cite{2018PASJ...70S..14S} & $6.6$ & $\sim10$$^{\dagger5}$ &  24.9--25.6$^{\dagger4}$ & 21.2 & 1153  \\
This study & $6.6$ & $\sim10$$^{\dagger5}$ &  24.9--25.5 & 25.0 & 680  \\
\hline 
 \cite{2018ApJ...867...46I} & $7.0$ & $20$ &  25.0$^{\dagger4}$ & 3.1$^{\dagger6}$ & 34  \\
  This study & $7.0$ & $20$ &  24.9 & 1.6 & 40  
\enddata 
\tablecomments{
(1) References.
(2) Redshifts.  
(3) Ly$\alpha$ EW limits in the rest frame. 
(4) $5\sigma$ limiting magnitudes in NBs measured with $2\farcs0$ diameter circular apertures unless otherwise mentioned. 
(5) Effective area in deg$^2$. 
(6) Numbers of LAE candidates.}
\tablenotetext{$\dagger1$}{%
We consider their LAE catalogs for the SXDS-C, SXDS-N, SXDS-S, COSMOS, GOODS-N, and GOODS-S fields. 
We do not include the SXDS-W and SSA22 fields due to the shallowness of the $NB387$ data compared to the other fields. 
}
\tablenotetext{$\dagger2$}{%
Measured with $3\farcs0$ diameter circular apertures. 
}
\tablenotetext{$\dagger3$}{%
These numbers correspond to those of their photometrically selected LAE candidates. 
The numbers of LAEs in their spectroscopic samples are $41$, $26$, and $17$ at $z=3.1$, $z=3.7$, and $z=5.7$, respectively. 
}
\tablenotetext{$\dagger4$}{%
Measured with $1\farcs5$ diameter circular apertures. 
}
\tablenotetext{$\dagger5$}{%
See also \cite{2018PASJ...70S..16K}. 
}
\tablenotetext{$\dagger6$}{%
Although they observe the COSMOS and SXDS fields, 
whose effective areas are $1.64$ deg$^2$ and $1.50$ deg$^2$, respectively, 
only two candidates are selected in the SXDS field 
due to the shallower depth of the NB data by about $0.7$ mag. 
}
\end{deluxetable*} 
%ttttttttttttttttttttttttttttttttttttttttttttttttttttttttttttttttttttttttt%

%ttttttttttttttttttttttttttttttttttttttttttttttttttttttttttttttttttttttttt%
\capstartfalse
\begin{deluxetable}{ccc} 
\tablecolumns{3} 
\tablewidth{0pt} 
\tablecaption{Fitting results of the filter transmission functions for the redshift distributions
\label{tab:redshift_distribution_fitting_results}}
\tablehead{
    \colhead{Filter} 
    &  \colhead{$\alpha$}
    &  \colhead{$\beta$}
    \\
    \colhead{(1)}
    &  \colhead{(2)}
    &  \colhead{(3)}
}
\startdata 
$NB387$ & $41.64^{+11.08}_{-9.40}$ & $-0.002^{+0.003}_{-0.004}$ \\
$NB718$ & $4.58^{+2.00}_{-1.57}$ & $-0.030^{+0.016}_{-0.018}$ \\ 
$NB816$ & $32.90^{+3.70}_{-3.47}$ & $-0.020^{+0.002}_{-0.003}$ \\ 
$NB921$ & $12.00^{+3.11}_{-2.91}$ & $-0.023^{+0.014}_{-0.010}$  
\enddata 
\tablecomments{
(1) Filter. 
(2)--(3) Best-fit normalization constant and shift along the $x$-axis derived from the two parameter fitting. 
}
\end{deluxetable} 
%ttttttttttttttttttttttttttttttttttttttttttttttttttttttttttttttttttttttttt%

Figure \ref{fig:accuracy} shows the $8$-label classification results for the test data. 
We classify each image into a class where a highest predicted score is obtained. 
All the classes show highest fractions for the true classes. 
In particular, Classes 1, 4, 5, 6, 7, 8 achieve high accuracies of $\gtrsim 80$--$90${\%} for the true classes. 
However, Classes 2 and 3 show relatively low accuracies, 
because non-negligible fractions of images in Classes 2 and 3 
are classified into the other classes of LAEs with different S/Ns.

%FFFFFFFFFFFFFFFFFFFFFFFFFFFFFFFFFFFFFFFFFFFFFFFFFFFFFFFFFFFFFFFF%
\begin{figure*}
\begin{center}
   \includegraphics[scale=0.25]{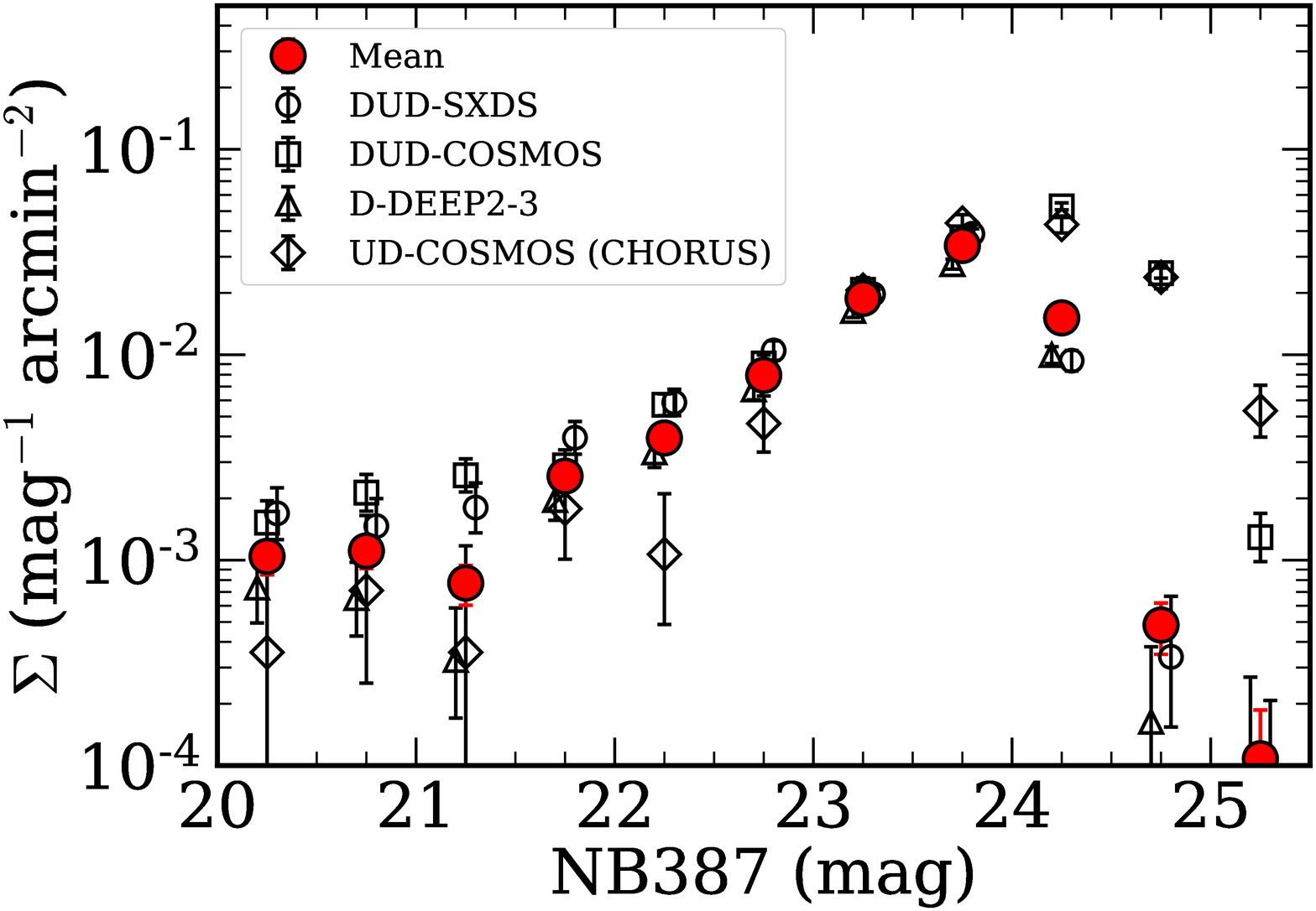}
   \includegraphics[scale=0.25]{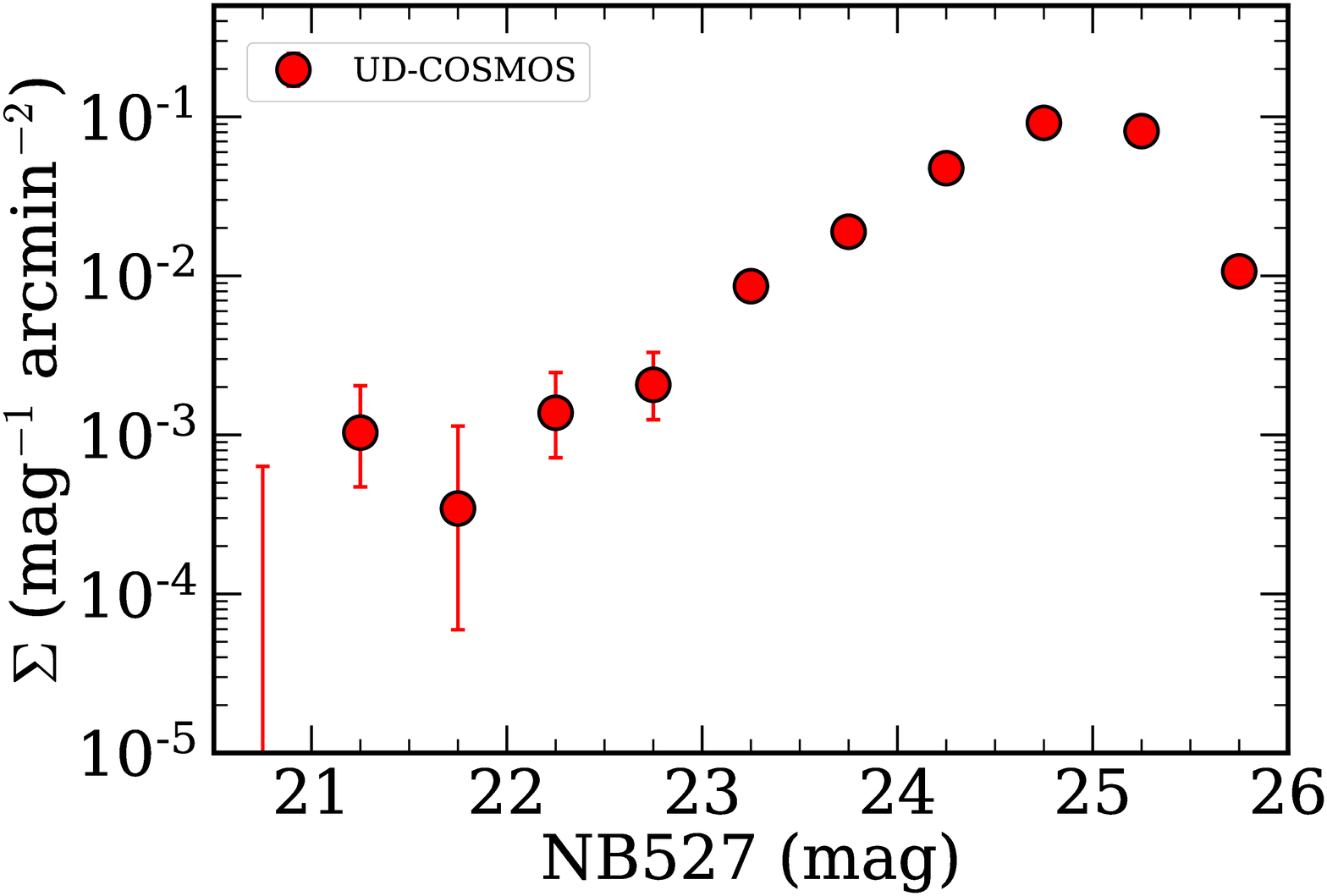}
   \includegraphics[scale=0.25]{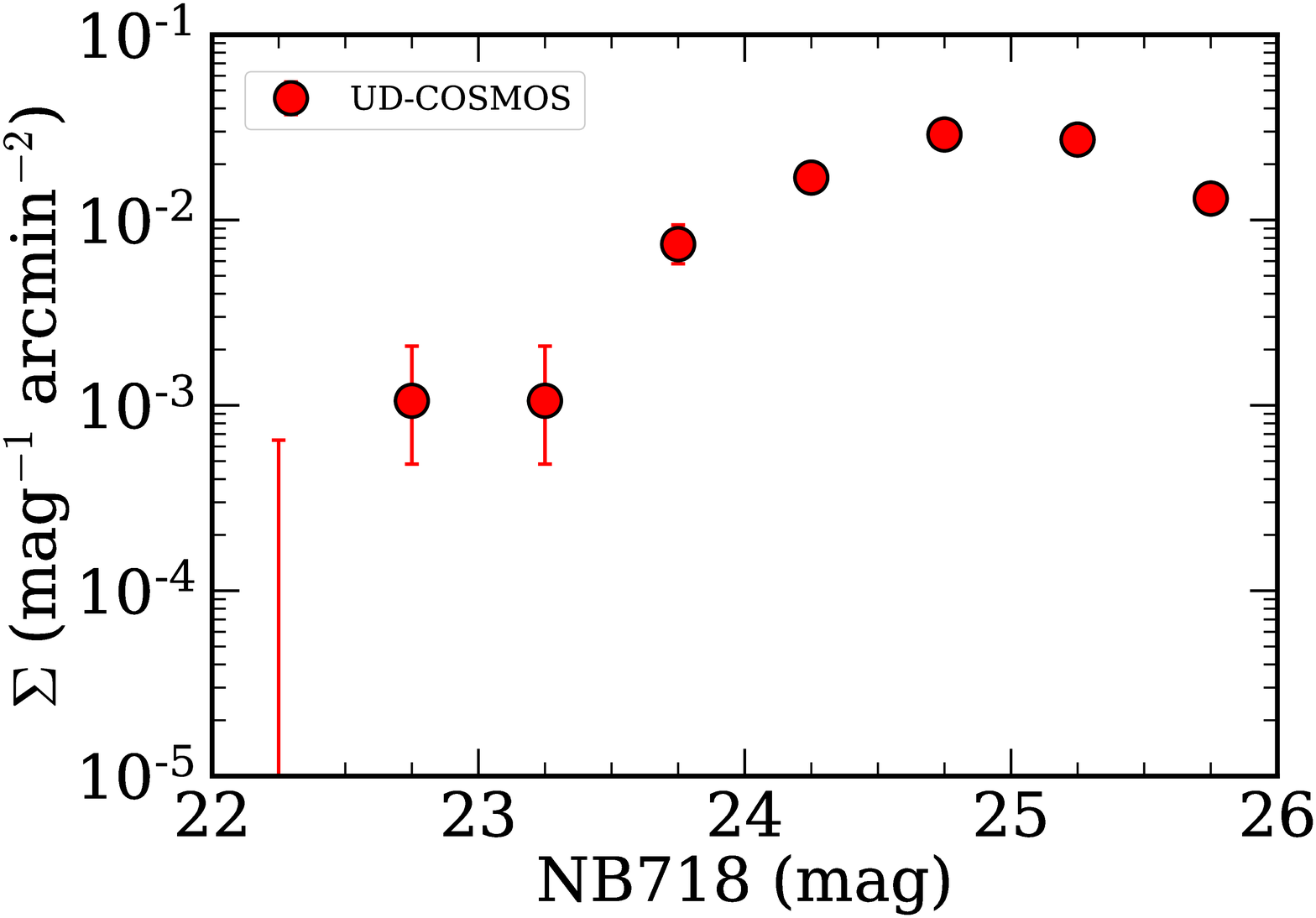}
   \includegraphics[scale=0.25]{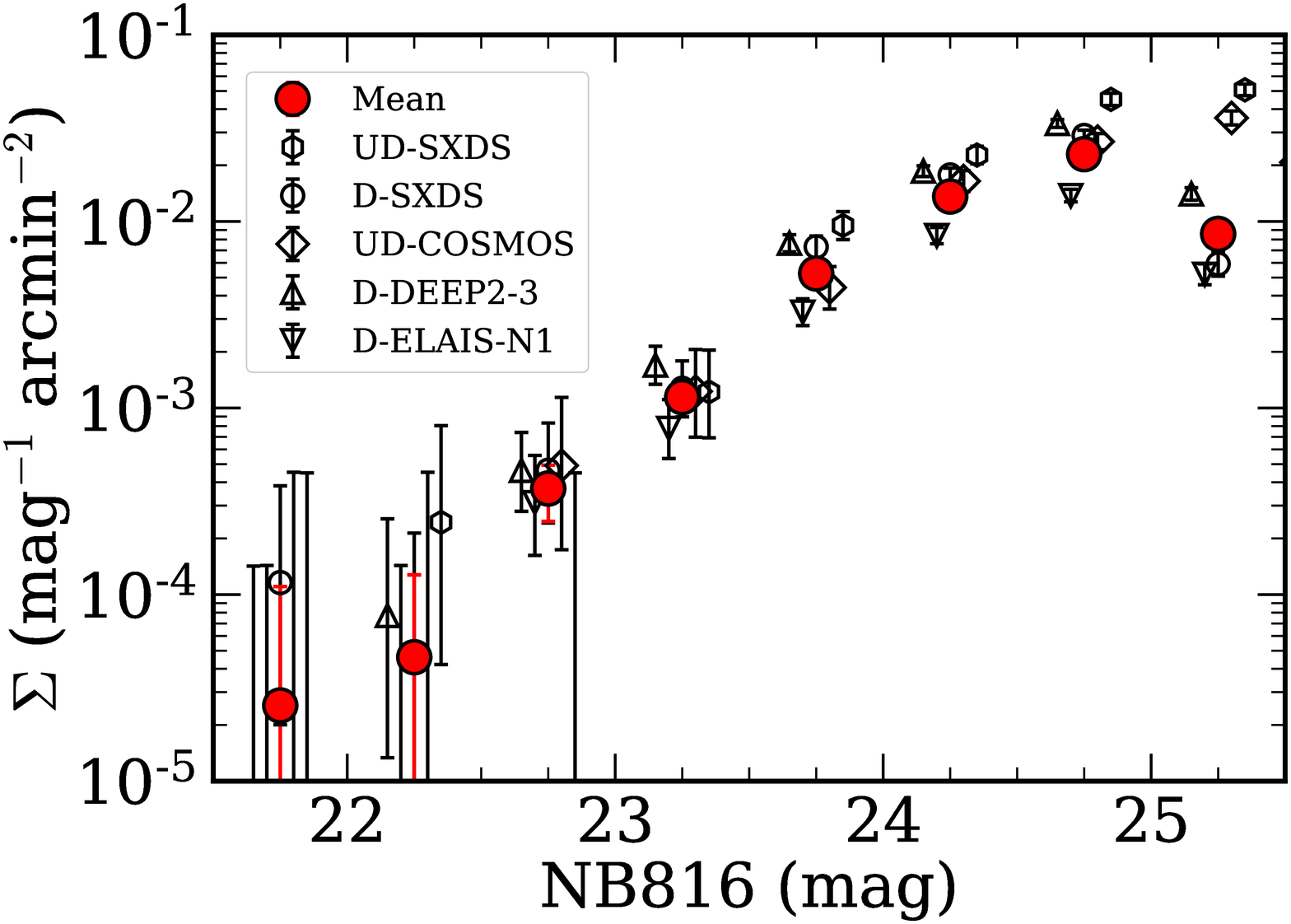}
   \includegraphics[scale=0.25]{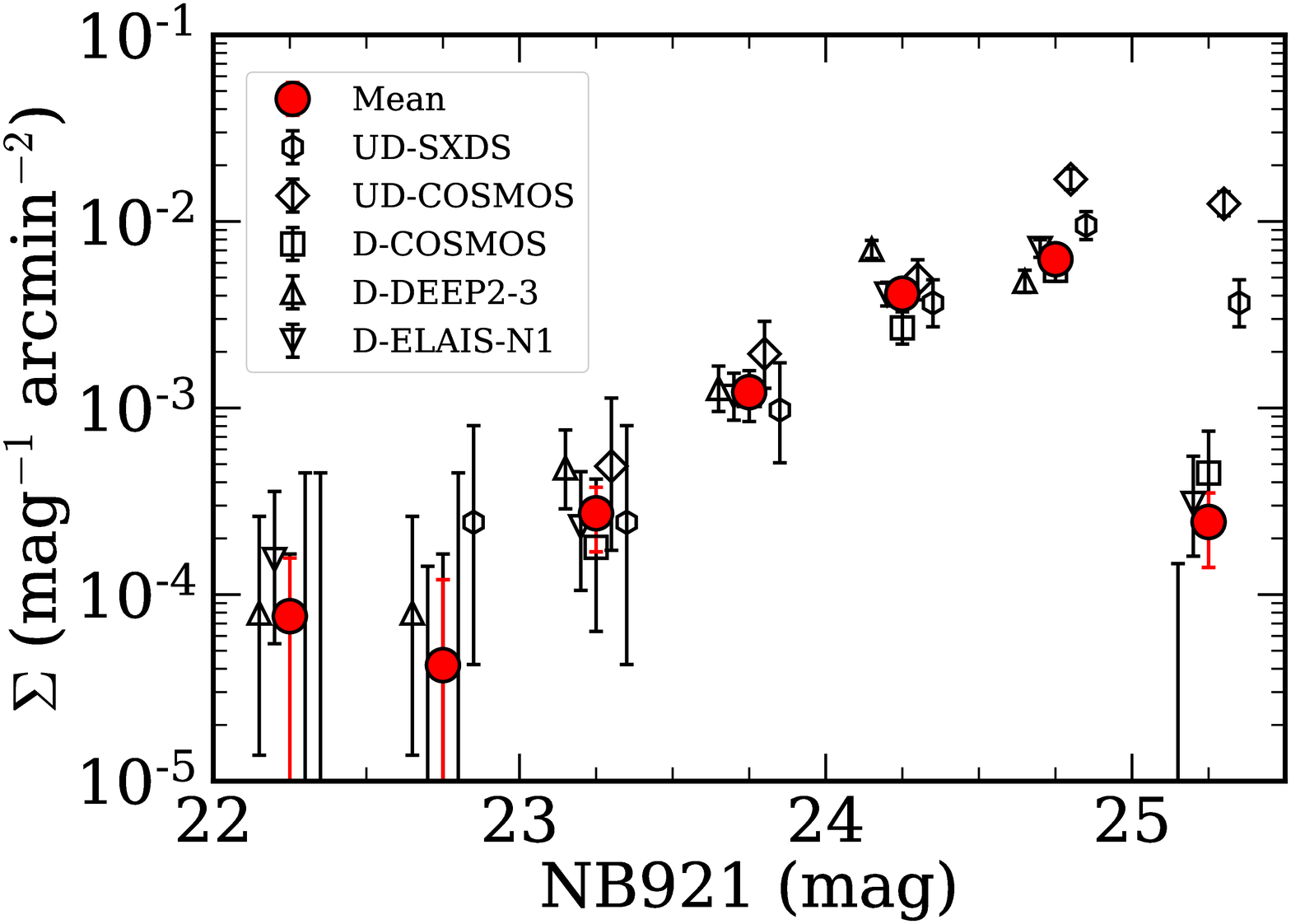}
   \includegraphics[scale=0.25]{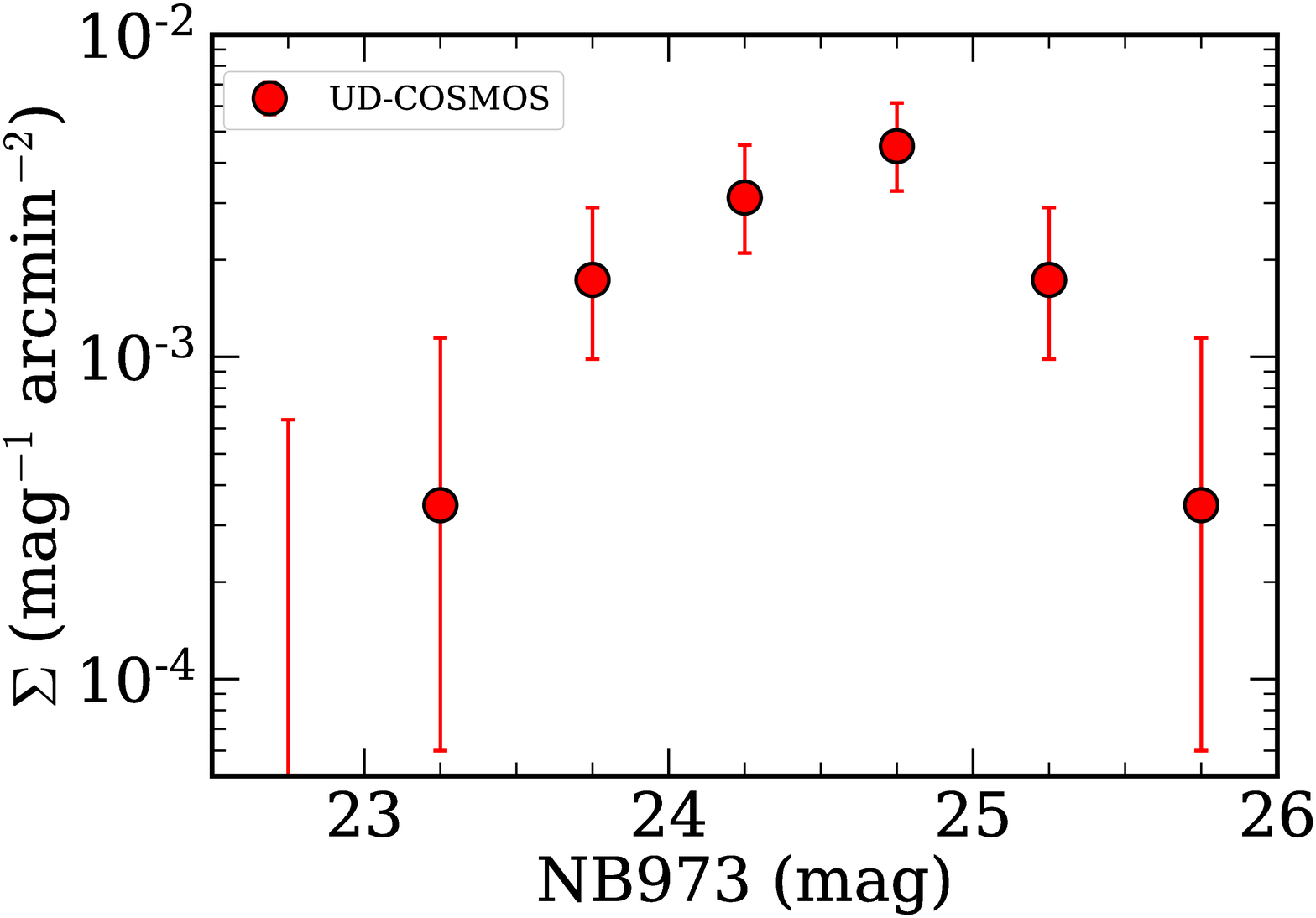}
\caption[]
{
Surface number density of LAEs at 
$z=2.2$ (top left), $z=3.3$ (top right), 
$z=4.9$ (middle left), $z=5.7$ (middle right), 
$z=6.6$ (bottom left), and $z=7.0$ (bottom right) 
as a function of total NB magnitude. 
For redshifts where multiple subfields are observed, 
we show 
the averaged results with red circles and 
the subfield results with black symbols.
These surface number density values are not corrected for completeness. 
The subfield results are slightly offset for clarity. 
}
\label{fig:surface_number_density_for_each_field}
\end{center}
\end{figure*}
%FFFFFFFFFFFFFFFFFFFFFFFFFFFFFFFFFFFFFFFFFFFFFFFFFFFFFFFFFFFFFFFF%

Although we divide the input data into the eight classes in total, 
our interest is basically whether an object is classified as 
positive (i.e., LAEs) 
or negative (i.e., contaminants such as satellite trails and randomly selected noise images). 
Thus, we use the CNN as a classifier that works on the binary classification problem. 
We define sources as positive 
if they show a high score of being an LAE,  
satisfying  
$p_1 + p_2 + p_3 + p_4 > 0.95$ 
and 
$p_2 + p_3 + p_4 > 0.5$, 
where $p_i$ denotes the 
score of Class $i$. 
Sources that do not satisfy these criteria are classified 
as negative ones.

Figure \ref{fig:confusion_matrix} shows the confusion matrices for our binary classification results. 
For binary classification problems, confusion matrices  
split predictions into true positives (TPs, i.e., simulated LAEs recovered by the CNN), 
false positives (FPs, i.e., contaminants that are selected as LAEs by the CNN), 
false negatives (FNs, i.e., simulated LAEs that are not selected by the CNN), 
and 
true negatives (TNs, i.e., contaminants recovered by the CNN).

%FFFFFFFFFFFFFFFFFFFFFFFFFFFFFFFFFFFFFFFFFFFFFFFFFFFFFFFFFFFFFFFF%
\begin{figure*}
\begin{center}
   \includegraphics[scale=0.25]{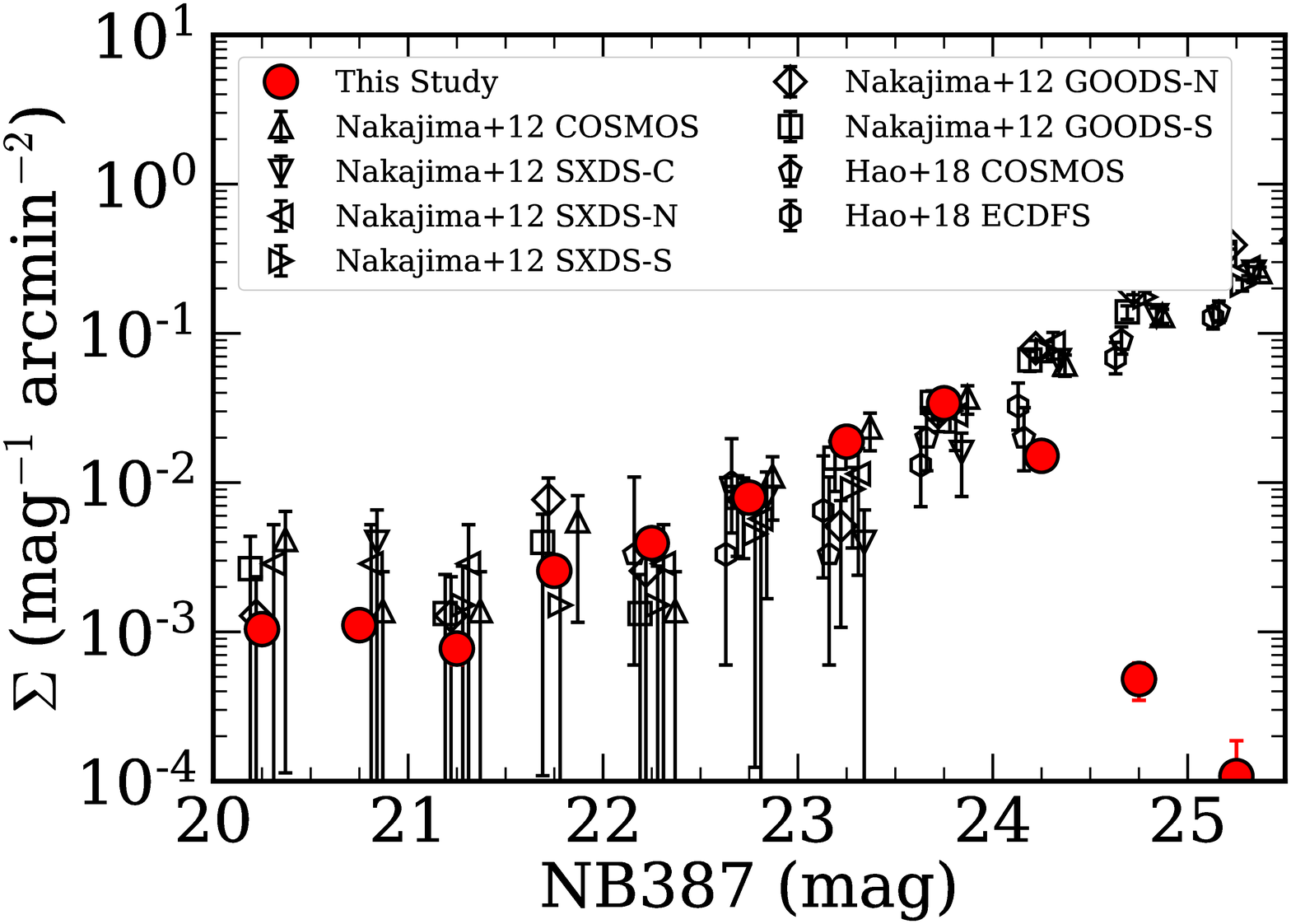}
   \includegraphics[scale=0.25]{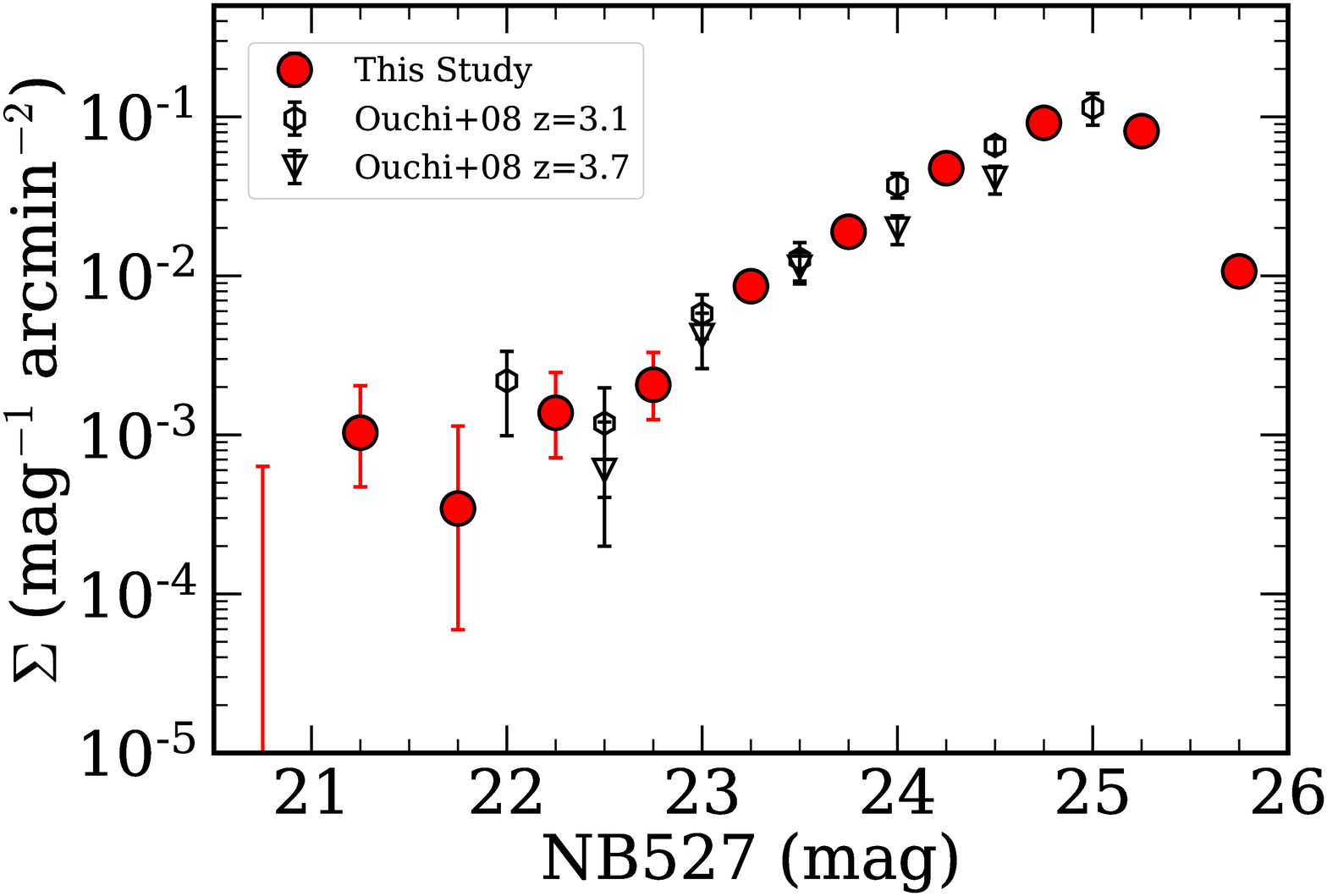}
   \includegraphics[scale=0.25]{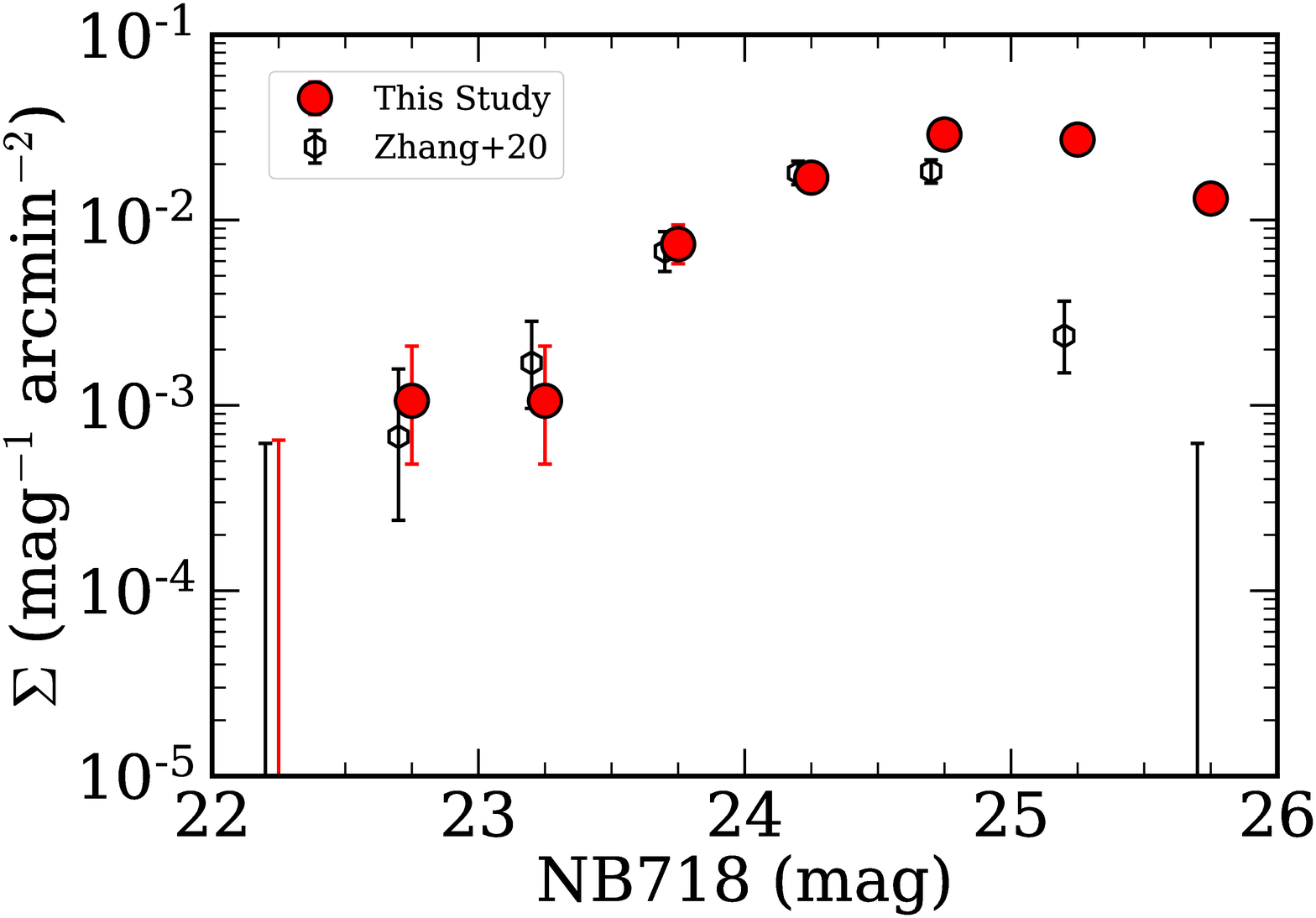}
   \includegraphics[scale=0.25]{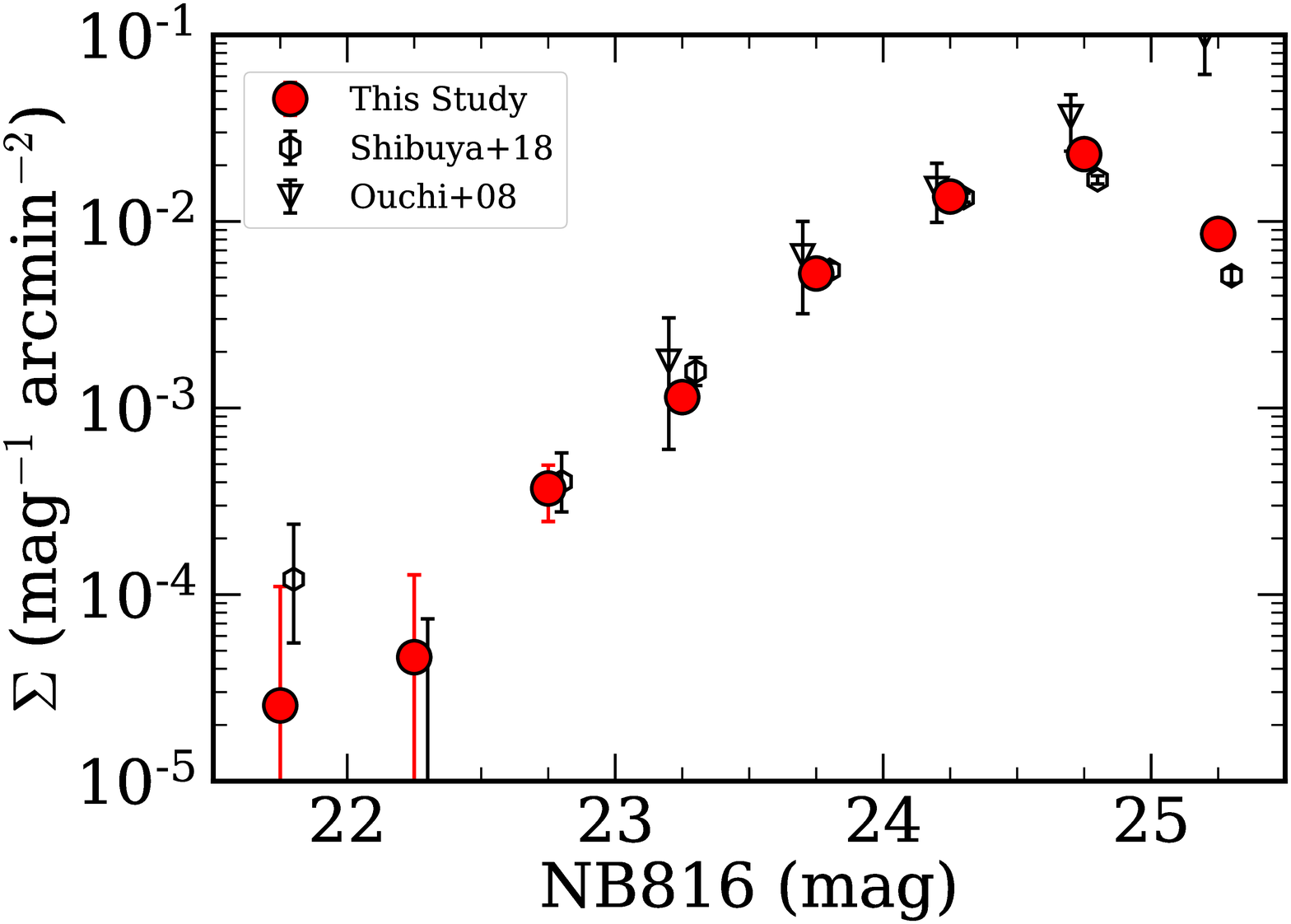}
   \includegraphics[scale=0.25]{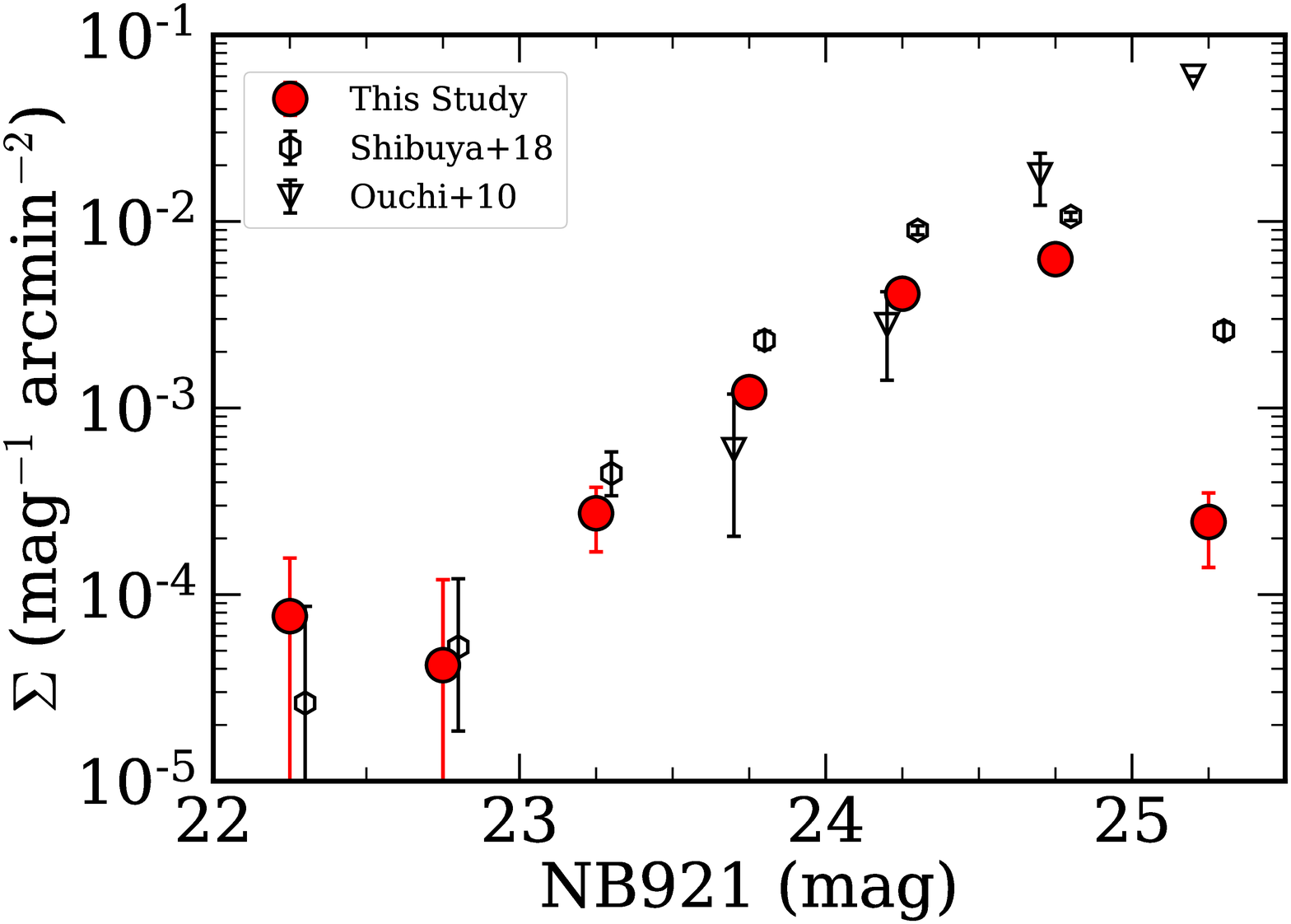}
   \includegraphics[scale=0.25]{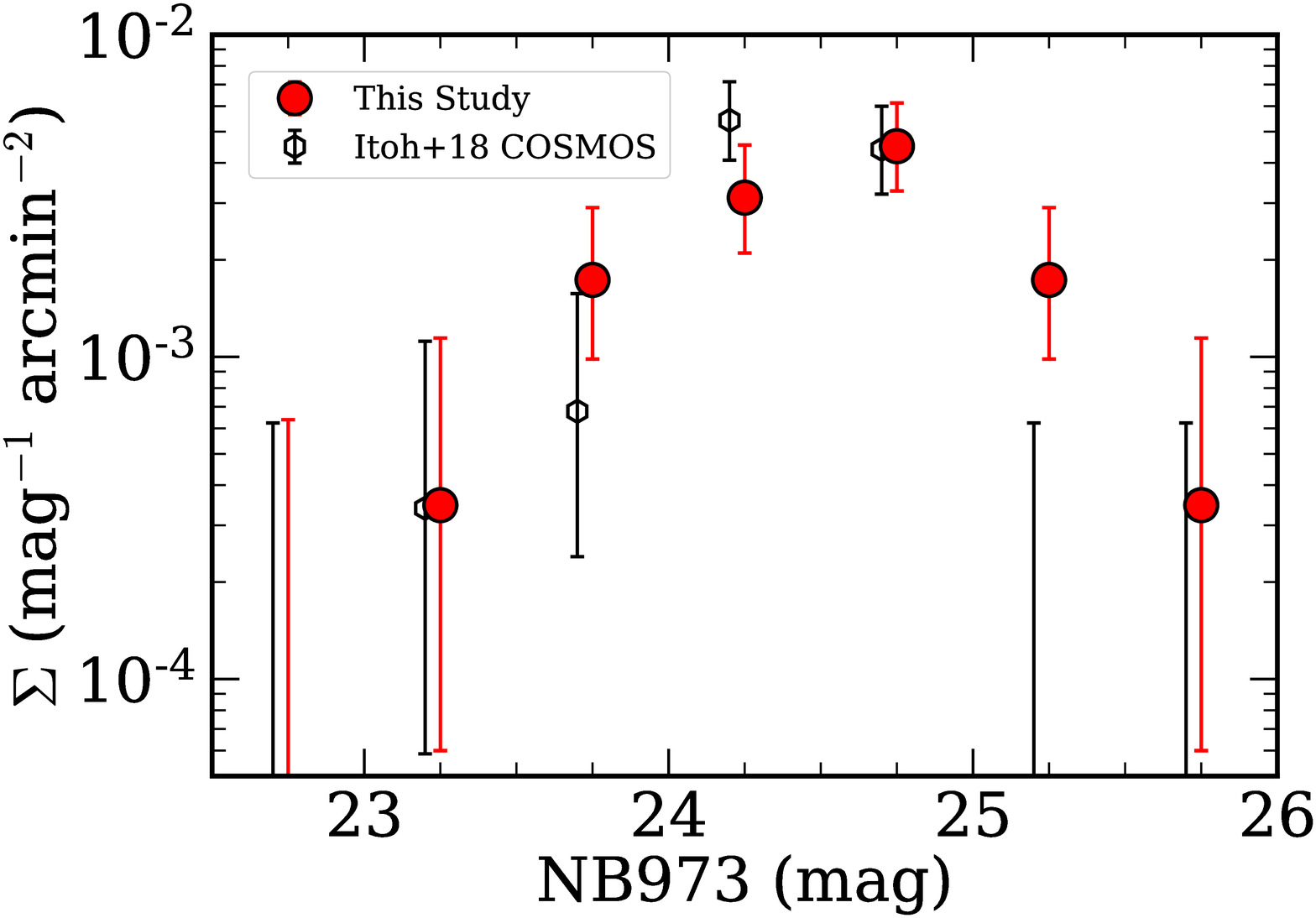}
\caption[]
{
Surface number density of LAEs at 
$z=2.2$ (top left), $z=3.3$ (top right), 
$z=4.9$ (middle left), $z=5.7$ (middle right), 
$z=6.6$ (bottom left), and $z=7.0$ (bottom right) 
as a function of total NB magnitude. 
The red circles are our results. 
These surface number density values are not corrected for completeness. 
For comparison, we also show previous results of 
\cite{2012ApJ...745...12N}, \cite{2013ApJ...769....3N}, and 
\cite{2018ApJ...864..145H} for $z = 2.2$ LAEs 
(see also \citealt{2016ApJ...823...20K}), 
\cite{2008ApJS..176..301O} for $z=3.1$, $z=3.7$, and $z = 5.7$ LAEs, 
\cite{2020ApJ...891..177Z} for $z=4.9$ LAEs, 
\cite{2010ApJ...723..869O} for $z = 6.6$ LAEs, 
\cite{2018PASJ...70S..14S} for $z = 5.7$ and $6.6$ LAEs, 
and 
 \cite{2018ApJ...867...46I} for $z=7.0$ LAEs. 
These previous results are slightly offset for clarity. 
}
\label{fig:surface_number_density}
\end{center}
\end{figure*}
%FFFFFFFFFFFFFFFFFFFFFFFFFFFFFFFFFFFFFFFFFFFFFFFFFFFFFFFFFFFFFFFF%

Based on the results shown in the top panel of Figure \ref{fig:confusion_matrix}, 
the completeness of the classifier, $f_{\rm comp}$, can be obtained from 
\begin{equation} 
f_{\rm comp} 
	= \frac{n_{\rm TP}}{n_{\rm TP} + n_{\rm FN}},   
\end{equation} 
where $n_{\rm TP}$ denotes the number of true positives, 
and $n_{\rm FN}$ is the number of false negatives. 
We find that the $f_{\rm comp}$ value for the test data of the simulated LAEs 
in Classes 1--4 is not so high, about $69${\%}. 
This is because these classes include simulated LAEs 
with low S/Ns down to only $3$. 
Figure \ref{fig:completeness_SNR} shows the completeness as a function of input S/N. 
We find that the $f_{\rm comp}$ values are $\gtrsim 90${\%} for bright sources 
with input magnitudes of $\lesssim 24.5$ mag. 
To remove the effect from low-S/N sources, 
we show the confusion matrix in the bottom panel of Figure \ref{fig:confusion_matrix}  
where we focus on simulated LAEs with S/Ns $>5$ for the true labels. 
We find that the $f_{\rm comp}$ value 
for simulated LAEs with S/Ns $>5$ is high, $\simeq 94${\%}．

From the confusion matrix, we also calculate 
the contamination rate of the classifier, $f_{\rm cont}$,  
\begin{equation} 
f_{\rm cont} 
	= \frac{n_{\rm FP}}{n_{\rm TP} + n_{\rm FP}},   
\end{equation} 
where $n_{\rm FP}$ is the number of false positives. 
We confirm that the contamination rate is very low, only $\simeq 1${\%}, 
based on the bottom panel of Figure \ref{fig:confusion_matrix}.

It should be noted that 
these completeness and contamination rate estimates 
are the results for the test data, 
controlling the balance of the numbers in the classes. 
For the actual data, 
these values should be different, 
because the number ratios between the classes are different. 
However, we do not rely solely on the machine learning selection for contamination removal. 
As described above in this section, 
after applying the machine learning selection, 
we perform visual inspection to further remove possible contaminants 
that are not excluded with the CNN.

%%%%%%%%%%%%%%%%%%%%%%%%%%%%%%%%%%%%%%%%%%%%%%%%%%%%%%%%%%%%%%%%%
%%%%%%%%%%%%%%%%%%%%%%%%%%%%%%%%%%%%%%%%%%%%%%%%%%%%%%%%%%%%%%%%%
\section{Results: Application to the New HSC SSP and CHORUS Data} \label{sec:results}
%%%%%%%%%%%%%%%%%%%%%%%%%%%%%%%%%%%%%%%%%%%%%%%%%%%%%%%%%%%%%%%%%
%%%%%%%%%%%%%%%%%%%%%%%%%%%%%%%%%%%%%%%%%%%%%%%%%%%%%%%%%%%%%%%%%

We apply the trained CNN to our photometrically selected LAE candidates 
prepared in Section \ref{sec:data} 
to efficiently remove contaminants such as satellite trails and low-$z$ interlopers  
by making use of our HSC NB and BB images.

Our machine learning selection consists of two steps. 
First we use the NB images of the photometrically selected LAE candidates 
as input data for the CNN 
and obtain their score for each class  
to select sources with a high score of being a positive (an LAE). 
In this process, we remove contaminants of satellite trails 
from the photometrically selected LAE candidates 
based on their morphologies in the NB images.  
Next, we use their BB images 
whose wavelength coverages are basically shorter than the Lyman limit wavelengths 
for targeted redshifts. 
We require that the candidates have $p_7 + p_8 > 0.5$, 
i.e., a high score of null detection, 
which enables us to remove low-$z$ interlopers 
from the photometrically selected LAE candidates.   
For this purpose, we use 
$g$-band images for $z=4.9$ and $z=5.7$ LAEs, 
$g$- and $r$-band images for $z=6.6$ LAEs, 
and 
$g$-, $r$-, and $i$-band images for $z=7.0$ LAEs. 
We do not use BB images for $z=2.2$ and $z=3.3$ LAEs, 
because their Lyman limit wavelengths are still shorter than the $g$-band wavelength coverage. 
Note that, 
because the absolute value of noise is different in each band, 
the input NB and BB images are scaled so that the absolute noise values are comparable to those of the training data  
in these two steps.

After these two steps, 
the numbers of LAE candidates in our samples are significantly reduced. 
The numbers of the remaining LAE candidates 
at $z=2.2$, $3.3$, $4.9$, $5.7$, $6.6$, and $7.0$ 
are 
about $8000$, $2600$, $1500$, $10000$, $5000$, and $380$, 
which correspond to about $3$, $18$, $58$, $6$, $3$, and $2$ {\%} of the photometrically selected LAE candidates, respectively.  
Although the difference of these fractions 
is due to the different fractions of obvious contaminants in the photometrically selected LAE candidates, 
which are likely to depend on the strictness of the photometric selection criteria, 
these low values indicate that 
our CNN removes a large number of contaminants in the photometrically selected LAE samples. 
Thanks to the CNN application, 
the human resources that are needed for the visual inspection 
are greatly reduced compared to our previous work 
(see Sections \ref{sec:data} and \ref{subsec:training}).

We then perform visual inspection of the NB and BB images of 
the LAE candidates that satisfy our machine learning selection criteria  
to further remove possible contaminants. 
Most contaminants excluded in this visual inspection process 
are the ones that are not considered in the training data for the CNN as expected, 
such as cosmic rays, diffuse artifacts, 
and noisy images around the edge of the survey area where the image qualities are poor.
Our careful selection yields samples of 
4409, 
959, 
349, 
2881, 
680, 
and 
40  
LAE candidates at $z=2.2$, $3.3$, $4.9$, $5.7$, $6.6$, and $7.0$, 
which correspond to about 
$60$, $40$, $20$, $30$, $10$, and $10$ {\%} of the LAE candidates that satisfy the machine learning selection criteria, 
respectively. 
Some of the fractions are still not so high. 
However, these fractions decrease without the machine learning selection 
to $1.7$, $6.9$, $14$, $1.5$, $0.4$, and $0.2${\%} 
for the $z=2.2$, $3.3$, $4.9$, $5.7$, $6.6$, and $7.0$ LAE samples, respectively, 
which indicates a significant improvement thanks to the machine learning selection, 
although it also depends on the strictness of the photometric selection criteria.
Note that the score of a spectroscopically confirmed spatially extended LAE, Himiko (\citealt{2009ApJ...696.1164O}; \citealt{2013ApJ...778..102O}),  
for its NB data is slightly smaller than the criterion of $p_1 + p_2 + p_3 + p_4 > 0.95$.    
This is probably because its morphology is significantly more extended 
than those adopted for the training data, 
although other extended LAEs such as CR7 (\citealt{2015ApJ...808..139S}) satisfy the NB criterion.  
Because Himiko has already been spectroscopically confirmed, 
we add it to the final catalog. 
In addition, we notice that the score of another spectroscopically confirmed LAE, 
COLA-1 (\citealt{2016ApJ...825L...7H}),  
for the BB data is higher than the criterion  
probably due to the presence of a foreground object at $z=2.142$ in its vicinity on the sky 
(\citealt{2018A&A...619A.136M}).
Because COLA-1 is also a spectroscopically confirmed LAE, 
it is also included in the final catalog.
The number of our LAE candidates in the final catalog with each NB for each field 
is summarized in Table \ref{tab:LAE_number}.

Figure \ref{fig:surface_number_density_for_each_field} shows 
the number counts of our LAE candidates in the final catalogs  
at $z=2.2$, $3.3$, $4.9$, $5.7$, $6.6$, and $7.0$ 
as a function of total NB magnitude. 
To calculate the $1\sigma$ uncertainties, 
we take account of Poisson confidence limits \citep{1986ApJ...303..336G}
on the number of the selected LAE candidates in each NB magnitude bin. 
For the LAE samples selected from the CHORUS data, 
we apply the masks that are provided by \cite{2020PASJ...72..101I} 
for conservative estimates, 
although the number count results are mostly consistent with the results without the masks. 
At $z=2.2$, $5.7$, and $6.6$, 
we also show the number counts of the LAE candidates for the survey fields separately. 
We confirm that these number counts are broadly consistent with each other at each redshift.

In Figure \ref{fig:surface_number_density}, 
we compare our average number counts 
with previous results where the selection criteria are similar to ours 
(\citealt{2008ApJS..176..301O}; \citealt{2010ApJ...723..869O}; 
\citealt{2012ApJ...745...12N}; \citealt{2013ApJ...769....3N}; 
\citealt{2016ApJ...823...20K}; 
\citealt{2018PASJ...70S..14S}; 
\citealt{2018ApJ...864..145H}; \citealt{2018ApJ...867...46I}; 
\citealt{2020ApJ...891..177Z}; 
see also Table \ref{tab:previou_number_count_work}). 
We confirm that, 
in the magnitude ranges that have been probed in previous studies, 
our results are in good agreement with the previous results. 
On the other hand, 
comparisons of faint end results are not simple, 
because not only NBs but also BBs are involved, 
whose limiting magnitudes are not completely uniform between and within the survey fields. 
Because the $5\sigma$ limiting magnitudes of our $NB387$ data 
for most of our survey fields are $\simeq 24.3$--$24.7$ mag, 
the incompleteness effects in our number counts of $z=2.2$ LAEs are non-negligible 
at faint magnitudes of $NB387 \gtrsim 24$ mag, 
compared to the previous results of \cite{2012ApJ...745...12N} and \cite{2018ApJ...864..145H}, 
whose $5\sigma$ limiting magnitudes are $\sim 26$ mag.  
In contrast, although the $5\sigma$ limiting magnitudes of our $NB527$ data 
are deeper than those of the previous work at similar redshifts, 
the incompleteness effect in our number counts is significant at around $NB527 \simeq 25$ mag, 
which is comparable to those of the previous results. 
This is probably because our BB data are relatively shallow. 
Similar discussion can be made for the results at higher redshifts.

Most of our survey fields have already been observed in our previous work \citep{2018PASJ...70S..14S}. 
We compare our LAE catalogs 
with those constructed in our previous work at $z=5.7$ and $z=6.6$ 
and calculate the object-matching rates as a function of NB magnitudes. 
We find that the object-matching rates are $\simeq 80$--$100${\%} at bright NB magnitudes of $\lesssim 24$ mag 
for the UD fields where the effect of photometric uncertainties is not relatively significant. 
The high object-matching rates indicate that we adequately select LAE candidates 
in our machine learning selection processes. 
Note that the object-matching rates decrease to $\simeq 60$--$70${\%} 
at the fainter magnitudes of $\simeq 24.5$--$25$ mag, 
mainly due to the increase of the effect of photometric uncertainties. 
We confirm that these object-matching rates are comparable to 
those obtained in \cite{2018PASJ...70S..14S}.

We also compare our LAE catalogs with 
the results of our previous spectroscopic observations 
for high-$z$ galaxies in our survey fields 
including the ones with Magellan/IMACS 
(PI: M. Ouchi; 
\textcolor{blue}{M. Ouchi in prep.}; 
\citealt{2018PASJ...70S..10O}; \citealt{2019ApJ...883..142H}), 
as well as the publicly available spectroscopic catalogs shown in previous studies 
(\citealt{2008ApJ...675.1076S}; 
\citealt{2008ApJS..176..301O}; 
\citealt{2010AJ....140..546W}; %2010.08
\citealt{2010ApJ...723..869O}; %2010.11
\citealt{2012ApJ...755..169M}; %2012.08
\citealt{2012ApJ...760..128M}; %2012.12
\citealt{2013AJ....145....4W}; %2013.01
\citealt{2013ApJ...768..105M}; %2013.05
\citealt{2013A&A...559A..14L}; %2013.11
\citealt{2014ApJ...788...74S}; 
\citealt{2015ApJS..218...15K}; %2015.06
\citealt{2015ApJ...808..139S}; %2015.08
\citealt{2016ApJ...819...24W}; %2016.03
\citealt{2016ApJ...825L...7H}; %2016.07
\citealt{2016ApJ...826..114T}; %2016.08
\citealt{2016ApJS..225...27M}; %2016.08
\citealt{2016ApJS..227...11B}; %2016.11
\citealt{2017A&A...600A.110T}; %2017.04
\citealt{2017AJ....153..184Y}; %2017.04
\citealt{2017ApJ...841..111M}; %2017.06
\citealt{2017ApJ...845L..16H}; %2017.08
\citealt{2017ApJ...846...57I}; %2017.09 
\citealt{2017ApJ...846..134J}; %2017.09
\citealt{2017ApJ...849...39S}; %2017.11
\citealt{2018PASJ...70S..15S}; %2018.01
\citealt{2018MNRAS.477.2817S}; %2018.06
\citealt{2018ApJS..237...31L}; %2018.08
\citealt{2018A&A...616A.174P}; %2018.09
\citealt{2018NatAs...2..962J}; %2018.10
\citealt{2019ApJ...877...81M}; %2019.06
\citealt{2019ApJ...883..142H}; %2019.10
\citealt{2019MNRAS.489.3294C}; %2019.11
\citealt{2020ApJ...891..177Z}; %2020.03
\citealt{2020ApJS..250....8L}; %2020.09
\citealt{2020arXiv201007545O}). 
We adopt their classifications between galaxies and AGNs in their catalogs if available.  
For the catalogs of 
the VIMOS VLT Deep Survey (VVDS: \citealt{2013A&A...559A..14L}) and 
the VIMOS Ultra Deep Survey (VUDS: \citealt{2017A&A...600A.110T}), 
we take into account sources whose redshifts are 
$>70${\%}--$75${\%} correct, 
i.e., sources with redshift reliability flags of 2, 3, 4, 9, 12, 13, 14, and 19.

\begin{figure*}
\begin{center}
   \includegraphics[scale=0.28]{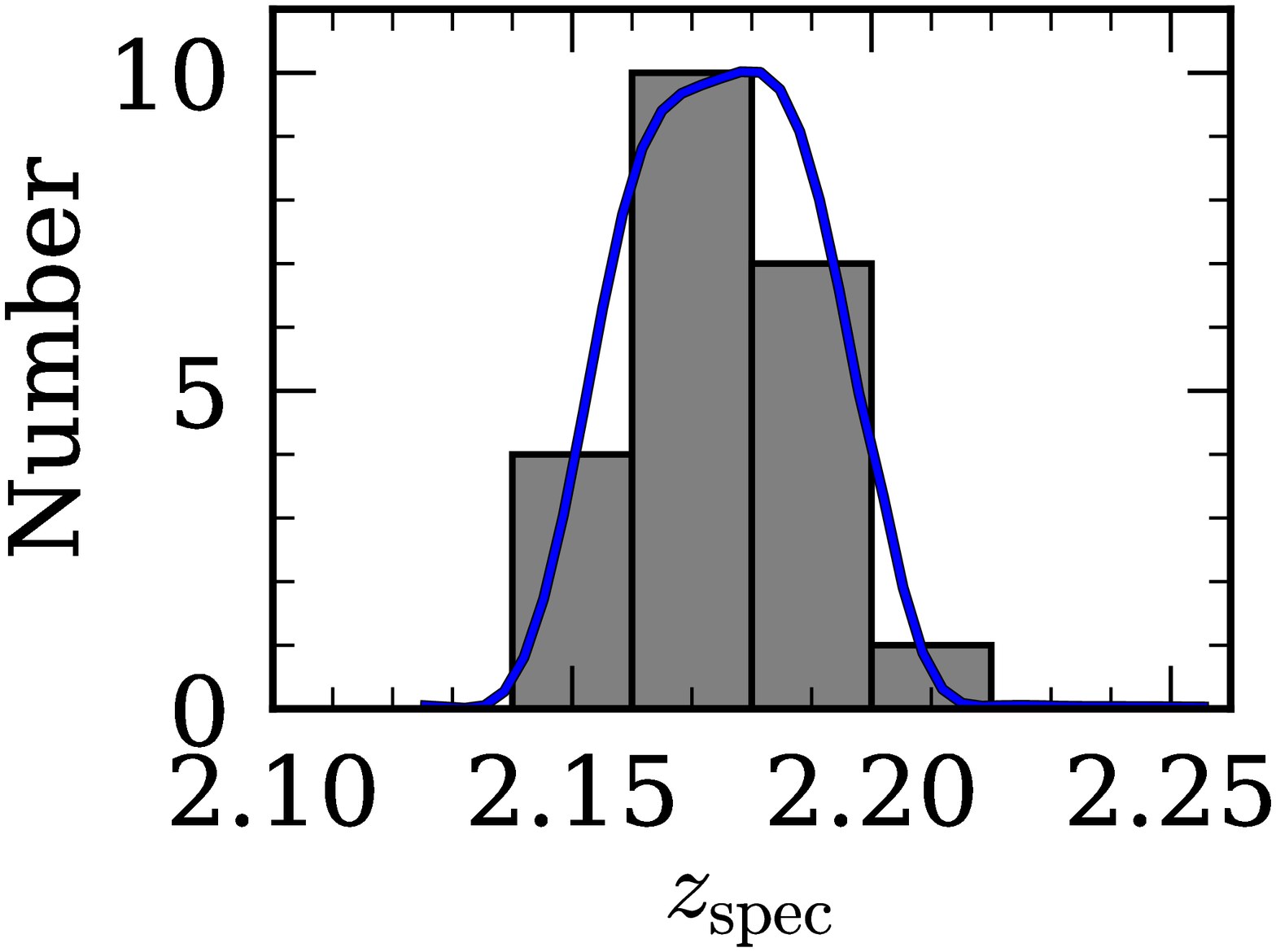}
   \includegraphics[scale=0.28]{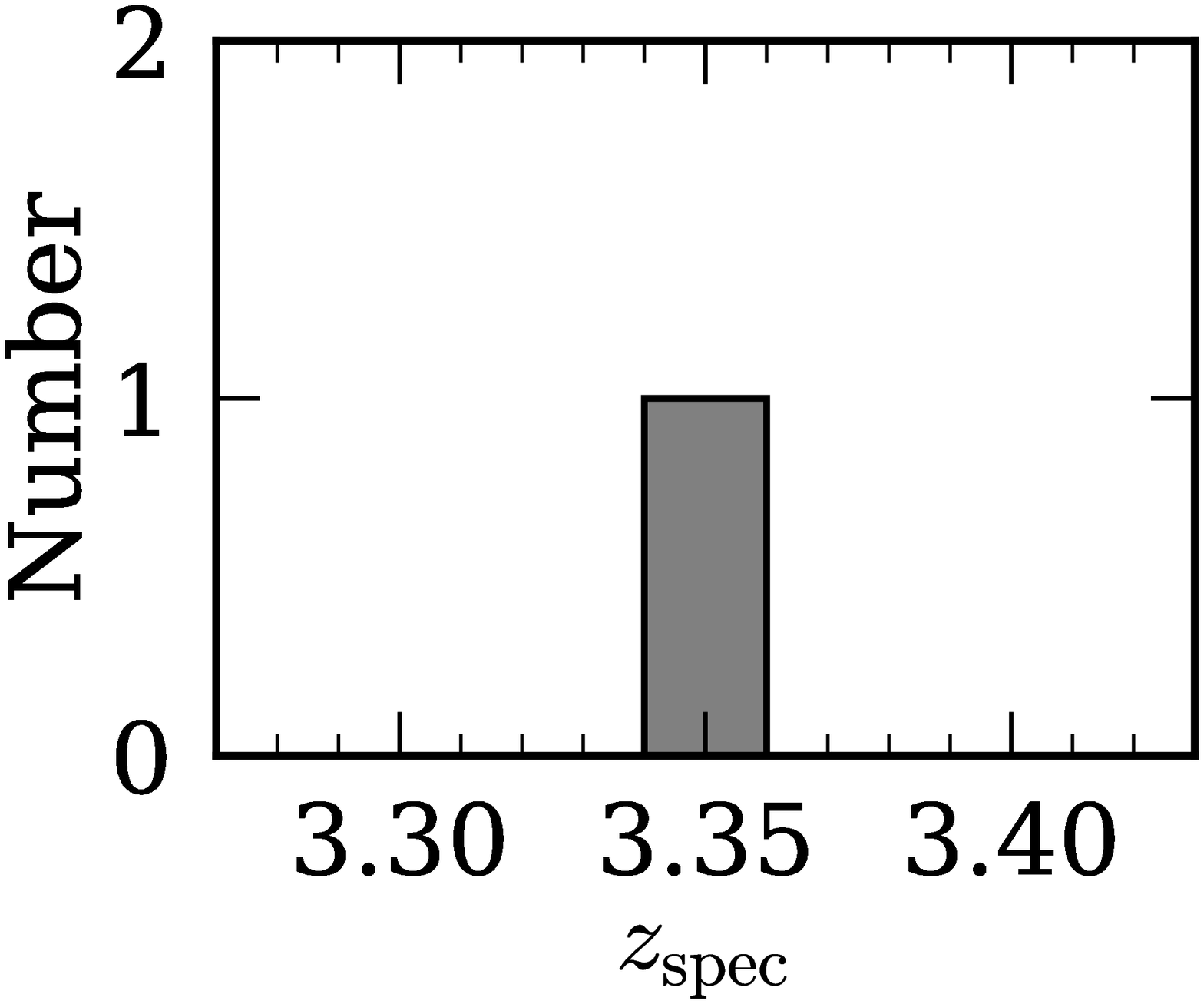}
   \includegraphics[scale=0.28]{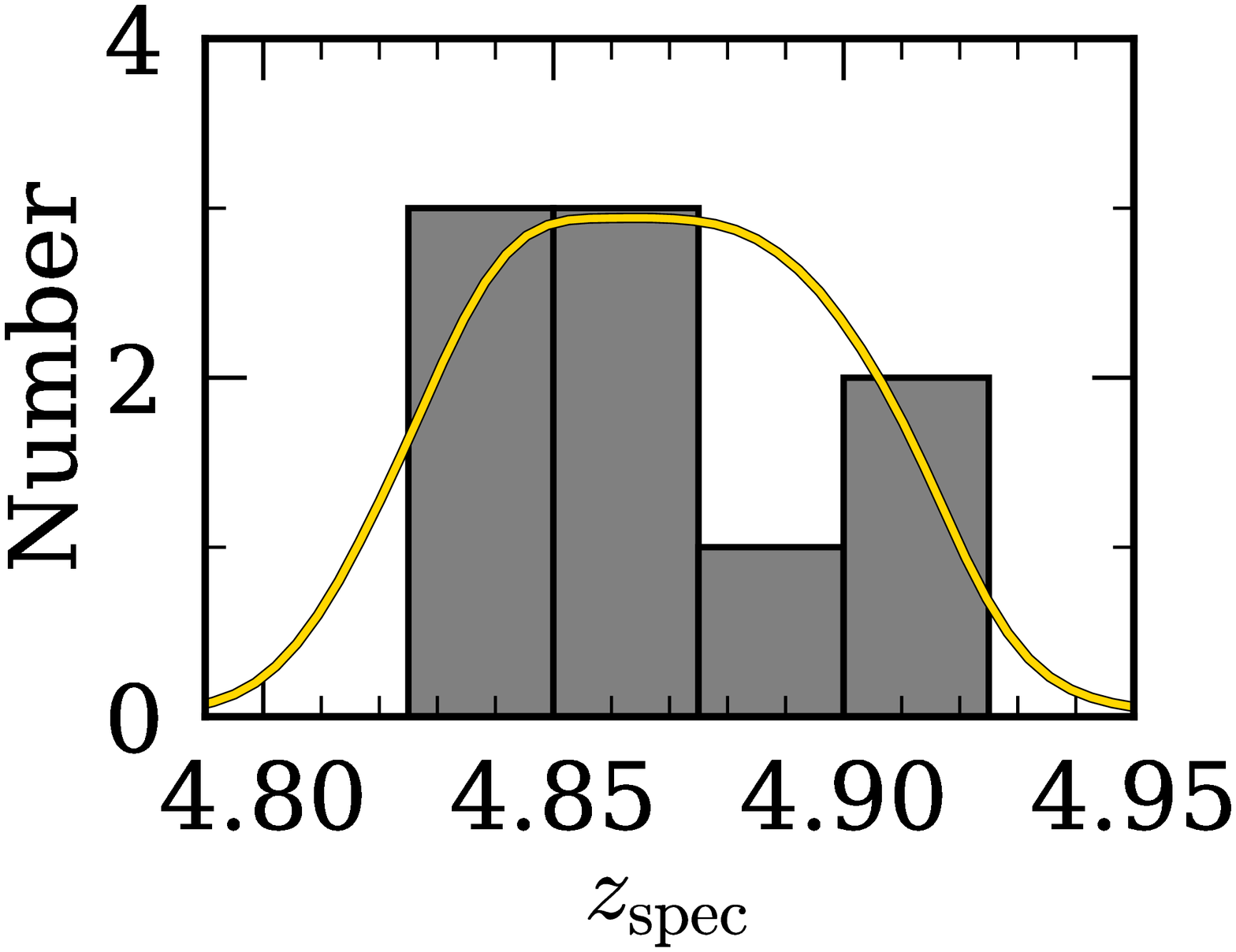}
   \includegraphics[scale=0.28]{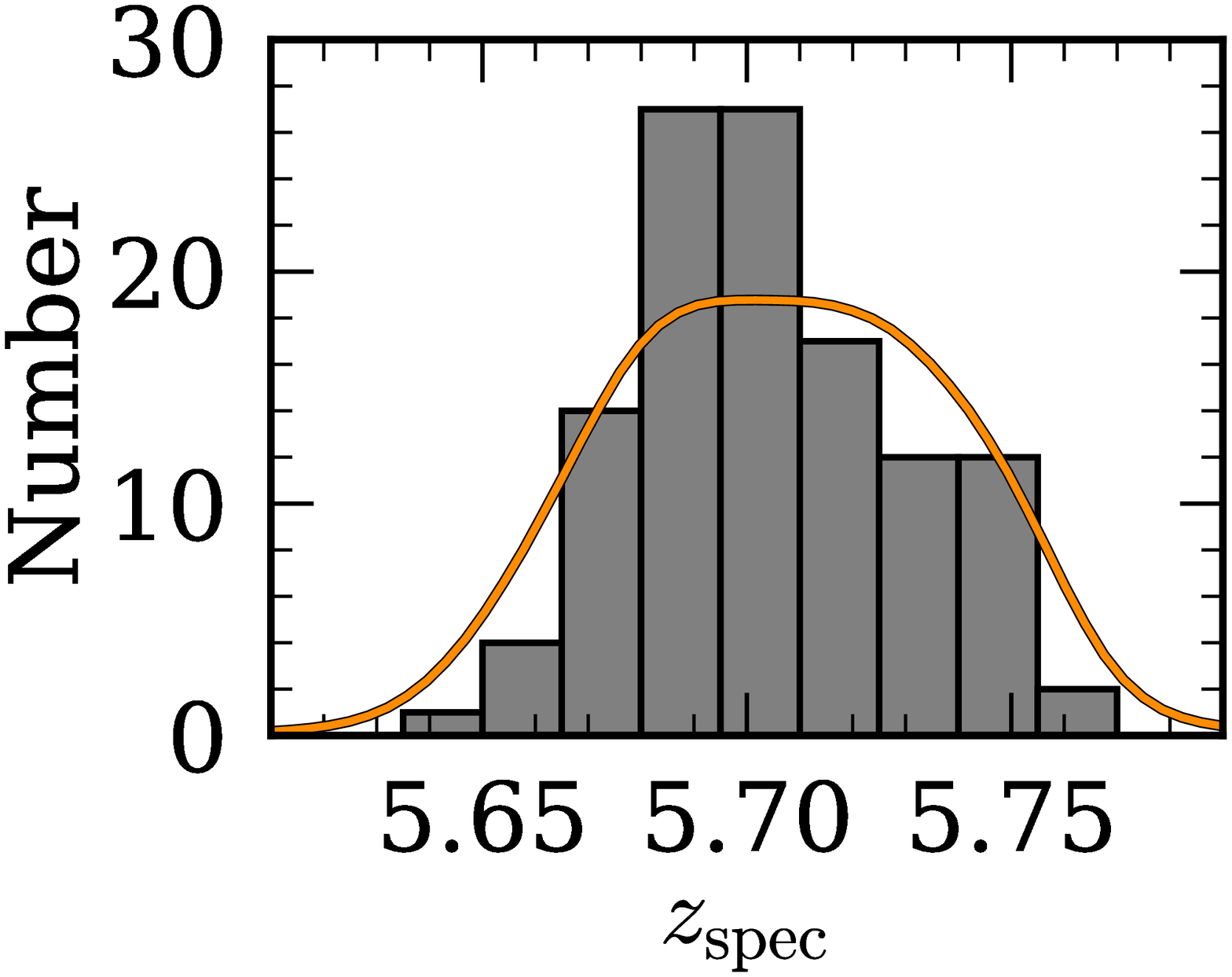}
   \includegraphics[scale=0.28]{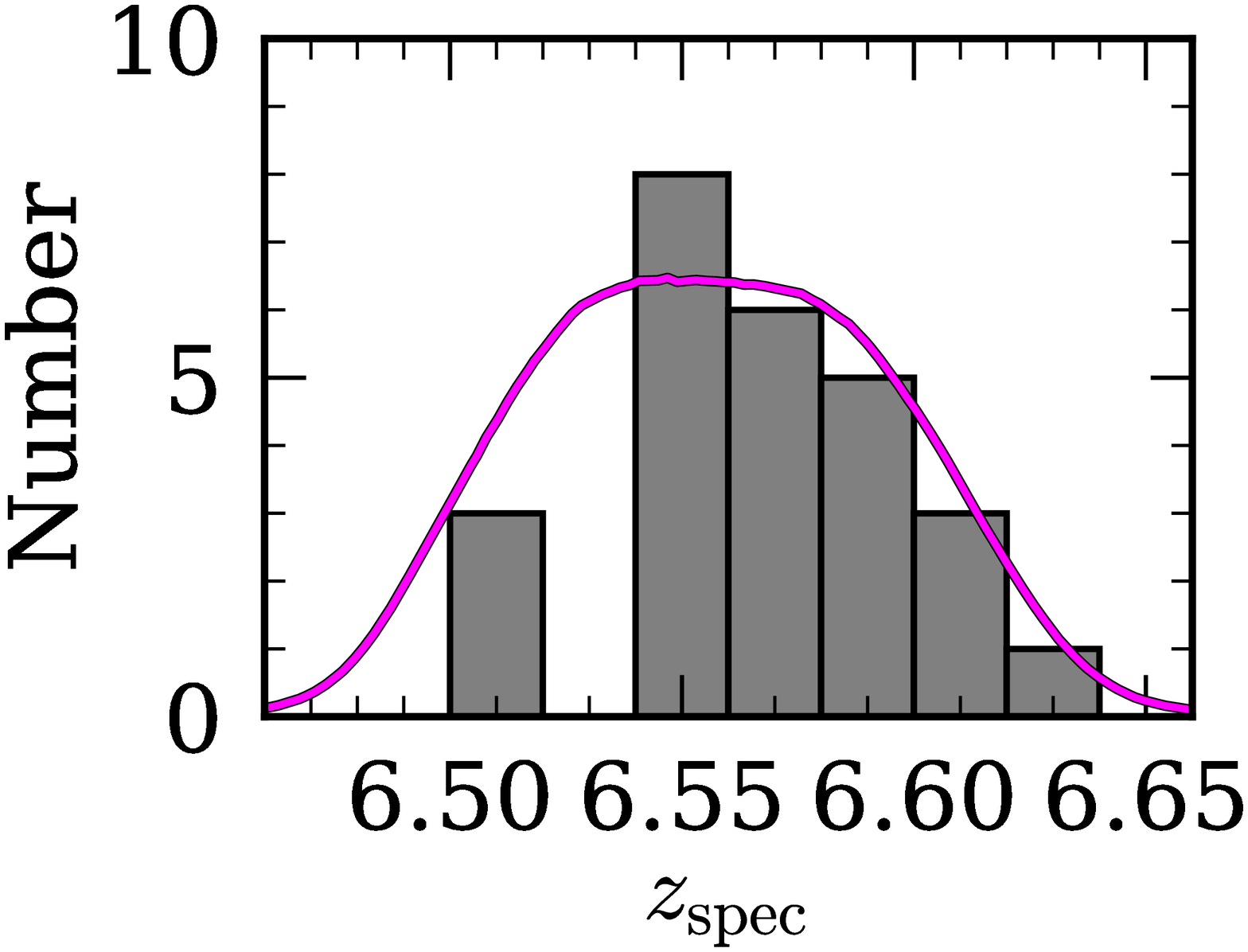}
   \includegraphics[scale=0.28]{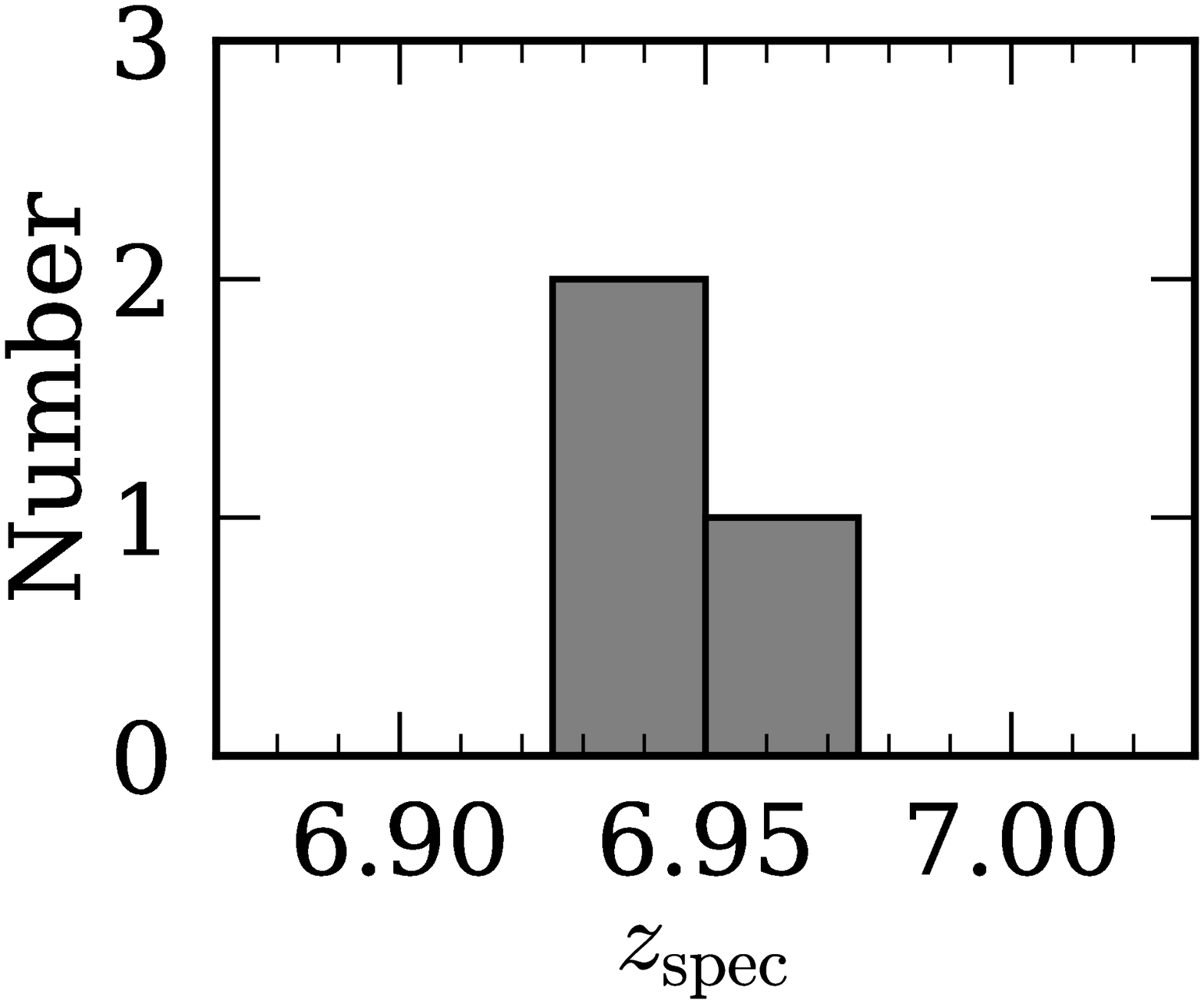}
\caption[]
{
Redshift distribution of spectroscopically confirmed LAEs in our samples (gray histograms). 
The colored curves are the best-fit results with the transmission curves of the NB filters shown in Figure \ref{fig:filters}.
The blue, yellow, orange, and magenta curves 
correspond to 
the results with the $NB387$, $NB718$, $NB816$, and $NB921$ filter transmission curves, respectively. 
We do not perform the fitting for $z=3.3$ and $z=7.0$  
because of the limited number of the data points. 
}
\label{fig:zspec_distribution}
\end{center}
\end{figure*}

We find that 
177 LAEs in our samples have been spectroscopically confirmed 
by our observations and in the previous studies. 
Among them, 
155 sources are found to be Ly$\alpha$-emitting galaxies, 
and the other 22 sources are Ly$\alpha$-emitting AGNs. 
The catalog of these spectroscopically confirmed LAEs is presented in Table \ref{tab:LAE_specz}. 
We confirm that their spectroscopic redshifts are consistent with 
those expected from the redshift coverages of the NBs used in this study. 
Their redshift distributions around the redshift ranges probed with the NBs 
are shown in Figure \ref{fig:zspec_distribution}.

%FFFFFFFFFFFFFFFFFFFFFFFFFFFFFFFFFFFFFFFFFFFFFFFFFFFFFFFFFFFFFFFF%
%FFFFFFFFFFFFFFFFFFFFFFFFFFFFFFFFFFFFFFFFFFFFFFFFFFFFFFFFFFFFFFFF%

To characterize these redshift distributions,  
we fit the transmission curves of the NBs to them. 
Specifically, 
we derive the best-fit NB transmission curves by a $\chi^2$ minimization fit 
with a normalization constant and a redshift shift as free parameters, 
i.e., $N(z) = \alpha T(z+\beta)$, 
where 
$N(z)$ is the redshift distribution, 
$T(z)$ is the NB transmission curve, 
$\alpha$ is the  normalization constant, 
and $\beta$ is the shift along the $x$-axis. 
The observed wavelengths of the NB transmission curves are converted to 
the Ly$\alpha$ redshifts,  $z_{\rm Ly\alpha}$, by 
$z_{\rm Ly\alpha} = \lambda_{\rm obs}/\lambda_{{\rm Ly}\alpha} - 1$,  
where $\lambda_{\rm obs}$ is the observed wavelength 
and $\lambda_{{\rm Ly}\alpha} = 1215.67${\AA} is the Ly$\alpha$ wavelength. 
The best-fit results are shown in Figure \ref{fig:zspec_distribution}, 
and the best-fit parameters are presented in Table \ref{tab:redshift_distribution_fitting_results}.  
We do not perform the fitting for $z=3.3$ and $z=7.0$, 
because of the limited number of the data points. 
We find that the solid curves fit the histograms well.  
We also find that, 
the best-fit $\beta$ values for $NB718$, $NB816$, and $NB921$ are negative. 
This is probably because, for high redshift LAEs, 
although a high sensitivity for Ly$\alpha$ emission is necessary, 
a high sensitivity for continuum emission also makes it easier for the NBs to detect LAEs. 
Another possible reason is that 
the majority of the spectroscopic redshifts used for the compilation 
are based on the catalogs whose sources are found with instruments other than HSC.

In addition to the 177 spectroscopically confirmed LAEs, 
we notice that 
4 sources in our sample have also been spectroscopically confirmed 
as lower-$z$ sources in the previous studies.  
In fact, these 4 source are AGNs at slightly lower redshift 
where strong {\sc Civ} emission is probed with our NBs.  
The catalog of these sources is presented in Table \ref{tab:lowz_LAE_specz}. 
Although it is difficult to make a fair comparison based on the available catalogs, 
the small number of the spectroscopic identifications at low redshifts 
may suggest that the contamination rate of low-$z$ interlopers is low. 
However, it should be noted that, because 
our selections do not use BB data whose wavelengths are bluer than the Lyman limit 
for our $z=2.2$ and $z=3.3$ LAE samples, 
it is not surprising that 
these two samples are relatively contaminated by lower-$z$ interlopers 
such as strong [{\sc Oii}] and [{\sc Oiii}] emitters. 
Quantitative discussion for such possibilities 
is waiting to be examined with further spectroscopic observations.

%%%%%%%%%%%%%%%%%%%%%%%%%%%%%%%%%%%%%%%%%%%%%%%%%%%%%%%%%%%%%%%%%
%%%%%%%%%%%%%%%%%%%%%%%%%%%%%%%%%%%%%%%%%%%%%%%%%%%%%%%%%%%%%%%%%
\section{Summary} \label{sec:summary}
%%%%%%%%%%%%%%%%%%%%%%%%%%%%%%%%%%%%%%%%%%%%%%%%%%%%%%%%%%%%%%%%%
%%%%%%%%%%%%%%%%%%%%%%%%%%%%%%%%%%%%%%%%%%%%%%%%%%%%%%%%%%%%%%%%%

We have constructed a catalog of $9$,$318$ LAE candidates 
at $z = 2.2$, $3.3$, $4.9$, $5.7$, $6.6$, and $7.0$ 
with the aid of the machine learning technique 
based on the multiwavelength data over the $1.6$--$25.0$ deg$^2$ sky 
taken with the six narrowband filters of $NB387$, $NB527$, $NB718$, $NB816$, $NB921$, and $NB973$, 
and the five broadband filters of $g$, $r$, $i$, $z$, and $y$ in the HSC SSP and CHORUS. 
We have first photometrically selected LAE candidates 
based on the NB-excess features of their SEDs. 
We have then created the CNN, 
which is recently frequently adopted for image classification problems, 
to distinguish between real LAEs and contaminants.
We have found that the completeness and contamination rate of the trained CNN 
are $94${\%} and $1${\%}, respectively.  
By taking advantage of the trained CNN, 
we have efficiently removed contaminants such as satellite trails and low-$z$ interlopers 
from the photometrically selected LAE candidates. 
Specifically, for all the photometrically selected LAE candidates, 
the amount of human effort to conduct visual inspection is estimated to be about 11 person-months, 
but we have significantly reduced it thanks to the CNN application. 
We have confirmed that 
our machine learning selected LAE samples are reliable 
on the basis of 
$177$ 
LAEs that have already been spectroscopically identified 
by our SILVERRUSH programs and previous studies.  
In addition, 
we have found that the object-matching rates between our LAE catalogs and our previous results 
are $\simeq 80$--$100${\%} at bright NB magnitudes of $\lesssim 24$ mag, 
which also supports the validity of our selection method. 
We have also confirmed that 
the surface number densities of our LAE candidates 
are consistent with previous results. 
Our LAE catalogs will become available online.

%%%%%%%%%%%%%%%%%%%%%%%%%%%%%%%%%%%%%%%%%%%%%%%%%%%%%%%%%%%%%%%%%
%%%%%%%%%%%%%%%%%%%%%%%%%%%%%%%%%%%%%%%%%%%%%%%%%%%%%%%%%%%%%%%%%
\section*{Acknowledgements}
%%%%%%%%%%%%%%%%%%%%%%%%%%%%%%%%%%%%%%%%%%%%%%%%%%%%%%%%%%%%%%%%%
%%%%%%%%%%%%%%%%%%%%%%%%%%%%%%%%%%%%%%%%%%%%%%%%%%%%%%%%%%%%%%%%%

The HSC collaboration includes the astronomical communities of Japan and Taiwan, and Princeton University.  
The HSC instrumentation and software were developed by the National Astronomical Observatory of Japan (NAOJ), 
the Kavli Institute for the Physics and Mathematics of the Universe (Kavli IPMU), 
the University of Tokyo, the High Energy Accelerator Research Organization (KEK), 
the Academia Sinica Institute for Astronomy and Astrophysics in Taiwan (ASIAA), and Princeton University.  
Funding was contributed by the FIRST program from the Japanese Cabinet Office, 
the Ministry of Education, Culture, Sports, Science and Technology (MEXT), 
the Japan Society for the Promotion of Science (JSPS), Japan Science and Technology Agency  (JST), 
the Toray Science  Foundation, NAOJ, Kavli IPMU, KEK, ASIAA, and Princeton University.

This paper makes use of software developed for the Large Synoptic Survey Telescope. 
We thank the LSST Project for making their code available as free software at 
\url{http://dm.lsst.org}. 

This paper is based on data collected at the Subaru Telescope and retrieved from the HSC data archive system, 
which is operated by Subaru Telescope and Astronomy Data Center (ADC) at NAOJ. 
Data analysis was in part carried out with the cooperation of Center for Computational Astrophysics (CfCA), NAOJ.

The Pan-STARRS1 Surveys (PS1) and the PS1 public science archive have been made possible 
through contributions by the Institute for Astronomy, the University of Hawaii, the Pan-STARRS Project Office, 
the Max Planck Society and its participating institutes, the Max Planck Institute for Astronomy, Heidelberg, 
and the Max Planck Institute for Extraterrestrial Physics, Garching, The Johns Hopkins University, Durham University, 
the University of Edinburgh, the Queen's University Belfast, the Harvard-Smithsonian Center for Astrophysics, 
the Las Cumbres Observatory Global Telescope Network Incorporated, the National Central University of Taiwan, 
the Space Telescope Science Institute, the National Aeronautics and Space Administration 
under grant No. NNX08AR22G issued through the Planetary Science Division of the NASA Science Mission Directorate, 
the National Science Foundation grant No. AST-1238877, the University of Maryland, Eotvos Lorand University (ELTE), 
the Los Alamos National Laboratory, and the Gordon and Betty Moore Foundation.

The NB387 filter was supported by KAKENHI (23244022) Grant-in-Aid for Scientific Research (A) through the JSPS. 
The NB527 filter was supported by KAKENHI (24244018) Grant-in-Aid for Scientific Research (A) through the JSPS. 
The NB718 and NB816 filters were supported by Ehime University. 
The NB921 and NB973 filters were supported by KAKENHI (23244025) Grant-in-Aid for Scientific Research (A) through the JSPS. 

This work was partially performed using the computer facilities of
the Institute for Cosmic Ray Research, The University of Tokyo. 
This work is supported by JSPS KAKENHI Grant Numbers 
15K17602, %Y. Ono
15H02064, % M. Ouchi
17H01110, %N. Sugiyama
17H01114, %A. Inoue
19K14752, %Y. Ono
and 
20H00180. %M. Ouchi

%%%%%%%%%%%%%%%%%%%%%%%%%%%%%%%%%%%%%%%%%%%%%%%%%%%%%%%%%%%%%%%%%
%%%%%%%%%%%%%%%%%%%%%%%%%%%%%%%%%%%%%%%%%%%%%%%%%%%%%%%%%%%%%%%%%

\bibliographystyle{aasjournal}
\bibliography{ref}

%%%%%%%%%%%%%%%%%%%%%%%%%%%%%%%%%%%%%%%%%%%%%%%%%%%%%%%%%%%%%%%%%
%%%%%%%%%%%%%%%%%%%%%%%%%%%%%%%%%%%%%%%%%%%%%%%%%%%%%%%%%%%%%%%%%

\appendix

\setcounter{table}{0}
\renewcommand{\thetable}{A\arabic{table}}

%%%%%%%%%%%%%%%%%%%%%%%%%%%%%%%%%%%%%%%%%%%%%%%%%%%%%%%%%%%%%%%%%
%%%%%%%%%%%%%%%%%%%%%%%%%%%%%%%%%%%%%%%%%%%%%%%%%%%%%%%%%%%%%%%%%
\section{A. Catalog of Spectroscopically Confirmed LAEs in Our Samples} 
%%%%%%%%%%%%%%%%%%%%%%%%%%%%%%%%%%%%%%%%%%%%%%%%%%%%%%%%%%%%%%%%%
%%%%%%%%%%%%%%%%%%%%%%%%%%%%%%%%%%%%%%%%%%%%%%%%%%%%%%%%%%%%%%%%%

In this paper, we present a catalog of spectroscopically confirmed galaxies and AGNs 
in our LAE samples in Table \ref{tab:LAE_specz}. 
In addition, we also present a catalog of spectroscopically confirmed AGNs 
that are selected with our NBs because of their strong {\sc Civ} emission 
in Table \ref{tab:lowz_LAE_specz}. 
Our new SILVERRUSH LAE catalogs based on new HSC SSP and CHORUS data 
will be made public on our project webpage as described in Section \ref{sec:introduction}.

%ttttttttttttttttttttttttttttttttttttttttttttttttttttttttttttttttttttttttt%
\LongTables
\capstartfalse
\begin{deluxetable*}{cccccccccccccc} 
\tablecolumns{14} 
\tablewidth{0pt} 
\tablecaption{Spectroscopically confirmed galaxies and AGNs in our LAE samples
\label{tab:LAE_specz}}
\tablehead{
    \colhead{ID}     
    &  \colhead{R.A.}
    &  \colhead{Decl.}
    &  \colhead{$z_{\rm spec}$}
    &  \colhead{$g_{\rm ap}$}
    &  \colhead{$r_{\rm ap}$}
    &  \colhead{$i_{\rm ap}$}
    &  \colhead{$z_{\rm ap}$}
    &  \colhead{$y_{\rm ap}$}
    &  \colhead{$NB_{\rm tot}$}
    &  \colhead{Sample}
    &  \colhead{Field}
    &  \colhead{Flag}
    &  \colhead{Reference}
    \\
    \colhead{(1)}
    &  \colhead{(2)}
    &  \colhead{(3)}
    &  \colhead{(4)}
    &  \colhead{(5)}
    &  \colhead{(6)}
    &  \colhead{(7)}
    &  \colhead{(8)}
    &  \colhead{(9)}
    &  \colhead{(10)}
    &  \colhead{(11)}
    &  \colhead{(12)}
    &  \colhead{(13)}
    &  \colhead{(14)}
}
\startdata
HSC J021525--045918 & 02:15:25.24 & $-$04:59:18.12 & $5.671$ & $99.0$ & $32.7$ & $26.6$ & $26.0$ & $25.5$ & $24.2$ & NB816 & UDSXDS & 1 & This study \\
HSC J021526--045229 & 02:15:26.23 & $-$04:52:29.83 & $5.655$ & $99.0$ & $29.6$ & $26.2$ & $25.3$ & $24.9$ & $24.6$ & NB816 & UDSXDS & 1 & This study \\
HSC J021533--050137 & 02:15:33.24 & $-$05:01:37.38 & $5.671$ & $29.2$ & $28.1$ & $26.3$ & $25.6$ & $25.4$ & $24.4$ & NB816 & UDSXDS & 1 & This study \\
HSC J021551--045325 & 02:15:51.35 & $-$04:53:25.47 & $5.710$ & $31.5$ & $28.8$ & $99.0$ & $27.3$ & $28.0$ & $25.2$ & NB816 & UDSXDS & 1 & This study \\
HSC J021555--045318 & 02:15:55.65 & $-$04:53:18.83 & $5.738$ & $28.9$ & $28.6$ & $28.0$ & $28.0$ & $99.0$ & $25.4$ & NB816 & UDSXDS & 1 & This study \\
HSC J021558--045301 & 02:15:58.50 & $-$04:53:01.82 & $5.718$ & $99.0$ & $99.0$ & $27.4$ & $28.3$ & $99.0$ & $24.6$ & NB816 & UDSXDS & 1 & This study \\
HSC J021611--045633 & 02:16:11.37 & $-$04:56:33.13 & $6.549$ & $30.0$ & $99.0$ & $99.0$ & $27.2$ & $27.0$ & $25.0$ & NB921 & UDSXDS & 1 & This study \\
HSC J021617--045419 & 02:16:17.15 & $-$04:54:19.37 & $5.707$ & $99.0$ & $99.0$ & $29.2$ & $29.4$ & $29.2$ & $25.5$ & NB816 & UDSXDS & 1 & This study \\
HSC J021624--045516 & 02:16:24.70 & $-$04:55:16.61 & $5.706$ & $99.0$ & $99.0$ & $26.4$ & $25.9$ & $26.9$ & $23.5$ & NB816 & UDSXDS & 1 & This study \\
HSC J021625--045237 & 02:16:25.64 & $-$04:52:37.24 & $5.730$ & $99.0$ & $31.9$ & $28.9$ & $30.0$ & $99.0$ & $25.0$ & NB816 & UDSXDS & 1 & This study \\
HSC J021628--050103 & 02:16:28.06 & $-$05:01:03.96 & $5.692$ & $30.9$ & $99.0$ & $28.3$ & $27.3$ & $27.3$ & $25.0$ & NB816 & UDSXDS & 1 & This study \\
HSC J021636--044723 & 02:16:36.44 & $-$04:47:23.69 & $5.718$ & $99.0$ & $99.0$ & $27.0$ & $27.0$ & $27.5$ & $24.9$ & NB816 & UDSXDS & 1 & This study \\
HSC J021654--045556 & 02:16:54.54 & $-$04:55:56.99 & $6.617$ & $99.0$ & $99.0$ & $99.0$ & $27.0$ & $26.9$ & $25.0$ & NB921 & UDSXDS & 1 & O10 \\
HSC J021654--052155 & 02:16:54.61 & $-$05:21:55.74 & $5.712$ & $30.2$ & $99.0$ & $27.1$ & $26.9$ & $26.8$ & $24.3$ & NB816 & UDSXDS & 1 & H19 \\
HSC J021657--052117 & 02:16:57.89 & $-$05:21:17.06 & $5.667$ & $99.0$ & $99.0$ & $26.8$ & $26.3$ & $29.1$ & $25.2$ & NB816 & UDSXDS & 1 & H19 \\
HSC J021702--050604 & 02:17:02.57 & $-$05:06:04.71 & $6.545$ & $99.0$ & $27.9$ & $28.9$ & $26.4$ & $28.3$ & $24.6$ & NB921 & UDSXDS & 1 & O10 \\
HSC J021703--045619 & 02:17:03.47 & $-$04:56:19.14 & $6.589$ & $99.0$ & $99.0$ & $99.0$ & $27.0$ & $26.3$ & $24.7$ & NB921 & UDSXDS & 1 & O10 \\
HSC J021704--052714 & 02:17:04.30 & $-$05:27:14.43 & $5.686$ & $99.0$ & $28.2$ & $26.4$ & $26.2$ & $26.5$ & $24.2$ & NB816 & UDSXDS & 1 & H19 \\
HSC J021707--053426 & 02:17:07.86 & $-$05:34:26.78 & $5.678$ & $99.0$ & $99.0$ & $27.1$ & $26.5$ & $99.0$ & $24.2$ & NB816 & UDSXDS & 1 & H19 \\
HSC J021709--050329 & 02:17:09.78 & $-$05:03:29.34 & $5.709$ & $29.3$ & $29.2$ & $27.4$ & $27.0$ & $99.0$ & $24.6$ & NB816 & UDSXDS & 1 & This study \\
HSC J021709--052646 & 02:17:09.97 & $-$05:26:46.76 & $5.689$ & $99.0$ & $99.0$ & $28.5$ & $27.1$ & $99.0$ & $25.2$ & NB816 & UDSXDS & 1 & H19 \\
HSC J021719--043150 & 02:17:19.13 & $-$04:31:50.71 & $5.735$ & $99.0$ & $99.0$ & $27.9$ & $27.5$ & $28.7$ & $24.3$ & NB816 & UDSXDS & 1 & This study \\
HSC J021724--053309 & 02:17:24.04 & $-$05:33:09.78 & $5.707$ & $31.1$ & $99.0$ & $25.9$ & $25.4$ & $26.5$ & $23.3$ & NB816 & UDSXDS & 1 & S18 \\
HSC J021725--050737 & 02:17:25.90 & $-$05:07:37.65 & $5.701$ & $99.0$ & $99.0$ & $27.9$ & $27.2$ & $26.6$ & $24.7$ & NB816 & UDSXDS & 1 & This study \\
HSC J021726--045126 & 02:17:26.74 & $-$04:51:26.79 & $5.709$ & $99.0$ & $99.0$ & $27.5$ & $99.0$ & $99.0$ & $24.9$ & NB816 & UDSXDS & 1 & This study \\
HSC J021729--053028 & 02:17:29.19 & $-$05:30:28.61 & $5.746$ & $99.0$ & $28.9$ & $28.0$ & $28.5$ & $99.0$ & $25.0$ & NB816 & UDSXDS & 1 & H19 \\
HSC J021734--053452 & 02:17:34.16 & $-$05:34:52.80 & $5.708$ & $99.0$ & $99.0$ & $99.0$ & $27.9$ & $99.0$ & $25.2$ & NB816 & UDSXDS & 1 & H19 \\
HSC J021734--044559 & 02:17:34.58 & $-$04:45:59.04 & $5.701$ & $28.7$ & $29.4$ & $26.6$ & $25.9$ & $25.8$ & $24.3$ & NB816 & UDSXDS & 1 & This study \\
HSC J021736--052701 & 02:17:36.39 & $-$05:27:01.90 & $5.672$ & $28.2$ & $99.0$ & $26.8$ & $26.4$ & $99.0$ & $24.1$ & NB816 & UDSXDS & 1 & H19 \\
HSC J021736--053027 & 02:17:36.68 & $-$05:30:27.60 & $5.686$ & $99.0$ & $99.0$ & $27.9$ & $26.8$ & $99.0$ & $25.4$ & NB816 & UDSXDS & 1 & H19 \\
HSC J021737--043943 & 02:17:37.97 & $-$04:39:43.12 & $5.755$ & $29.5$ & $99.0$ & $26.9$ & $26.2$ & $25.8$ & $24.5$ & NB816 & UDSXDS & 1 & This study \\
HSC J021739--043837 & 02:17:39.26 & $-$04:38:37.38 & $5.721$ & $99.0$ & $99.0$ & $28.6$ & $26.8$ & $99.0$ & $24.6$ & NB816 & UDSXDS & 1 & This study \\
HSC J021742--052810 & 02:17:42.18 & $-$05:28:10.64 & $5.679$ & $99.0$ & $28.9$ & $27.2$ & $26.3$ & $26.0$ & $25.1$ & NB816 & UDSXDS & 1 & H19 \\
HSC J021743--052807 & 02:17:43.34 & $-$05:28:07.05 & $5.685$ & $28.1$ & $27.4$ & $26.2$ & $26.0$ & $26.1$ & $24.0$ & NB816 & UDSXDS & 1 & O08 \\
HSC J021745--052842 & 02:17:45.03 & $-$05:28:42.58 & $5.751$ & $99.0$ & $28.9$ & $26.7$ & $25.6$ & $25.9$ & $24.4$ & NB816 & UDSXDS & 1 & O08 \\
HSC J021745--052936 & 02:17:45.25 & $-$05:29:36.17 & $5.688$ & $99.0$ & $99.0$ & $26.9$ & $26.4$ & $27.4$ & $24.2$ & NB816 & UDSXDS & 1 & O08 \\
HSC J021745--044129 & 02:17:45.75 & $-$04:41:29.34 & $5.674$ & $28.1$ & $99.0$ & $26.9$ & $29.4$ & $27.9$ & $24.9$ & NB816 & UDSXDS & 1 & This study \\
HSC J021748--053127 & 02:17:48.47 & $-$05:31:27.17 & $5.690$ & $99.0$ & $29.3$ & $27.4$ & $26.5$ & $26.3$ & $24.6$ & NB816 & UDSXDS & 1 & O08 \\
HSC J021749--052854 & 02:17:49.12 & $-$05:28:54.24 & $5.696$ & $28.6$ & $28.6$ & $26.4$ & $25.6$ & $25.8$ & $24.0$ & NB816 & UDSXDS & 1 & O08 \\
HSC J021749--052708 & 02:17:49.99 & $-$05:27:08.36 & $5.693$ & $99.0$ & $99.0$ & $27.8$ & $26.7$ & $25.9$ & $24.2$ & NB816 & UDSXDS & 1 & O08 \\
HSC J021750--050203 & 02:17:50.87 & $-$05:02:03.33 & $5.708$ & $28.8$ & $27.9$ & $27.1$ & $26.7$ & $27.5$ & $24.7$ & NB816 & UDSXDS & 1 & This study \\
HSC J021751--053003 & 02:17:51.15 & $-$05:30:03.78 & $5.712$ & $99.0$ & $29.6$ & $27.9$ & $26.4$ & $25.9$ & $25.1$ & NB816 & UDSXDS & 1 & O08 \\
HSC J021752--053511 & 02:17:52.64 & $-$05:35:11.79 & $5.756$ & $28.2$ & $27.5$ & $25.3$ & $24.7$ & $24.6$ & $23.1$ & NB816 & UDSXDS & 1 & S18 \\
HSC J021755--043251 & 02:17:55.41 & $-$04:32:51.63 & $5.690$ & $99.0$ & $30.2$ & $27.1$ & $26.9$ & $26.7$ & $24.6$ & NB816 & UDSXDS & 1 & This study \\
HSC J021757--050844 & 02:17:57.58 & $-$05:08:44.70 & $6.595$ & $99.0$ & $99.0$ & $99.0$ & $25.6$ & $25.6$ & $23.3$ & NB921 & UDSXDS & 1 & O10 \\
HSC J021757--053309 & 02:17:57.67 & $-$05:33:09.50 & $5.749$ & $99.0$ & $29.0$ & $27.1$ & $26.6$ & $25.6$ & $24.3$ & NB816 & UDSXDS & 1 & H19 \\
HSC J021758--043030 & 02:17:58.92 & $-$04:30:30.46 & $5.689$ & $99.0$ & $99.0$ & $26.7$ & $26.1$ & $25.8$ & $24.2$ & NB816 & UDSXDS & 1 & This study \\
HSC J021800--053518 & 02:18:00.73 & $-$05:35:18.99 & $5.673$ & $99.0$ & $30.9$ & $26.4$ & $25.3$ & $25.0$ & $24.8$ & NB816 & UDSXDS & 1 & H19 \\
HSC J021802--052011 & 02:18:02.19 & $-$05:20:11.58 & $5.718$ & $99.0$ & $29.8$ & $28.5$ & $28.6$ & $99.0$ & $25.3$ & NB816 & UDSXDS & 1 & H19 \\
HSC J021803--052643 & 02:18:03.89 & $-$05:26:43.51 & $5.747$ & $99.0$ & $28.5$ & $27.5$ & $27.1$ & $29.3$ & $24.7$ & NB816 & UDSXDS & 1 & H19 \\
HSC J021804--052147 & 02:18:04.18 & $-$05:21:47.31 & $5.734$ & $99.0$ & $29.1$ & $26.9$ & $25.6$ & $26.3$ & $24.4$ & NB816 & UDSXDS & 1 & H19 \\
HSC J021805--052704 & 02:18:05.19 & $-$05:27:04.20 & $5.746$ & $99.0$ & $99.0$ & $26.9$ & $27.3$ & $26.6$ & $24.8$ & NB816 & UDSXDS & 1 & H19 \\
HSC J021805--052027 & 02:18:05.30 & $-$05:20:27.04 & $5.742$ & $99.0$ & $99.0$ & $28.6$ & $26.8$ & $99.0$ & $25.0$ & NB816 & UDSXDS & 1 & H19 \\
HSC J021806--042847 & 02:18:06.16 & $-$04:28:47.66 & $5.726$ & $31.6$ & $99.0$ & $28.1$ & $27.9$ & $26.9$ & $25.3$ & NB816 & UDSXDS & 1 & This study \\
HSC J021806--044510 & 02:18:06.23 & $-$04:45:10.90 & $6.576$ & $28.1$ & $28.8$ & $27.8$ & $26.7$ & $25.4$ & $24.4$ & NB921 & UDSXDS & 1 & This study \\
HSC J021810--053707 & 02:18:10.68 & $-$05:37:07.93 & $5.747$ & $99.0$ & $28.1$ & $26.8$ & $27.1$ & $26.4$ & $24.3$ & NB816 & UDSXDS & 1 & This study \\
HSC J021814--053249 & 02:18:14.41 & $-$05:32:49.15 & $5.673$ & $29.3$ & $30.8$ & $26.9$ & $25.9$ & $25.9$ & $24.9$ & NB816 & UDSXDS & 1 & O08 \\
HSC J021819--050355 & 02:18:19.58 & $-$05:03:55.46 & $5.743$ & $99.0$ & $99.0$ & $27.0$ & $26.2$ & $25.8$ & $24.7$ & NB816 & UDSXDS & 1 & This study \\
HSC J021820--051109 & 02:18:20.70 & $-$05:11:09.95 & $6.575$ & $99.0$ & $28.8$ & $99.0$ & $27.4$ & $26.8$ & $24.7$ & NB921 & UDSXDS & 1 & O10 \\
HSC J021822--042926 & 02:18:22.92 & $-$04:29:26.05 & $5.697$ & $99.0$ & $99.0$ & $29.7$ & $99.0$ & $99.0$ & $25.4$ & NB816 & UDSXDS & 1 & This study \\
HSC J021823--043524 & 02:18:23.54 & $-$04:35:24.03 & $6.519$ & $99.0$ & $99.0$ & $99.0$ & $26.5$ & $25.5$ & $24.6$ & NB921 & UDSXDS & 1 & O10 \\
HSC J021825--050700 & 02:18:25.43 & $-$05:07:00.63 & $5.683$ & $29.2$ & $99.0$ & $99.0$ & $29.3$ & $99.0$ & $25.5$ & NB816 & UDSXDS & 1 & This study \\
HSC J021827--050726 & 02:18:27.01 & $-$05:07:26.98 & $6.554$ & $29.0$ & $99.0$ & $28.0$ & $27.7$ & $99.0$ & $25.0$ & NB921 & UDSXDS & 1 & O10 \\
HSC J021827--043508 & 02:18:27.02 & $-$04:35:08.06 & $6.511$ & $29.4$ & $29.5$ & $27.7$ & $25.9$ & $25.5$ & $24.5$ & NB921 & UDSXDS & 1 & O10 \\
HSC J021827--044737 & 02:18:27.45 & $-$04:47:37.13 & $5.703$ & $99.0$ & $99.0$ & $26.6$ & $26.5$ & $26.1$ & $23.7$ & NB816 & UDSXDS & 1 & This study \\
HSC J021828--051423 & 02:18:28.89 & $-$05:14:23.09 & $5.737$ & $28.5$ & $99.0$ & $26.8$ & $26.7$ & $26.3$ & $23.9$ & NB816 & UDSXDS & 1 & H19 \\
HSC J021830--051457 & 02:18:30.53 & $-$05:14:57.79 & $5.688$ & $99.0$ & $28.5$ & $26.3$ & $26.5$ & $26.7$ & $23.9$ & NB816 & UDSXDS & 1 & H19 \\
HSC J021830--052950 & 02:18:30.76 & $-$05:29:50.39 & $5.706$ & $99.0$ & $28.5$ & $26.6$ & $26.7$ & $25.5$ & $24.1$ & NB816 & UDSXDS & 1 & This study \\
HSC J021835--042321 & 02:18:35.95 & $-$04:23:21.69 & $5.755$ & $99.0$ & $99.0$ & $25.7$ & $25.4$ & $25.8$ & $23.2$ & NB816 & UDSXDS & 1 & S18 \\
HSC J021836--053528 & 02:18:36.38 & $-$05:35:28.14 & $5.698$ & $28.6$ & $99.0$ & $26.2$ & $26.0$ & $25.1$ & $23.6$ & NB816 & UDSXDS & 1 & This study \\
HSC J021842--043011 & 02:18:42.59 & $-$04:30:11.40 & $6.548$ & $30.4$ & $29.8$ & $99.0$ & $26.7$ & $26.8$ & $24.6$ & NB921 & UDSXDS & 1 & This study \\
HSC J021843--050915 & 02:18:43.64 & $-$05:09:15.71 & $6.512$ & $28.7$ & $28.5$ & $28.1$ & $26.0$ & $25.8$ & $23.8$ & NB921 & UDSXDS & 1 & S18 \\
HSC J021844--043636 & 02:18:44.66 & $-$04:36:36.27 & $6.621$ & $30.0$ & $99.0$ & $99.0$ & $26.2$ & $27.3$ & $24.4$ & NB921 & UDSXDS & 1 & O10 \\
HSC J021845--052915 & 02:18:45.03 & $-$05:29:15.91 & $5.660$ & $99.0$ & $28.9$ & $26.7$ & $25.9$ & $26.0$ & $24.9$ & NB816 & UDSXDS & 1 & This study \\
HSC J021846--043722 & 02:18:46.66 & $-$04:37:22.48 & $5.735$ & $99.0$ & $99.0$ & $27.6$ & $28.1$ & $26.8$ & $24.7$ & NB816 & UDSXDS & 1 & This study \\
HSC J021848--051715 & 02:18:48.23 & $-$05:17:15.65 & $5.741$ & $27.9$ & $99.0$ & $26.8$ & $26.1$ & $25.7$ & $24.5$ & NB816 & UDSXDS & 1 & This study \\
HSC J021849--052235 & 02:18:49.01 & $-$05:22:35.37 & $5.718$ & $99.0$ & $99.0$ & $26.5$ & $26.4$ & $26.4$ & $24.4$ & NB816 & UDSXDS & 1 & This study \\
HSC J021850--053007 & 02:18:50.99 & $-$05:30:07.79 & $5.705$ & $99.0$ & $99.0$ & $27.1$ & $26.3$ & $25.4$ & $24.2$ & NB816 & UDSXDS & 1 & This study \\
HSC J021857--045648 & 02:18:57.32 & $-$04:56:48.95 & $5.681$ & $99.0$ & $99.0$ & $28.0$ & $28.0$ & $99.0$ & $25.3$ & NB816 & UDSXDS & 1 & This study \\
HSC J021859--052916 & 02:18:59.93 & $-$05:29:16.77 & $5.674$ & $99.0$ & $28.1$ & $25.7$ & $24.8$ & $24.6$ & $24.1$ & NB816 & UDSXDS & 1 & This study \\
HSC J021901--045859 & 02:19:01.43 & $-$04:58:59.06 & $6.553$ & $30.8$ & $31.4$ & $29.6$ & $26.8$ & $26.4$ & $24.6$ & NB921 & UDSXDS & 1 & This study \\
HSC J021911--045707 & 02:19:11.04 & $-$04:57:07.60 & $5.704$ & $99.0$ & $99.0$ & $27.8$ & $29.0$ & $99.0$ & $25.2$ & NB816 & UDSXDS & 1 & This study \\
HSC J021933--050820 & 02:19:33.13 & $-$05:08:20.83 & $6.590$ & $99.0$ & $99.0$ & $99.0$ & $27.2$ & $26.8$ & $24.9$ & NB921 & UDSXDS & 1 & This study \\
HSC J021943--044914 & 02:19:43.92 & $-$04:49:14.43 & $5.684$ & $99.0$ & $99.0$ & $27.9$ & $27.0$ & $99.0$ & $25.5$ & NB816 & UDSXDS & 1 & This study \\
HSC J022001--045839 & 02:20:01.06 & $-$04:58:39.73 & $5.682$ & $99.0$ & $99.0$ & $27.5$ & $28.6$ & $26.5$ & $25.4$ & NB816 & UDSXDS & 1 & This study \\
HSC J022001--051637 & 02:20:01.11 & $-$05:16:37.43 & $5.711$ & $99.0$ & $29.2$ & $26.2$ & $25.8$ & $25.5$ & $23.4$ & NB816 & UDSXDS & 1 & S18 \\
HSC J022003--045416 & 02:20:03.20 & $-$04:54:16.58 & $5.697$ & $99.0$ & $99.0$ & $26.4$ & $25.3$ & $25.1$ & $23.9$ & NB816 & UDSXDS & 1 & This study \\
HSC J022012--044950 & 02:20:12.14 & $-$04:49:50.97 & $5.681$ & $30.7$ & $28.2$ & $26.9$ & $27.5$ & $99.0$ & $24.8$ & NB816 & UDSXDS & 1 & O08 \\
HSC J022013--045109 & 02:20:13.33 & $-$04:51:09.49 & $5.744$ & $27.9$ & $27.5$ & $26.3$ & $25.9$ & $25.4$ & $24.0$ & NB816 & UDSXDS & 1 & O08 \\
HSC J022026--050542 & 02:20:26.83 & $-$05:05:42.43 & $6.566$ & $99.0$ & $99.0$ & $99.0$ & $27.0$ & $99.0$ & $25.1$ & NB921 & UDSXDS & 1 & This study \\
HSC J022026--045217 & 02:20:26.88 & $-$04:52:17.86 & $5.720$ & $99.0$ & $28.8$ & $26.5$ & $26.1$ & $25.5$ & $24.8$ & NB816 & UDSXDS & 1 & This study \\
HSC J095919+020322  & 09:59:19.74 &    02:03:22.02 & $5.704$ & $99.0$ & $99.0$ & $28.0$ & $28.3$ & $27.4$ & $24.5$ & NB816 & UDCOSMOS & 1 & M12 \\
HSC J095922+021029  & 09:59:22.27 &    02:10:29.28 & $5.716$ & $29.2$ & $29.0$ & $27.2$ & $28.6$ & $27.5$ & $25.2$ & NB816 & UDCOSMOS & 1 & This study \\
HSC J095929+022950  & 09:59:29.35 &    02:29:50.17 & $4.840$ & $99.0$ & $28.0$ & $25.8$ & $26.0$ & $25.9$ & $24.7$ & NB718 & UDCOSMOS & 1 & M12 \\
HSC J095930+021642  & 09:59:30.08 &    02:16:42.79 & $5.680$ & $28.0$ & $27.7$ & $27.0$ & $26.7$ & $26.8$ & $25.1$ & NB816 & UDCOSMOS & 1 & M12 \\
HSC J095933+024955  & 09:59:33.43 &    02:49:55.93 & $5.724$ & $29.7$ & $27.8$ & $26.6$ & $27.0$ & $28.4$ & $24.1$ & NB816 & UDCOSMOS & 1 & M12 \\
HSC J095944+020050  & 09:59:44.06 &    02:00:50.62 & $5.688$ & $31.6$ & $30.0$ & $26.9$ & $26.2$ & $26.9$ & $24.4$ & NB816 & UDCOSMOS & 1 & M12 \\
HSC J095945+022807  & 09:59:45.97 &    02:28:07.84 & $2.174$ & $24.9$ & $24.9$ & $24.8$ & $24.7$ & $24.9$ & $24.1$ & NB387 & DCOSMOS & 1 & S14 \\
HSC J095946+014353  & 09:59:46.13 &    01:43:53.09 & $5.717$ & $99.0$ & $99.0$ & $27.4$ & $27.3$ & $27.0$ & $25.1$ & NB816 & UDCOSMOS & 1 & M12 \\
HSC J095952+015005  & 09:59:52.02 &    01:50:05.75 & $5.744$ & $99.0$ & $29.5$ & $26.4$ & $25.5$ & $25.4$ & $24.1$ & NB816 & UDCOSMOS & 1 & M12 \\
HSC J095952+013723  & 09:59:52.13 &    01:37:23.18 & $5.724$ & $99.0$ & $30.6$ & $26.5$ & $26.0$ & $26.0$ & $24.1$ & NB816 & UDCOSMOS & 1 & M12 \\
HSC J095953+020705  & 09:59:53.25 &    02:07:05.35 & $5.692$ & $29.2$ & $28.4$ & $25.8$ & $24.6$ & $24.5$ & $23.6$ & NB816 & UDCOSMOS & 1 & M12 \\
HSC J095954+022629  & 09:59:54.39 &    02:26:29.99 & $2.194$ & $25.0$ & $24.6$ & $24.3$ & $24.0$ & $23.7$ & $23.1$ & NB387 & UDCOSMOS & 1 & S14 \\
HSC J095954+021516  & 09:59:54.52 &    02:15:16.56 & $5.688$ & $99.0$ & $99.0$ & $27.1$ & $26.9$ & $27.6$ & $24.8$ & NB816 & UDCOSMOS & 1 & M12 \\
HSC J095954+021039  & 09:59:54.77 &    02:10:39.26 & $5.662$ & $33.7$ & $28.6$ & $26.7$ & $26.5$ & $26.4$ & $24.9$ & NB816 & UDCOSMOS & 1 & M12 \\
HSC J095955+014720  & 09:59:55.00 &    01:47:20.65 & $5.715$ & $99.0$ & $99.0$ & $26.7$ & $26.0$ & $25.9$ & $24.1$ & NB816 & UDCOSMOS & 1 & M12 \\
HSC J100004+020845  & 10:00:04.17 &    02:08:45.65 & $4.840$ & $99.0$ & $26.7$ & $25.0$ & $25.2$ & $25.2$ & $24.0$ & NB718 & UDCOSMOS & 1 & M12 \\
HSC J100005+020717  & 10:00:05.05 &    02:07:17.00 & $5.704$ & $99.0$ & $99.0$ & $27.9$ & $27.2$ & $27.9$ & $24.3$ & NB816 & UDCOSMOS & 1 & M12 \\
HSC J100019+020103  & 10:00:19.98 &    02:01:03.21 & $5.645$ & $99.0$ & $99.0$ & $27.2$ & $26.9$ & $26.6$ & $25.6$ & NB816 & UDCOSMOS & 1 & M12 \\
HSC J100029+024115  & 10:00:29.12 &    02:41:15.66 & $5.735$ & $99.0$ & $27.3$ & $26.5$ & $26.4$ & $26.8$ & $23.4$ & NB816 & UDCOSMOS & 1 & M12 \\
HSC J100029+015000  & 10:00:29.58 &    01:50:00.69 & $5.707$ & $30.6$ & $29.4$ & $27.3$ & $26.5$ & $26.9$ & $24.9$ & NB816 & UDCOSMOS & 1 & M12 \\
HSC J100030+021714  & 10:00:30.40 &    02:17:14.78 & $5.695$ & $99.0$ & $99.0$ & $27.6$ & $26.8$ & $26.9$ & $24.4$ & NB816 & UDCOSMOS & 1 & M12 \\
HSC J100030+013621  & 10:00:30.43 &    01:36:21.70 & $4.844$ & $29.4$ & $27.7$ & $25.8$ & $26.0$ & $26.1$ & $24.8$ & NB718 & UDCOSMOS & 1 & M12 \\
HSC J100034+013616  & 10:00:34.61 &    01:36:16.15 & $4.902$ & $28.8$ & $27.1$ & $25.5$ & $25.4$ & $25.2$ & $24.4$ & NB718 & UDCOSMOS & 1 & M12 \\
HSC J100040+021903  & 10:00:40.23 &    02:19:03.66 & $5.719$ & $99.0$ & $30.5$ & $27.3$ & $26.8$ & $27.4$ & $24.5$ & NB816 & UDCOSMOS & 1 & M12 \\
HSC J100041+022637  & 10:00:41.08 &    02:26:37.33 & $4.867$ & $99.0$ & $27.4$ & $25.7$ & $25.8$ & $26.0$ & $24.0$ & NB718 & UDCOSMOS & 1 & M12 \\
HSC J100042+022019  & 10:00:42.19 &    02:20:19.55 & $5.661$ & $28.1$ & $27.9$ & $26.7$ & $26.2$ & $25.9$ & $24.8$ & NB816 & UDCOSMOS & 1 & This study \\
HSC J100055+021309  & 10:00:55.43 &    02:13:09.13 & $4.872$ & $99.0$ & $29.4$ & $26.2$ & $26.4$ & $28.4$ & $23.9$ & NB718 & UDCOSMOS & 1 & M12 \\
HSC J100055+013630  & 10:00:55.52 &    01:36:30.84 & $5.670$ & $29.8$ & $28.6$ & $26.8$ & $26.4$ & $26.4$ & $25.2$ & NB816 & UDCOSMOS & 1 & M12 \\
HSC J100058+014815  & 10:00:58.00 &    01:48:15.08 & $6.604$ & $99.0$ & $30.9$ & $29.8$ & $25.5$ & $24.8$ & $23.1$ & NB921 & UDCOSMOS & 1 & S15 \\
HSC J100058+013642  & 10:00:58.41 &    01:36:42.77 & $5.688$ & $32.0$ & $28.1$ & $27.0$ & $27.2$ & $26.1$ & $24.7$ & NB816 & UDCOSMOS & 1 & M12 \\
HSC J100102+015144  & 10:01:02.96 &    01:51:44.74 & $5.666$ & $31.9$ & $28.1$ & $26.6$ & $26.2$ & $26.2$ & $24.7$ & NB816 & UDCOSMOS & 1 & M12 \\
HSC J100107+015222  & 10:01:07.35 &    01:52:22.69 & $5.668$ & $99.0$ & $99.0$ & $27.4$ & $27.0$ & $99.0$ & $25.5$ & NB816 & UDCOSMOS & 1 & M12 \\
HSC J100109+021513  & 10:01:09.72 &    02:15:13.47 & $5.712$ & $28.1$ & $29.1$ & $25.9$ & $25.9$ & $25.8$ & $23.1$ & NB816 & UDCOSMOS & 1 & M12 \\
HSC J100110+022829  & 10:01:10.06 &    02:28:29.03 & $5.681$ & $99.0$ & $28.0$ & $26.1$ & $25.0$ & $25.0$ & $24.2$ & NB816 & UDCOSMOS & 1 & M12 \\
HSC J100122+022249  & 10:01:22.45 &    02:22:49.83 & $4.871$ & $29.8$ & $29.0$ & $26.5$ & $26.6$ & $26.5$ & $24.5$ & NB718 & UDCOSMOS & 1 & M12 \\
HSC J100123+015600  & 10:01:23.84 &    01:56:00.29 & $5.726$ & $99.0$ & $99.0$ & $26.5$ & $25.9$ & $25.9$ & $23.6$ & NB816 & UDCOSMOS & 1 & M12 \\
HSC J100124+023145  & 10:01:24.79 &    02:31:45.48 & $6.541$ & $29.9$ & $28.5$ & $27.5$ & $25.6$ & $25.6$ & $23.5$ & NB921 & UDCOSMOS & 1 & S15 \\
HSC J100126+014430  & 10:01:26.89 &    01:44:30.15 & $5.686$ & $99.0$ & $99.0$ & $26.7$ & $26.3$ & $26.0$ & $24.6$ & NB816 & UDCOSMOS & 1 & M12 \\
HSC J100127+023005  & 10:01:27.76 &    02:30:05.89 & $5.696$ & $29.4$ & $28.4$ & $26.7$ & $26.0$ & $26.1$ & $24.2$ & NB816 & UDCOSMOS & 1 & M12 \\
HSC J100129+014929  & 10:01:29.08 &    01:49:29.79 & $5.707$ & $99.0$ & $29.7$ & $26.2$ & $25.6$ & $25.5$ & $23.1$ & NB816 & UDCOSMOS & 1 & M12 \\
HSC J100131+023105  & 10:01:31.07 &    02:31:05.81 & $5.690$ & $30.5$ & $99.0$ & $26.9$ & $26.9$ & $26.8$ & $24.4$ & NB816 & UDCOSMOS & 1 & M12 \\
HSC J100131+014320  & 10:01:31.12 &    01:43:20.31 & $5.728$ & $29.7$ & $28.4$ & $26.8$ & $26.3$ & $26.5$ & $24.4$ & NB816 & UDCOSMOS & 1 & M12 \\
HSC J100145+015712  & 10:01:45.12 &    01:57:12.23 & $4.909$ & $99.0$ & $26.6$ & $25.1$ & $25.0$ & $24.9$ & $23.6$ & NB718 & UDCOSMOS & 1 & M12 \\
HSC J100146+014827  & 10:01:46.64 &    01:48:27.07 & $5.704$ & $99.0$ & $28.3$ & $27.0$ & $27.0$ & $27.6$ & $24.4$ & NB816 & UDCOSMOS & 1 & This study \\
HSC J100153+020459  & 10:01:53.45 &    02:04:59.79 & $6.931$ & $30.4$ & $28.8$ & $27.7$ & $27.2$ & $25.5$ & $24.3$ & NB973 & UDCOSMOS & 1 & H17 \\
HSC J100205+020646  & 10:02:05.96 &    02:06:46.15 & $6.936$ & $99.0$ & $99.0$ & $99.0$ & $99.0$ & $25.3$ & $24.0$ & NB973 & UDCOSMOS & 1 & H17 \\
HSC J100207+023217  & 10:02:07.81 &    02:32:17.18 & $6.616$ & $99.0$ & $30.7$ & $30.5$ & $27.0$ & $26.6$ & $24.9$ & NB921 & UDCOSMOS & 1 & This study \\
HSC J100208+015445  & 10:02:08.80 &    01:54:45.04 & $5.676$ & $29.4$ & $27.8$ & $26.1$ & $26.0$ & $25.7$ & $24.3$ & NB816 & UDCOSMOS & 1 & M12 \\
HSC J100214+021242  & 10:02:14.21 &    02:12:42.92 & $5.731$ & $31.7$ & $99.0$ & $27.4$ & $26.6$ & $26.8$ & $24.6$ & NB816 & UDCOSMOS & 1 & This study \\
HSC J100215+024033  & 10:02:15.51 &    02:40:33.39 & $6.965$ & $27.5$ & $27.3$ & $27.2$ & $27.4$ & $25.2$ & $23.4$ & NB973 & UDCOSMOS & 1 & Z20 \\
HSC J100235+021213  & 10:02:35.37 &    02:12:13.89 & $6.593$ & $27.0$ & $26.8$ & $26.4$ & $25.1$ & $25.2$ & $23.1$ & NB921 & UDCOSMOS & 1 & H16 \\
HSC J100301+020235  & 10:03:01.14 &    02:02:35.99 & $5.682$ & $99.0$ & $28.8$ & $26.1$ & $25.0$ & $25.0$ & $24.1$ & NB816 & UDCOSMOS & 1 & M12 \\
HSC J100301+015011  & 10:03:01.81 &    01:50:11.15 & $5.695$ & $28.3$ & $27.0$ & $25.5$ & $25.1$ & $24.6$ & $23.5$ & NB816 & UDCOSMOS & 1 & M12 \\
HSC J100305+015141  & 10:03:05.33 &    01:51:41.09 & $5.694$ & $30.6$ & $29.0$ & $27.2$ & $26.5$ & $26.0$ & $24.7$ & NB816 & UDCOSMOS & 1 & M12 \\
HSC J100306+014742  & 10:03:06.12 &    01:47:42.69 & $5.680$ & $99.0$ & $99.0$ & $26.9$ & $26.9$ & $27.0$ & $24.7$ & NB816 & UDCOSMOS & 1 & M12 \\
HSC J100334+024546  & 10:03:34.67 &    02:45:46.58 & $6.586$ & $28.3$ & $28.1$ & $99.0$ & $25.6$ & $25.3$ & $23.8$ & NB921 & DCOSMOS & 1 & S18 \\
HSC J160107+550720  & 16:01:07.45 &    55:07:20.58 & $6.573$ & $28.9$ & $99.0$ & $99.0$ & $26.0$ & $25.5$ & $23.7$ & NB921 & DELAISN1 & 1 & S18 \\
HSC J160707+555347  & 16:07:07.45 &    55:53:47.82 & $6.564$ & $99.0$ & $99.0$ & $99.0$ & $26.2$ & $99.0$ & $24.0$ & NB921 & DELAISN1 & 1 & S18 \\
HSC J160940+541409  & 16:09:40.24 &    54:14:09.06 & $6.575$ & $27.8$ & $27.6$ & $27.1$ & $25.6$ & $26.0$ & $23.9$ & NB921 & DELAISN1 & 1 & S18 \\
HSC J162126+545719  & 16:21:26.50 &    54:57:19.09 & $6.545$ & $28.3$ & $99.0$ & $27.9$ & $24.5$ & $24.0$ & $22.4$ & NB921 & DELAISN1 & 1 & S18 \\
HSC J232558--002557 & 23:25:58.44 & $-$00:25:57.56 & $5.703$ & $99.0$ & $28.1$ & $25.6$ & $25.2$ & $25.5$ & $23.3$ & NB816 & DDEEP23 & 1 & S18 \\
HSC J233125--005216 & 23:31:25.36 & $-$00:52:16.49 & $6.559$ & $99.0$ & $99.0$ & $99.0$ & $25.5$ & $99.0$ & $23.1$ & NB921 & DDEEP23 & 1 & S18 \\
HSC J233408--004403 & 23:34:08.80 & $-$00:44:03.69 & $5.707$ & $28.1$ & $99.0$ & $25.5$ & $25.9$ & $26.6$ & $22.7$ & NB816 & DDEEP23 & 1 & S18 \\
HSC J233454--003603 & 23:34:54.95 & $-$00:36:03.97 & $5.732$ & $31.2$ & $27.5$ & $25.5$ & $25.2$ & $24.9$ & $23.0$ & NB816 & DDEEP23 & 1 & S18 \\
HSC J022028--045802 & 02:20:28.96 & $-$04:58:02.81 & $2.157$ & $20.6$ & $20.7$ & $20.4$ & $20.1$ & $20.2$ & $19.1$ & NB387 & DSXDS & 2 & L20 \\
HSC J022120--032353 & 02:21:20.93 & $-$03:23:53.52 & $2.183$ & $22.9$ & $22.6$ & $22.6$ & $22.3$ & $99.0$ & $21.2$ & NB387 & DSXDS & 2 & L20 \\
HSC J022121--034338 & 02:21:21.16 & $-$03:43:38.10 & $2.176$ & $23.2$ & $23.2$ & $23.1$ & $22.7$ & $22.7$ & $21.8$ & NB387 & DSXDS & 2 & L20 \\
HSC J022202--050351 & 02:22:02.59 & $-$05:03:51.98 & $2.161$ & $21.8$ & $21.7$ & $21.6$ & $21.3$ & $21.6$ & $20.4$ & NB387 & DSXDS & 2 & L20 \\
HSC J022219--052231 & 02:22:19.95 & $-$05:22:31.54 & $2.204$ & $20.4$ & $20.1$ & $20.2$ & $19.9$ & $20.1$ & $19.3$ & NB387 & DSXDS & 2 & L20 \\
HSC J022230--033545 & 02:22:30.42 & $-$03:35:45.37 & $2.172$ & $22.1$ & $22.2$ & $22.0$ & $21.5$ & $21.8$ & $19.9$ & NB387 & DSXDS & 2 & L20 \\
HSC J022312--050625 & 02:23:12.46 & $-$05:06:25.07 & $2.190$ & $20.1$ & $20.1$ & $20.1$ & $19.8$ & $19.9$ & $19.1$ & NB387 & DSXDS & 2 & L20 \\
HSC J022349--033930 & 02:23:49.43 & $-$03:39:30.65 & $2.141$ & $21.1$ & $20.8$ & $20.8$ & $20.5$ & $20.8$ & $19.9$ & NB387 & DSXDS & 2 & L20 \\
HSC J022351--044730 & 02:23:51.07 & $-$04:47:30.05 & $2.167$ & $20.7$ & $20.6$ & $20.7$ & $20.3$ & $20.4$ & $19.5$ & NB387 & DSXDS & 2 & L20 \\
HSC J095508+011205  & 09:55:08.45 &    01:12:05.71 & $2.166$ & $21.9$ & $21.9$ & $21.7$ & $21.3$ & $21.5$ & $19.9$ & NB387 & DCOSMOS & 2 & L20 \\
HSC J095539+011316  & 09:55:39.50 &    01:13:16.30 & $2.142$ & $21.6$ & $21.6$ & $21.6$ & $21.1$ & $21.4$ & $20.3$ & NB387 & DCOSMOS & 2 & L20 \\
HSC J095714+013145  & 09:57:14.02 &    01:31:45.51 & $2.142$ & $21.7$ & $21.6$ & $21.4$ & $21.0$ & $21.2$ & $20.0$ & NB387 & DCOSMOS & 2 & L20 \\
HSC J095822+010806  & 09:58:22.03 &    01:08:06.39 & $2.171$ & $20.8$ & $20.7$ & $20.7$ & $20.3$ & $20.4$ & $19.4$ & NB387 & DCOSMOS & 2 & L20 \\
HSC J095930+024124  & 09:59:30.21 &    02:41:24.95 & $2.187$ & $22.0$ & $22.0$ & $21.6$ & $21.2$ & $21.4$ & $20.6$ & NB387 & DCOSMOS & 2 & L20 \\
HSC J100057+023932  & 10:00:57.79 &    02:39:32.48 & $3.360$ & $23.6$ & $22.9$ & $22.9$ & $22.8$ & $22.8$ & $22.0$ & NB527 & UDCOSMOS & 2 & Mas12 \\
HSC J100145+020244  & 10:01:45.97 &    02:02:44.35 & $4.888$ & $30.5$ & $25.8$ & $24.1$ & $24.1$ & $24.2$ & $22.6$ & NB718 & UDCOSMOS & 2 & Z20 \\
HSC J232459--001451 & 23:24:59.71 & $-$00:14:51.21 & $2.172$ & $21.4$ & $21.2$ & $21.2$ & $20.8$ & $21.0$ & $19.6$ & NB387 & DDEEP23 & 2 & L20 \\
HSC J232506--012203 & 23:25:06.73 & $-$01:22:03.29 & $2.181$ & $20.4$ & $20.4$ & $20.3$ & $20.1$ & $20.2$ & $19.1$ & NB387 & DDEEP23 & 2 & L20 \\
HSC J232619--000152 & 23:26:19.35 & $-$00:01:52.72 & $2.173$ & $22.1$ & $22.1$ & $21.6$ & $21.5$ & $21.7$ & $20.8$ & NB387 & DDEEP23 & 2 & L20 \\
HSC J232855--004212 & 23:28:55.78 & $-$00:42:12.22 & $2.198$ & $20.5$ & $20.4$ & $20.1$ & $19.8$ & $19.9$ & $19.1$ & NB387 & DDEEP23 & 2 & L20 \\
HSC J233159--000856 & 23:31:59.69 & $-$00:08:56.47 & $2.184$ & $20.8$ & $20.6$ & $20.6$ & $20.3$ & $20.4$ & $19.6$ & NB387 & DDEEP23 & 2 & L20 \\
HSC J233217--011416 & 23:32:17.04 & $-$01:14:16.90 & $2.168$ & $99.0$ & $21.1$ & $20.9$ & $20.6$ & $20.7$ & $19.8$ & NB387 & DDEEP23 & 2 & L20 
\enddata 
\tablecomments{
(1) Object ID. 
(2) Right Ascension. 
(3) Declination. 
(4) Spectroscopic redshift. 
(5)--(9) Apparent magnitudes with $2\farcs0$ diameter circular apertures in $g$, $r$, $i$, $z$, and $y$. 
(10) Apparent total $NB$ magnitude.  
(11) The LAE sample in which the source is selected. 
(12) Survey field. 
(13) Galaxy/AGN flag ($1 =$ galaxy; $2 =$ AGN). 
(14) Reference for spectroscopic redshifts: 
O08 $=$ \cite{2008ApJS..176..301O}; 
O10 $=$ \cite{2010ApJ...723..869O}; %2010.11
Mas12 $=$ \cite{2012ApJ...755..169M}; %2012.08
M12 $=$ \cite{2012ApJ...760..128M}; %2012.12
S14 $=$ \cite{2014ApJ...788...74S}; 
S15 $=$ \cite{2015ApJ...808..139S}; %2015.08
H16 $=$ \cite{2016ApJ...825L...7H}; %2016.07
T17 $=$ \cite{2017A&A...600A.110T}; %2017.04
H17 $=$ \cite{2017ApJ...845L..16H}; %2017.08
J17 $=$ \cite{2017ApJ...846..134J}; %2017.09
S18 $=$ \cite{2018PASJ...70S..15S}; %2018.01
H19 $=$ \cite{2019ApJ...883..142H}; %2019.10
Z20 $=$ \cite{2020ApJ...891..177Z}; %2020.03 
L20 $=$ \cite{2020ApJS..250....8L} %2020.09
}
\end{deluxetable*} 
%ttttttttttttttttttttttttttttttttttttttttttttttttttttttttttttttttttttttttt%

%ttttttttttttttttttttttttttttttttttttttttttttttttttttttttttttttttttttttttt%
\LongTables
\capstartfalse
\begin{deluxetable*}{cccccccccccccc} 
\tablecolumns{14} 
\tablewidth{0pt} 
\tablecaption{Same as Table \ref{tab:LAE_specz}, but for lower-$z$ AGNs 
whose strong {\sc Civ} emission is probed with our NBs
\label{tab:lowz_LAE_specz}}
\tablehead{
    \colhead{ID}     
    &  \colhead{R.A.}
    &  \colhead{Decl.}
    &  \colhead{$z_{\rm spec}$}
    &  \colhead{$g_{\rm ap}$}
    &  \colhead{$r_{\rm ap}$}
    &  \colhead{$i_{\rm ap}$}
    &  \colhead{$z_{\rm ap}$}
    &  \colhead{$y_{\rm ap}$}
    &  \colhead{$NB_{\rm tot}$}
    &  \colhead{Sample}
    &  \colhead{Field}
    &  \colhead{Flag}
    &  \colhead{Reference}
    \\
    \colhead{(1)}
    &  \colhead{(2)}
    &  \colhead{(3)}
    &  \colhead{(4)}
    &  \colhead{(5)}
    &  \colhead{(6)}
    &  \colhead{(7)}
    &  \colhead{(8)}
    &  \colhead{(9)}
    &  \colhead{(10)}
    &  \colhead{(11)}
    &  \colhead{(12)}
    &  \colhead{(13)}
    &  \colhead{(14)}
}
\startdata
HSC J022528--043642 & 02:25:28.08 & $-$04:36:42.00 & $1.504$ & $23.5$ & $23.5$ & $23.3$ & $23.0$ & $23.2$ & $22.4$ & NB387 & UDSXDS & 2 & L13 \\
HSC J022718--043134 & 02:27:18.86 & $-$04:31:34.01 & $1.503$ & $22.0$ & $21.6$ & $21.4$ & $21.2$ & $21.2$ & $20.3$ & NB387 & UDSXDS & 2 & L13 \\
HSC J095815+014923 & 09:58:15.50 & 01:49:23.03 & $1.507$ & $20.7$ & $20.3$ & $20.1$ & $20.1$ & $20.1$ & $19.2$ & NB387 & UDCOSMOS & 2 & L09 \\
HSC J095801+014832 & 09:58:01.45 & 01:48:32.86 & $2.402$ & $22.8$ & $22.7$ & $22.4$ & $22.1$ & $21.8$ & $21.1$ & NB527 & UDCOSMOS & 2 & L09 
\enddata 
\tablecomments{
(1) Object ID. 
(2) Right Ascension. 
(3) Declination. 
(4) Spectroscopic redshift. 
(5)--(9) Apparent magnitudes with $2\farcs0$ diameter circular apertures in $g$, $r$, $i$, $z$, and $y$. 
(10) Apparent total $NB$ magnitude.  
(11) The LAE sample in which the source is selected. 
(12) Survey field. 
(13) Galaxy/AGN flag ($1 =$ galaxy; $2 =$ AGN). 
(14) Reference for spectroscopic redshifts: 
L09 $=$ \cite{2009ApJS..184..218L}; 
L13  $=$ \cite{2013A&A...559A..14L}
}
\end{deluxetable*} 
%ttttttttttttttttttttttttttttttttttttttttttttttttttttttttttttttttttttttttt%

\end{document}